\documentclass[a4paper,11pt]{article}
\usepackage{jheppub}
\usepackage[utf8x]{inputenc}
\usepackage{graphicx}
\usepackage{float}
\usepackage[T1]{fontenc}
\usepackage{epsfig}
\usepackage{color}
\usepackage{amsmath,mathtools}
\usepackage{subfloat}
\usepackage{amsfonts}
\usepackage{braket}
\usepackage{cleveref}
\usepackage{epstopdf}
\usepackage{caption}
\usepackage{subcaption}
\usepackage{enumitem}
\usepackage{comment}
\usepackage[titletoc,toc,title]{appendix}
\usepackage{hyperref}
\hypersetup{
	colorlinks=true,
	linkcolor=blue,
	filecolor=red,      
	urlcolor=blue,
	citecolor=red
}

\DeclareMathOperator{\sech}{sech}

\title{Bridging Boundaries: $T\bar{T}$, Double Holography, and Reflected Entropy}

\author[]{Debarshi Basu,$^{a,b}$}
\author[]{Himanshu Chourasiya,$^{b}$}
\author[]{Ankur Dey,$^{b}$ and}
\author[]{Vinayak Raj$^{b,c}$}

\affiliation[]{
	$^a$Shing-Tung Yau Center and School of Physics, Southeast University,\\
	Nanjing 210096, China\\
	$^b$Department of Physics, Indian Institute of Technology,\\
	 Kanpur 208016, India\\
	$^c$School of Science, Huzhou University,\\ Huzhou 313000, Zhejiang, China}
\emailAdd{debarshi.128@gmail.com}
\emailAdd{chim@iitk.ac.in}
\emailAdd{ankurd21@iitk.ac.in}
\emailAdd{vinayak.hep.th@gmail.com}
\date{}

\abstract{
We investigate the reflected entropy for bipartite mixed state configurations in a $T\bar{T}$ deformed boundary conformal field theory in $2$ dimensions (BCFT$_2$). The bulk dual is described by asymptotically AdS$_3$ geometries with the cut off surface pushed deeper into the bulk and truncated by an end of the world brane.
We obtain the reflected entropy up to a linear order in the radial cut-off for static and time dependent configurations involving an eternal black hole, from the island and defect extremal surface (DES) prescriptions in the context of the deformed AdS/BCFT. 
We observe agreement of the leading order correction for all cases between the two prescriptions.
We also obtain the analogous of the Page curves for the reflected entropy and investigate the modification due to the $T\bar{T}$ deformation.
}

\begin{document}

\maketitle
	
\section{Introduction}\label{sec:intro}
The black hole information loss problem has been a central topic of research for several decades, spurring significant progress in semi-classical and quantum gravity. Recent studies have made impactful strides towards resolving this problem through the \textit{island} framework that provide access to regions inside black hole geometries at late times \cite{Almheiri:2019hni, Almheiri:2019psf, Almheiri:2019qdq, Almheiri:2019psy, Almheiri:2019yqk, Almheiri:2020cfm}. Through this framework, the Page Curve \cite{Page:1993wv, Page:1993df, Page:2013dx} for the von Neumann entropy of the Hawking radiation was reproduced for a CFT coupled to a theory of gravity by including the ``island'' inside the gravitational region in the entanglement wedge of the radiation bath. The emergence of these islands, is attributed to the late time dominance of \textit{replica wormhole} saddles in the gravitational path integral for the R\'enyi entanglement entropy. The so called island formula was inspired by the quantum extremal surface (QES) prescription \cite{Engelhardt:2014gca} which describes the quantum corrections to the holographic entanglement entropy \cite{Ryu:2006bv, Hubeny:2007xt, Faulkner:2013ana,Geng:2024xpj}.

A more elegant and insightful interpretation of the island formula may be provided by the \textit{double holography} framework \cite{Almheiri:2019hni, Sully:2020pza, Rozali:2019day, Chen:2020uac, Chen:2020hmv, Grimaldi:2022suv, Suzuki:2022xwv, Geng:2020qvw, Geng:2020fxl, Geng:2021iyq, Geng:2021mic, Geng:2021hlu,Geng:2024xpj,Karch:2022rvr}. This involves considering the CFT$_d$ coupled to the semiclassical gravity to be holographic and be described by a $(d+1)$-dimensional gravity theory. In the literature, this higher dimensional picture have been termed as the bulk description for the original setup which is termed as the brane description. This gravity-plus-matter system in the brane description may, in turn be described through a quantum mechanical system residing on the boundary of the CFT$_d$ which was dubbed as the boundary description. In this double holographic setup, the computation of the entanglement entropy through the island formula in the brane description translates to its holographic characterization through the Ryu-Takayanagi (RT) prescription in the bulk AdS$_{d+1}$ geometry.

Diverging from the above, boundary conformal field theories (BCFTs) \cite{Cardy:2004hm, Takayanagi:2011zk, Fujita:2011fp, Sully:2020pza, Kastikainen:2021ybu}, defined as CFTs on a manifold with a boundary, have garnered significant interest in recent years. The holographic dual to a BCFT$_2$ is described by an asymptotically AdS$_3$ spacetime bounded by an end-of-the-world (EOW) brane with Neumann boundary conditions. The holographic entanglement entropy formula \cite{Takayanagi:2011zk, Fujita:2011fp} for this framework involves modification of the homology condition to incorporate extremal surfaces anchored to the EOW brane as well. In \cite{Deng:2020ent}, the authors explore an extension of this AdS$_3$/BCFT$_2$ duality, where the Neumann boundary condition on the EOW brane is modified by addition of \textit{defect} conformal matter.\footnote{The term ``defect matter'' refers to degrees of freedom that are localized on a defect, a lower dimensional submanifold, within the full spacetime.} The works \cite{Deng:2020ent, Chu:2021gdb} introduce the defect extremal surface (DES) formula which extends the RT prescription by incorporating contributions from defect matter fields in the context of defect AdS$_3$/BCFT$_2$. Notably, the DES formula in this defect AdS$_3$/BCFT$_2$ has been proposed and verified as the doubly holographic counterpart of the island formula in the lower dimensional effective description.\footnote{Note that in \cite{Deng:2020ent, Chu:2021gdb, Li:2021dmf, Basu:2022reu, Shao:2022wrm}, the brane perspective described in the double holographic framework has been termed as the \textit{lower dimensional effective description} as the authors obtained the brane perspective through partial dimensional reduction of the bulk AdS$_3$ geometry with the EOW brane. We will use these two terminologies interchangeably in this article.}

While entanglement entropy is a suitable metric for investigating the entanglement structure of bipartite pure states, it fails to accurately characterize the mixed state entanglement structure. For systems involving bipartite mixed states, it is preferable to utilize alternative (computable) measures such as reflected entropy \cite{Dutta:2019gen, Akers:2021pvd}, entanglement negativity \cite{Vidal:2002zz, Plenio:2005cwa}, entanglement of purification \cite{Takayanagi:2017knl, Nguyen:2017yqw}, and balanced partial entanglement entropy \cite{Wen:2021qgx}. These measures offer a more nuanced characterization of the complex entanglement structure in mixed states.

While the original island framework was primarily developed for pure states, many physically relevant situations—such as subsystems within the Hawking radiation or black holes coupled to non-isolated environments—often lead to mixed states. As stated above, in such scenarios, it becomes imperative to generalize the island framework to accommodate mixed states, enabling a more detailed description of entanglement structures in gravitational settings. The authors in \cite{Li:2020ceg, Chandrasekaran:2020qtn} provided this generalization in terms of the reflected entropy \cite{Dutta:2019gen, Jeong:2019xdr}. Reflected entropy is based on canonical purification of the given mixed state and is described holographically in terms of the area of the bulk entanglement wedge cross-section (EWCS) \cite{Takayanagi:2017knl, Nguyen:2017yqw}. Subsequently, in \cite{Li:2021dmf}, a DES formula for the reflected entropy was forwarded which was shown to be the bulk counterpart of the island formula for the reflected entropy in the defect AdS$_3$/BCFT$_2$ scenario.\footnote{Similar DES formulas for entanglement negativity were also forwarded in \cite{Basu:2022reu, Shao:2022wrm} which provide double holographic description for the island formula for the entanglement negativity \cite{KumarBasak:2020ams}.} This equivalence provides a robust framework for studying the entanglement in mixed states within these holographic settings.

Motivated by the desire to test the robustness of the island framework for the reflected entropy and its doubly holographic generalizations, it is imperative to investigate them in less conventional gravitational settings. In this context, CFTs deformed by an irrelevant operator composed of the stress-energy tensor \cite{Zamolodchikov:2004ce, Cavaglia:2016oda, Smirnov:2016lqw} provide a fertile testing ground due to their solvable nature. These irrelevant deformations introduce a finite cutoff in the dual AdS$_3$ geometries \cite{McGough:2016lol} satisfying Dirichlet boundary conditions,\footnote{Note that alternative proposals for the holographic duality of these $T \bar{T}$ deformed theories have also been proposed in the literature, such as \cite{Hirano:2020nwq, Kraus:2018xrn, Guica:2019nzm,Dubovsky:2018bmo, Tolley:2019nmm}. It is expected that they will coincide under certain limits, however the exact matching for general cases remains a non-trivial open issue.} which have been studied extensively in the literature \cite{Shyam:2017znq, Kraus:2018xrn, Cottrell:2018skz, Taylor:2018xcy, Hartman:2018tkw, Shyam:2018sro, Caputa:2019pam, Giveon:2017myj, Asrat:2017tzd, Donnelly:2018bef, Lewkowycz:2019xse, Chen:2018eqk, Banerjee:2019ewu, Jeong:2019ylz, Murdia:2019fax, Park:2018snf, Asrat:2019end, He:2019vzf, Grieninger:2019zts, Khoeini-Moghaddam:2020ymm, Basu:2024bal, Basu:2023aqz, Basu:2023bov, Basu:2024enr,Grieninger:2023knz,Chang:2024voo}. We here consider such deformations of BCFTs which provide a natural toy model to investigate the island formalism. The dual AdS geometry with an EOW brane now also involves a finite cut-off, which adds subtlety to the standard AdS/BCFT correspondence and demands a careful analysis. While the island formula and its reflected entropy analogue have been primarily studied in standard asymptotically AdS spacetimes, it is not a priori clear whether their structure survives in such geometries with finite radial cutoff and modified boundary conditions. Investigating mixed state entanglement in such setups, particularly through computable measures like reflected entropy, therefore provides a test and extends the applicability of the island paradigm in a more intricate setting.

The two kind of boundaries discussed earlier, namely the EOW brane in the context of AdS/BCFT, and the asymptotic boundary at the finite cut-off surfaces in holographic dual of $T \bar{T}$ deformed theories, have very different nature because of distinct boundary conditions governing them. In order to elucidate these differences, the authors in \cite{Deng:2023pjs,Deng:2024dct} investigated a spacetime with both of these boundaries. In order to have a transparent boundary between the two surfaces, they introduced the same matter field on the EOW as is present at the asymptotic boundary in the dual picture. In the boundary description, this corresponds to a $T \bar T$ deformation of a BCFT$_2$, which may be obtained through a partial dimensional reduction of a braneworld gravity theory glued to a non-conformal bath with $T\bar{T}$ deformation \cite{Deng:2023pjs}. The authors investigated the entanglement structure of bipartite pure states in such a model through the entanglement entropy.

The above advancements underscore the importance of characterizing mixed state entanglement in AdS geometries that incorporate both an EOW brane and a finite radial cut-off surface. To this end, we utilize the reflected entropy to study bipartite mixed states in asymptotically AdS$_3$ geometries with a finite radial cut-off, which originates from $T\bar{T}$ deformation, and is truncated by an EOW brane. We utilize the DES formula for the reflected entropy to compute bulk reflected entropy through the EWCS and compare these with island computation in the lower dimensional effective description. Additionally, we extend our analysis to time-dependent configurations, involving black hole and radiation subsystems in an effective lower-dimensional theory.

The remainder of this article is structured as follows. In \cref{sec:review}, we briefly review various aspects of $T\bar{T}$ deformation, reflected entropy, and the island and DES formulas that are essential for computing reflected entropy. In \cref{sec:Zero_Temp}, we introduce the DES formula for calculating reflected entropy in asymptotically AdS$_3$ geometries with a finite radial cut-off bounded by an EOW brane. We compute the reflected entropy for both adjacent and disjoint intervals, considering various possible entanglement entropy phases in a static time slice, and demonstrate the equivalence of the DES and the island formula. In \cref{sec:Finite_Temp}, we extend our analysis to time-dependent configurations. We start by reviewing the essentials of the eternal black hole in the two-dimensional effective picture, followed by describing the DES and island computations for the reflected entropy between interior regions of the black hole, between the black hole and radiation in the bath system, and between radiation subsystems. Finally, we summarize our findings in \cref{sec:conc}.

\section{Review}\label{sec:review}

\subsection{$T \bar{T}$ deformed CFT}\label{TTbar-Review}
In this subsection we review the $T\bar{T}$ deformation in CFTs and its holographic dual. Consider 2d conformal field theory in a flat space given by the metric $ds^2=-dt^2+dx^2$. $T\bar{T}$ deformation of such a CFT$_2$ is defined by \cite{Smirnov:2016lqw,Cavaglia:2016oda}
\begin{align}
	\left. \frac{dI_{QFT}^{(\mu)}}{d\mu}=- 2 \pi \int d^2x (T\bar{T})_{\mu}, \ \  \ \ I_{QFT}^{(\mu)} \right|_{\mu=0}=I_{CFT},
\end{align}
where $I$ is the Lorentzian action and $\mu$ is the deformation parameter. $(T\bar{T})$ is the $T\bar{T}$ deformation operator, which is defined in terms of the components of the stress energy tensor $T^{ab}$ as
\begin{align}
	(T\bar{T})=\frac{1}{8} \left( T_{ab}T^{ab}-(T_a^a)^2 \right).
\end{align}
The author in \cite{Zamolodchikov:2004ce} have shown that the $T\bar{T}$ operator satisfies the factorization formula
\begin{align}
	\langle T\bar{T} \rangle=\frac{1}{8} \left( \langle T_{ab}\rangle \langle T^{ab}\rangle-\langle T_a^a\rangle ^2 \right).
\end{align}
In the bulk picture, the authors in \cite{McGough:2016lol,Lewkowycz:2019xse} have proposed the $T\bar{T}$ deformed CFT$_2$ in flat space to be dual to a quantum gravity in AdS$_3$ with a radial cut-off
\begin{align}
	ds^2=\frac{\ell^2}{z^2} \left( -dt^2+dx^2+dz^2 \right), \ \ \ \ z>z_c,
\end{align}
where Dirichlet boundary conditions are imposed at $z=z_c$. The deformation parameter is related to the radial cut-off as \cite{Lewkowycz:2019xse}
\begin{align}
	\mu=\frac{8G_N}{\ell} z_c^2=\frac{12}{c} z_c^2 .
\end{align}

\subsection{AdS$_3$/BCFT$_2$ and defect extremal surface}
 In this subsection, we briefly review the AdS$_3$/BCFT$_2$ correspondence, proposed in \cite{Takayanagi:2011zk,Fujita:2011fp}. A boundary conformal field theory (BCFT) is a conformal field theory (CFT) defined on a manifold $\mathcal{M}$ with boundary $\partial\mathcal{M}$, where conformal boundary conditions are specified. 
It was shown in \cite{Cardy:2004hm} that the holographic dual of a BCFT is given by an asymptotically AdS spacetime $\mathcal{N}$ truncated by an end-of-world (EOW) brane $\mathcal{Q}$ on which the Neumann boundary condition is imposed.
The bulk action is given by \cite{Takayanagi:2011zk,Fujita:2011fp}
\begin{align}\label{AdS/BCFT-action}
	I_0=-\frac{1}{16\pi G_N}\int_\mathcal{N}\sqrt{g}(R-2\Lambda)-\frac{1}{8\pi G_N}\int_\mathcal{Q}\sqrt{h}(K-(d-1)T)\,.
\end{align}
Here $K$ represents the trace of the extrinsic curvature on the EOW brane with tension $T$.
By varying the above action with respect to the induced metric on the brane $h_{ab}$, we can determine the brane trajectory, which consequently leads to the Neumann boundary conditions
\begin{align}
	K_{ab}=(K-T)h_{ab}\,.\label{NBC}
\end{align}
The AdS$_3$/BCFT$_2$ framework was further extended in \cite{Deng:2020ent,Chu:2021gdb} by incorporating conformal matter (essentially a two-dimensional CFT) on a tensionless EOW brane, which consequently induces a finite tension. As a result, the brane moves to a position with constant angle.
The corresponding bulk action in this setup is given by
\begin{align}\label{DES-action}
	I=I_0+\int_\mathcal{Q}\,\sqrt{h}\mathcal{L}_\mathcal{Q}
\end{align}
where $\mathcal{L}_\mathcal{Q}$ corresponds to the Lagrangian of a CFT$_2$ with large central charge; typical examples may be cooked up in terms a large number of massless scalar fields or in terms of the Liouville action \cite{Chen:2020uac}. The Neumann boundary condition \eqref{NBC} is modified through the expectation value of the stress tensor of this conformal matter theory as \cite{Takayanagi:2011zk,Deng:2020ent,Chu:2021gdb}
\begin{align}\label{Modify-NBC}
	K_{ab}=(K-T)h_{ab}+ 8\pi G_N \left<T_{ab}\right>~~,~~\left<T_{ab}\right>=\frac{2}{\sqrt{h}}\frac{\partial \mathcal{L}_\mathcal{Q}}{\partial h_{ab}}
\end{align}
and the EOW brane is essentially treated as a defect in the bulk geometry\footnote{Note that since the EOW brane has an AdS$_2$ profile, which is maximally symmetric, the stress tensor expectation value in the defect CFT$_2$ is proportional to induced metric $h_{ab}$ and only affects the tension on the EOW brane. In this sense, the CFT$_2$ degrees of freedom on the EOW brane do not really act as dynamical fields backreacting on the ambient geometry, and the EOW brane may be treated as a lower dimensional defect.}.

\begin{figure}[H]
	\centering
	\includegraphics[scale=0.55]{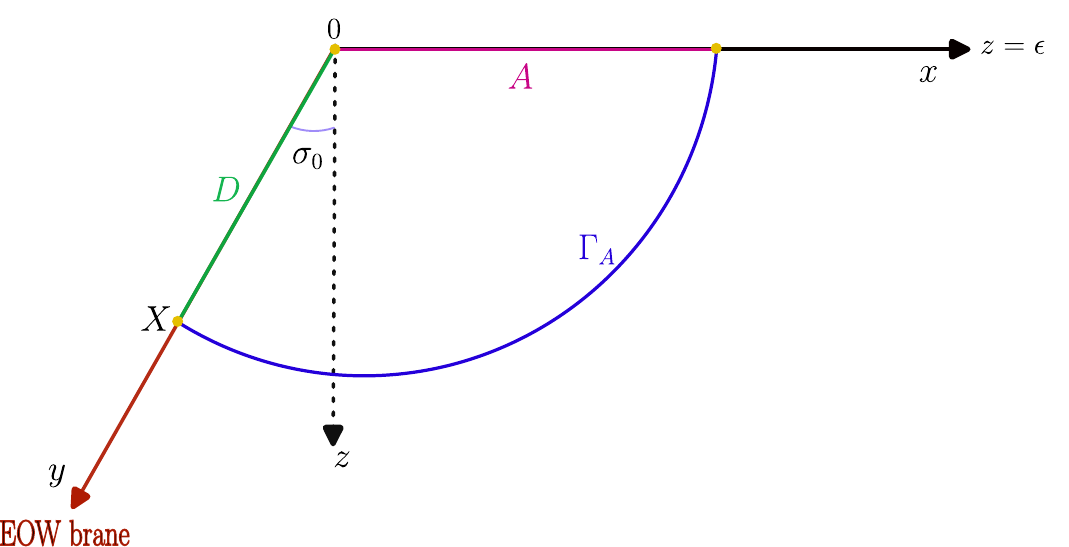}
	\caption{ Schematics of the defect extremal surface for the EE of a subsystem $A$.} 
	\label{fig:EEDES}
	\end{figure}

In the modified bulk scenario with defect conformal matter on the EOW brane, the entanglement entropy of an interval in the original BCFT includes contributions from the defect matter. This leads to a modification of the usual Ryu-Takayanagi (RT) formula to the defect extremal surface (DES) formula as
\begin{equation}\label{DES formula}
	S_\text{DES}= \text{min ext}_{\Gamma_A, X} \Bigl\{\frac{\text{Area}[\Gamma_A]}{4 G_N}+S_{\text{defect}}(D)\Bigl\}, \quad \quad X=\Gamma_{A} \cap D,
\end{equation}
where $\Gamma_{A}$ is the RT surface and $D$ is the defect region along the EOW brane as depicted in \cref{fig:EEDES}.

\subsection{Holographic $T\bar{T}$ with a boundary}\label{TTbar with boundary}
A $T \bar{T}$ deformed theory with a boundary can be realized by ensuring that the deformed theory preserves a space-time $Z_2$
symmetry with a fixed point at $x=0$ which corresponds to a tensionless brane in the holographic bulk. The holographic dual of a $T \bar{T}$
deformed CFT$_2$
in flat space with a boundary is still given by AdS$_3$ with a radial cut-off at $z=z_c$ and the introduction of the $Z_2$
quotient modifies the bulk metric as
\begin{align}\label{TTbar-metric}
	ds^2=\frac{\ell^2}{z^2} \left( -dt^2+dx^2+dz^2 \right), \ \ \ \ z>z_c, \ \ \ \ x\geq 0.
\end{align}
As stated in \cite{Aharony:2010ay,Kastikainen:2021ybu}, such a metric is equivalent to considering a tensionless EOW brane at $x=0$.
In the bulk, the EOW brane starts from $z= z_c$ as the boundary of the $Z_2$ quotient $T\bar{T}$ deformed CFT$_2$ moves from $z=\epsilon$ to $z=z_c$. Under the bulk coordinate transformation 
\begin{align}
	x=- y \tanh\frac{\sigma}{\ell}, ~~~~~ z= z_c+ y \sech\frac{\sigma}{\ell},
\end{align}
the metric \cref{TTbar-metric} becomes,
\begin{align}
	ds^2= \frac{\ell^2}{(z_c+ y \sech\frac{\sigma}{\ell})^2}\left(-dt^2+dy^2+ \frac{y^2 \sech\frac{\sigma}{\ell}}{\ell^2}d\sigma^2\right).
\end{align}
Assume that the EOW brane is located at $\sigma= \sigma_0$. Now solving the Neumann boundary condition on the EOW brane, one can easily find the relation between tension $T$ and brane location as
\begin{align}
	T= \frac{\tanh\frac{\sigma_0}{\ell}}{\ell}.
\end{align}
The metric induced on the brane may be written as follows
\begin{align}\label{brane-metric}
	ds^2=\Omega(y)^2\left(-dt^2+dy^2\right),
\end{align}
where $\Omega(y)= \frac{\ell}{(z_c+ y \sech\frac{\sigma_0}{\ell})}$ is the conformal factor.
As described in \cite{Deng:2020ent}, the tensionless brane is orthogonal to the asymptotic boundary and if we add the conformal matter on the brane, it turns to a finite tension and moves to a position with a constant $\sigma$ angle.
To impose transparent boundary conditions for the $T \bar{T}$ deformed bath, we include the same field theory on the EOW brane with an identical deformation parameter to that of the $T \bar{T}$-deformed CFT on the cut-off boundary.
This modified the Neumann boundary condition on the EOW brane is given by \cref{Modify-NBC}

\paragraph{On inclusion of matter in the bulk :} It is important to emphasize that incorporating conformal matter localized on the EOW brane does not require the introduction of additional sources in the holographic dictionary \cite{Deng:2020ent}. Rather, the conformal matter acts as a modification to the boundary conditions placed at the conformal boundary $x=0$ \cite{Deng:2023pjs}. As described in \cite{Deng:2023pjs}, the new EOW brane is still along the constant $\sigma_0$ slice with tension $T=\frac{1}{\ell}$. The conformal matter  is placed on the EOW brane in order to obtain a lower dimensional effective theory more suited to the island paradigm, where one requires the same quantum field theory to reside on the entire hybrid manifold. Furthermore, as the inclusion of such matter theory does not backreact on the ambient geometry, it is safely assumed that cut-off prescription for the holographic dual of $T \bar{T}$-deformed CFTs is still applicable in the large $N$ approximation \cite{Deng:2023pjs}.

\subsection{Entanglement entropy}
In this subsection, we will briefly review the computation of the EE between a subsystem $A=[0,L]$ and its complement in the above setup, both from the bulk (DES) and island perspective.

\subsection*{DES formula:} 	

Consider a subsystem $A=[0,L]$ in the bath region with a boundary, the generalized entropy is given as 
\begin{align}
	S_{\text{gen}}(a)=& S_{\text{RT}}(-a,L)+S_{\text{brane matter}}(a),\notag\\
	=& \frac{c}{6} \cosh ^{-1}\left(\frac{(L+a \tanh \frac{\sigma_0}{\ell})^2+(z_c +a \sech\frac{\sigma_0}{\ell})^2+z_c^2}{2 z_c (z_c+a \sech\frac{\sigma_0}{\ell})}\right)\notag\\
	&+ \frac{c}{6} \log\frac{2 a \ell}{(z_c+a \sech\frac{\sigma_0}{\ell})\epsilon_y},
\end{align}
where the first term is the usual RT surface length between point $(L,z_c)$ and $(-a \tanh\frac{\sigma_0}{\ell},z_c+a \sech\frac{\sigma_0}{\ell})$ and second term is the EE of the quantum matter on the defect brane. Here $a$ is the boundary position of island and $\epsilon_y$ is the UV cut-off for the $T \bar{T}$ deformed CFT on the brane.		
Now extremizing the above expression over $a$ we get the extremal value of $a$ 
\begin{align}\label{extremum-a}
	a= L-2 e^{\frac{\sigma_0}{\ell}}z_c+ \mathcal{O}({z_c^{2}}),
\end{align}  	
Substituting $a$ into $S_{\text{gen}}$ we obtain the EE for subsystem $A$ from the DES formula as
\begin{align}\label{EE-DES}
	S_{\text{DES}}= \frac{1}{4 G_N} \left(\frac{\sigma_0}{\ell}+\log\left[\frac{2 \ell \cosh \frac{\sigma_0}{\ell}}{\epsilon_y}\right]+\log\frac{2 L}{z_c}- \frac{e^{\frac{\sigma_0}{\ell}}}{L}z_c\right)+\mathcal{O}({z_c^{2}}).
\end{align}

\subsection*{Boundary island formula:} 
We now describe the lower-dimensional effective theory for the above bulk configuration, where the island formula is applicable. 
This can be achieved through a combination of a partial Randall-Sundrum reduction \cite{Karch:2000ct,Randall:1999vf} \textit{i.e.},  doing dimensional reduction for the wedge between EOW brane and tensionless brane and the usual AdS/CFT duality \cite{Maldacena:1997re} for the rest of the bulk. 
As described in \cite{Deng:2020ent,Li:2021dmf,Chu:2021gdb},
this procedure involves dividing the AdS$_3$ bulk into two regions by inserting an imaginary codimension-$1$ zero tension brane with transparent boundary condition. 
To obtain a $2d$ description of the region between the EOW brane and the zero tension brane, one can perform a partial Randall-Sundrum reduction along the $\sigma$ direction which leads to a gravity theory coupled with the $T\bar{T}$ deformed CFT$_2$ matter on the EOW brane. The rest of the bulk may be dualized by using the duality discussed in \cref{TTbar with boundary}. Eventually we have an effective theory including a gravity theory coupled with $T\bar{T}$ deformed CFT matter on EOW brane, and also a $T\bar{T}$ deformed CFT bath glued to EOW brane. 

In the effective theory one may utilize the island formula \cite{Almheiri:2019hni,Almheiri:2019qdq} to compute the EE for an interval in the bath.
In the boundary description, for a subsystem $A\equiv[0,L]$ in the bath region
an island region $I_A=[-a,0]$ appears in the gravitational sector on the EOW brane. 
The  $2d$ area term in the island formula may be obtained as a part of the $3d$ RT surface area. This is given by the length of the minimal surface (RT surface) that connects zero tension brane to the EOW brane at point $y=a$. 
The EE of the quantum matter presented in the region $[-a,L]$ may be computed by using the two-point twist field correlator $\langle \sigma_{g^{-1}_A}(a)\sigma_{g_A}(L) \rangle$. Utilizing these two, the generalized EE in the effective description may be written as \cite{Deng:2023pjs}
\begin{align}\label{S_gen-island}
	S_{\text{gen}}(a)&=S_{\text{area}}(a)+S_{\text{matter}}(-a,L),\notag\\
	& = \frac{c}{6}\cosh ^{-1}\left[\frac{\sqrt{a^2+2 a z_c \sech \frac{\sigma_0}{\ell}+z_{c}^2}}{(z_c+a \sech \frac{\sigma_0}{\ell})}\right]+\frac{c}{6} \log\left[\frac{(a+L)^2 \ell}{(z_c+a \sech \frac{\sigma_0}{\ell})\epsilon_y z_c}\right].
\end{align}
Now by extremizing the above expression with respect to $a$ which is only possible perturbatively, we get the extremum value of $a$ similar to \cref{extremum-a}. Now by plugging this extremum value into \cref{S_gen-island} and then expanding the expression linearly in $z_c$ we get the expression for the EE of a subsystem $A$, which is similar to \cref{EE-DES} upon utilizing the Brown-Henneaux relation\cite{Brown:1986nw}.

\subsection{Reflected entropy}
In this subsection we briefly review the reflected entropy and its computation in CFT$_2$ as described in \cite{Dutta:2019gen}. Consider a bipartite quantum system $A \cup B$ in a mixed state $\rho_{AB}$ whose canonical purification involves the doubling of its Hilbert space to define a pure state $\ket{\sqrt{\rho_{AB}}}_{A B A^* B^*}$. 
The reflected entropy $S_{R}(A:B)$ is then defined as the von Neumann entropy of the reduced density matrix $\rho_{AA^{*}}$ as follows 
\begin{equation}\label{def.}
	S_{R}(A:B) = S_{vN} (\rho^{}_{AA^{*}})_{\sqrt{\rho^{}_{AB}}} \,\,\,,
\end{equation}
where $\rho_{AA^{*}}$ may be obtained by tracing out the degree of freedom of $B$ and $B^*$ from the density matrix $\ket{\sqrt{\rho^{}_{AB}}}\bra{\sqrt{\rho^{}_{AB}}}$.
The authors in \cite{Dutta:2019gen} developed a novel replica technique to compute the reflected entropy between two disjoint subsystems $A \equiv[z_1, z_2]$ and $B \equiv [z_3, z_4]$ in a CFT$_2$. The reflected entropy may then be obtained in terms of a four-point twist field correlator as follows
\begin{equation}\label{defination}
	S_R(A:B)=\lim_{n,m \to1}S_{n}(AA^{*})_{\psi_{m}}=\lim_{n,m \to1}\frac{1}{1-n} \log \frac{\left<\sigma_{g_{A}}(z_{1})\sigma_{g_{A}^{-1}}(z_{2})\sigma_{g_{B}}(z_{3})\sigma_{g_{B}^{-1}}(z_{4})\right>_{\text{CFT}^{\otimes mn}}}{\left<\sigma_{g_{m}}(z_{1})\sigma_{g_{m}^{-1}}(z_{2})\sigma_{g_{m}}(z_{3})\sigma_{g_{m}^{-1}}(z_{4})\right>^{n}_{\text{CFT}^{\otimes m}}} \, ,
\end{equation}
where $m$, $n$ are the replica indices\footnote{As discussed in \cite{Kusuki:2019evw, Akers:2021pvd, Akers:2022max}, the two replica limits $n \to 1$ and $m \to 1$ are non-commuting. In this article, we compute the reflected entropy by first taking $n \to 1$ and subsequently $m \to 1$ as suggested in \cite{Kusuki:2019evw, Akers:2021pvd}.} and twist operators $\sigma_{g_A}$ and $\sigma_{g_B}$ are inserted at the end points of the subsystems. The conformal dimensions of the operators $\sigma_{g^{}_A}, \sigma_{g_B}$ and $\sigma_{g_{m}}$ are given as \cite{Dutta:2019gen}
\begin{equation}\label{conformal weights}
	h\equiv h_{A}=h_{B}=\frac{n c}{24} \left(m-\frac{1}{m}\right), \quad  h_{m}= \frac{c}{24}\left(m-\frac{1}{m}\right), \quad  h_{AB}= \frac{2c}{24}\left(n-\frac{1}{n}\right).
\end{equation} 
It was proved in \cite{Dutta:2019gen} that the reflected entropy for a bipartite state in a CFT$_{d}$ in the large central charge limit is dual to twice the minimal EWCS for the bulk static AdS$_{d+1}$ geometry. In the next subsection we review the island and DES formula for the reflected entropy.

\subsection{Island and DES formula for Reflected entropy}
\begin{figure}[h!]
	\centering
	\includegraphics[scale=0.5]{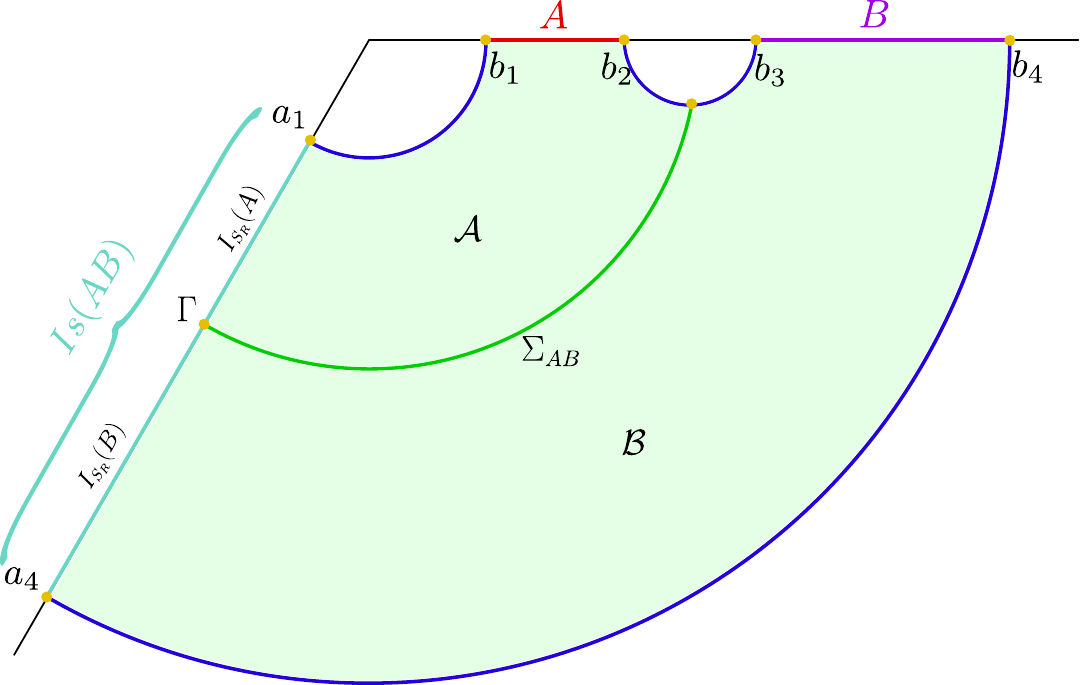}
	\caption{ Schematics of the island and DES formula for the reflected entropy. The entanglement wedge of $A \cup B$ is the green shaded region and green curve is the EWCS that splits the bulk into $\mathcal{A}$ and $\mathcal{B}$. The intersection of the EWCS and EOW brane is denoted by $\Gamma$. Figure modify from \cite{Li:2021dmf}.} 
	\label{fig:DES}
\end{figure}

As described in \cite{Li:2020ceg, Chandrasekaran:2020qtn}, the island formula for the reflected entropy between two subsystems $A$ and $B$ in asymptotic boundary is given as 
\begin{equation}\label{SR-bdy}
	S^{bdy}_{R}(A:B)= \text{min}~ \text{Ext}_{\Gamma} \Bigl\{S^{\text{eff}}_{R}(A\cup I_{S_{R}}(A):B \cup I_{S_{R}}(B))+ \frac{\text{Area}[\Gamma]}{2 G_N}\Bigl\},
\end{equation}
where the reflected entropy island $I_{S_{R}}(A)$ and $I_{S_{R}}(B)$ divide the entanglement island $I_{S}(AB)$ into two parts and $\Gamma$ is the island cross section of $I_{S_{R}}(A) \cap I_{S_{R}}(B)$, as shown in \cref{fig:DES}.

In the bulk description, the reflected entropy may be computed by defect extremal cross section formula as \cite{Li:2021dmf}
\begin{equation}\label{SR-bulk}
	S^{bulk}_{R}(\mathcal{A}:\mathcal{B})= \text{min}~ \text{Ext}_{\Sigma} \Bigl\{S^{\text{eff}}_{R}(\mathcal{A}:\mathcal{B})+ \frac{\text{Area}[\Sigma_{AB}]}{2 G_N}\Bigl\}.
\end{equation}
where the EWCS $(\Sigma_{AB})$ splits the entanglement wedge of $A \cup B$ into two parts $\mathcal{A}$ and $\mathcal{B}$ in the bulk, as depicted in \cref{fig:DES}.
In the above expression, since the conformal matter is only located on the EOW brane, the first term is equal to the reflected entropy between $I_A$ and $I_B$ on the brane and the second term is the area of the EWCS.

\section{Zero Temperature}\label{sec:Zero_Temp}
In this section, we analyse the reflected entropy for various bipartite mixed state
configurations defined on a fixed time slice of the $T\bar{T}$ deformed AdS$_3$ /BCFT$_2$
setup described in \cref{TTbar with boundary}. We describe the computation of the reflected
entropy from both the island and DES perspectives and verify the consistence
matching between them.

\subsection{Disjoint Subsystems}
In this subsection we investigate the reflected entropy corresponding to two disjoint subsystems $A\equiv(b_{1},b_{2})$ and $B\equiv (b_{3},b_{4})$ in a fixed time slice 
of the $T\bar{T}$ deformed AdS$_3$/BCFT$_2$ setup described in \cref{sec:review}. 
To obtain the various phases of the reflected entropy or the bulk EWCS, first it is required to determine the EE phases for two disjoint subsystems under consideration. Note that, depending on the subsystem size and its location we have two possible phases of the EE. In the following, we describe the computation of the reflected entropy from boundary and bulk perspective for these EE phases and show exact agreement between these two results.

\subsubsection{Entanglement Entropy phase-1}
\begin{figure}[h!]
	\centering
	\includegraphics[scale=0.65]{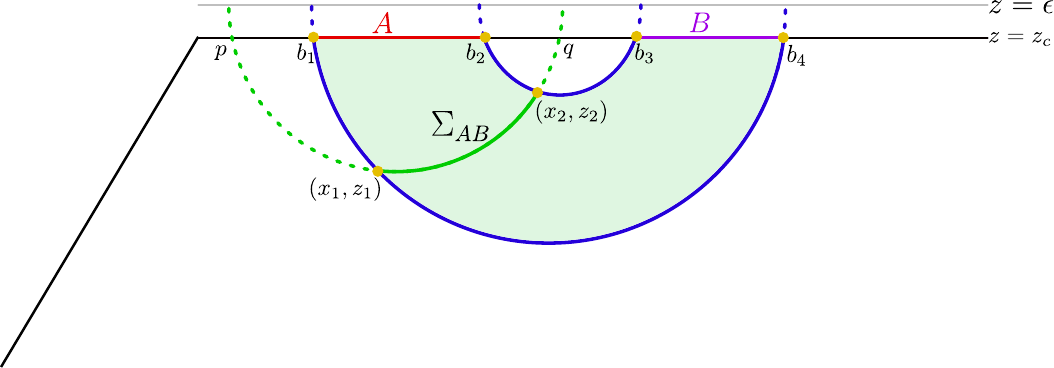}
	\caption{ Schematic diagram of the EE phase $1$ when the RT surfaces for $A \cup B$ and the EWCS are shown as solid blue and green curve respectively. } 
	\label{fig:dis-1}
\end{figure}
In this EE phase we consider that both the subsystems are small and close to each other away from the boundary. Hence the EE is sum of the lengths of two dome-type RT surfaces depicted as solid blue curve in \cref{fig:dis-1}. It is well known that the geodesic length between two points $(x_1, t_1, z_1)$ and $(x_2, t_2, z_2)$ on the Poincaré half plane is given as 
\begin{align}\label{geodesic-length}
	L= \cosh^{-1}\left[\frac{1}{2 z_1 z_2} \left((x_2-x_1)^2-(t_2-t_1)^2+z_1^2+z_2^2\right)\right].
\end{align}
Now by using the end-points of the subsystem in \cref{geodesic-length}, the EE for this configuration may be obtained as
\begin{equation}
	S_1= \frac{1}{2 G_N} \log\left[\frac{b_3-b_2}{z_c}\right]+\frac{1}{2 G_N} \log\left[\frac{b_4-b_1}{z_c}\right].
\end{equation}
For this EE phase, there is only one phase of the reflected entropy or the bulk EWCS as shown as green curve in \cref{fig:dis-1}. In the boundary description, the reflected entropy may be computed by utilizing the following four-point twist field correlator 
\begin{equation}
	S_R^{\text{bdy}}(A:B)= \lim_{{m,n} \to 1}\frac{1}{1-n}\log\frac{\langle \sigma_{g_A}(b_1)\sigma_{g_A^{-1}}(b_2)\sigma_{g_B}(b_3)\sigma_{g^{-1}_B}(b_4)\rangle}{\langle \sigma_{g_m}(b_1)\sigma_{g_m^{-1}}(b_2)\sigma_{g_m}(b_3)\sigma_{g^{-1}_m}(b_4)\rangle^n}.
\end{equation}
Now by utilizing the form of the four-point function in the large central charge limit \cite{Dutta:2019gen, Fitzpatrick:2014vua}, we may obtain the reflected entropy as follows
\begin{equation}\label{SR-dis-1}
	S_R^{\text{bdy}}(A:B)=\frac{c}{3}  \log \left[\frac{1+\sqrt{X}}{1-\sqrt{X}}\right],
\end{equation} 
where $X$ is the cross ratio defined as 
\begin{equation}\label{cross-ratio}
	X= \frac{(b_2-b_1)(b_4-b_3)}{(b_3-b_1) (b_4-b_2)}.
\end{equation}
In the bulk perspective, the EWCS for this phase is proportional to the length of the green geodesic $\Sigma_{AB}$ as shown in \cref{fig:dis-1}. 
Recall that, on a constant time slice, the RT surfaces/geodesics in the bulk geometry dual to a $T\bar{T}$ deformed CFT$_2$ are semi-circles centered at the original asymptotic boundary $z=\epsilon$ where the undeformed theory would live. 
Now utilizing the fact that the minimal curve $\Sigma_{AB}^{\text{min}}$ is perpendicular to both RT surfaces at their intersection points $(x_1,z_1)$ and $(x_2,z_2)$ leads to the following constraint equations 
\begin{align}
	&\left(x_1-\frac{b_1+b_4}{2}\right)^2+z_1^2=\left(\frac{b_4-b_1}{2}\right)^2+z_c^2~~,~~\left(x_1-\frac{p+q}{2}\right)^2+z_1^2=\left(\frac{q-p}{2}\right)^2+z_c^2,\notag\\
	&\left(x_2-\frac{b_2+b_3}{2}\right)^2+z_2^2=\left(\frac{b_3-b_2}{2}\right)^2+z_c^2~~,~~\left(x_2-\frac{p+q}{2}\right)^2+z_2^2=\left(\frac{q-p}{2}\right)^2+z_c^2,\notag\\&
	z_{1}^2=-\left(x_1-\frac{b_1+b_4}{2}\right) \left(x_1-\frac{p+q}{2}\right),~~ z_{2}^2=-\left(x_2-\frac{b_2+b_3}{2}\right) \left(x_2-\frac{p+q}{2}\right),
\end{align}
where, the extension of $\Sigma_{AB}^\textrm{min}$ intersects the boundary at the points $(p,z_c)$ and $(q,z_c)$. The above equations may be solved to fix the endpoints of the curve $\Sigma_{AB}^{\text{min}}$ as
\begin{align}
	&x_1= \frac{b_1^2 b_4+b_1 \left(-2 b_4 (b_2+b_3)+b_2 b_3+b_4^2-2 z_c^2\right)+2 z_c^2 (b_2+b_3-b_4)+b_2 b_3 b_4}{b_1^2-b_1 (b_2+b_3)-b_4 (b_2+b_3)+2 b_2 b_3+b_4^2}, \notag\\ &x_2=\frac{b_1 \left(b_4 (b_2+b_3)-2 b_2 b_3+2 z_c^2\right)-2 z_c^2 (b_2+b_3-b_4)+b_2 b_3 (b_2+b_3-2 b_4)}{-b_1 (b_2+b_3-2 b_4)+b_2^2-b_4 (b_2+b_3)+b_3^2},\notag\\& 
	z_1= \frac{\sqrt{(b_1-b_4)^2+4 z_c^2} \sqrt{(b_1-b_2) (b_1-b_3) (b_2-b_4) (b_3-b_4)-z_c^2 (b_1-b_2-b_3+b_4)^2}}{b_1^2-b_1 (b_2+b_3)-b_2 (b_4-2 b_3)-b_4 (b_3-b_4)},\notag\\& 
	z_2= \frac{\sqrt{(b_2-b_3)^2+4 z_c^2} \sqrt{(b_1-b_2) (b_1-b_3) (b_2-b_4) (b_3-b_4)-z_c^2 (b_1-b_2-b_3+b_4)^2}}{b_2^2-b_1 (b_2+b_3-2 b_4)-b_2 b_4-b_3 (b_4-b_3)}.
\end{align}
Now substituting the above values in \cref{geodesic-length} and then linearly expanding the expression in $z_c$, the bulk EWCS in this phase may be obtained as 
\begin{align}
	S_R^{\text{bulk}(\mathcal{A}:\mathcal{B})}=\frac{1}{4 G_N} \cosh^{-1}\left[\frac{b_1(2 b_4-b_3-b_2)+b_2(b_4-2b_3)-b_3 b_4}{(b_4-b_1)(b_3-b_2)}\right].
\end{align}
Note that the above expression of the bulk EWCS is  precisely equal to half of the reflected entropy computed in \cref{SR-dis-1} upon utilizing the Brown-Henneaux relation.

\subsubsection{Entanglement Entropy phase 2}
\begin{figure}[h!]
	\centering
	\includegraphics[scale=0.5]{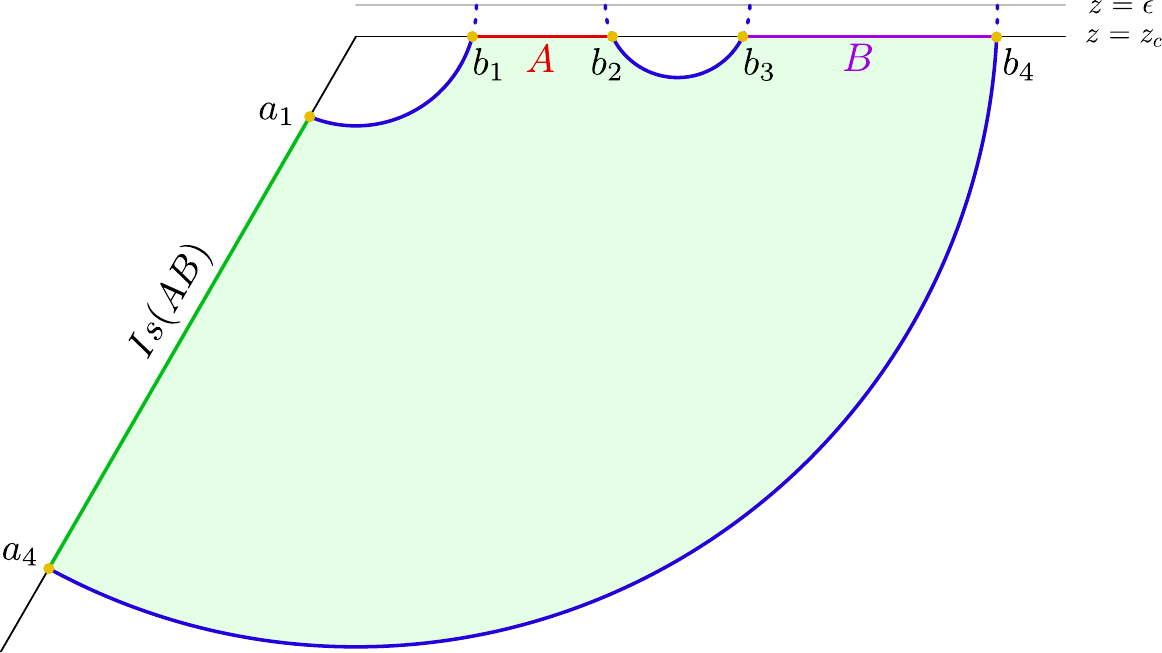}
	\caption{Schematic depicting the EE phase $2$ when the RT surface for $A \cup B$ is shown by the solid blue curves.  } 
	\label{fig:disjoint}
\end{figure}
In this EE phase we assume that the subsystem $A$ and $B$ both are close to the boundary, so the EE corresponds to the sum of the lengths of two island surfaces and a dome-type RT surface, shown as blue curves in \cref{fig:disjoint}. So the EE for this phase is given as
\begin{equation}
	S_2=  \frac{1}{2 G_N} \frac{\sigma_0}{\ell}+\frac{1}{2 G_N}\log\left[\frac{2 \ell \cosh \frac{\sigma_0}{\ell}}{\epsilon_y}\right] +\frac{1}{2 G_N} \log\left[\frac{b_3-b_2}{z_c}\right]+\frac{1}{4 G_N}\log\left[\frac{4 b_1 b_4}{z_c^{2}}\right]- \frac{e^{\frac{\sigma_0}{\ell}}}{4 G_N}\left(\frac{1}{ b_1}+\frac{1}{ b_4}\right)z_c,
\end{equation}
where $\epsilon_y$ is the finite cutoff for the $T\bar{T}$ deformed CFT on the brane.
In this EE phase we observe three distinct phases for the reflected entropy or the bulk EWCS. The computation of the reflected entropy for each phases from the boundary and bulk perspective is described in the following subsection.

\subsubsection*{Phase-I}\label{SR-dis2(i)}
\paragraph{The boundary perspective:} 
For this phase, consider that the subsystem $A$ is smaller than the subsystem $B$, hence the EWCS lands on the island surface corresponding to subsystem $A$. As there is no island cross section for this phase, the reflected entropy in the boundary description reduces to $S^{\text{(eff)}}_{R}(A:B\cup I_{S_{R}}(B))$ which may be computed as follows
\begin{equation}
	S^{\text{(bdy)}}_{R}(A:B)= S^{\text{(eff)}}_{R}(A:B\cup I_{S_{R}}(B))= \lim_{{m,n} \to 1}\frac{1}{1-n}\log\frac{\langle \sigma_{g^{-1}_B}(a_1) \sigma_{g_A}(b_1)\sigma_{g_A^{-1}}(b_2)\sigma_{g_B}(b_3)\rangle_{mn}}{\langle \sigma_{g^{-1}_m}(a_1) \sigma_{g_m}(b_1)\sigma_{g_m^{-1}}(b_2)\sigma_{g_m}(b_3)\rangle^n_{ m}}.
\end{equation}
Here $a_1$ is the intersection point between the island surface corresponding to subsystem $A$ and the EOW brane. As described in \cite{Deng:2023pjs}, the intersection point $a_1$ may be determined through the extremization of the EE of $A \cup B$ which is given as $a_1=  b_1- 2 z_c e^{\frac{\sigma_0}{\ell}}+\mathcal{O}(z_{c}^2)$. Utilizing this coordinate value of $a_1$ and the form of the four-point twist field correlator in the large central charge limit \cite{Dutta:2019gen, Fitzpatrick:2014vua} and then expanding the expression linearly in $z_c$, we may obtain the reflected entropy as follows
\begin{equation}\label{SR-dis-2(i)}
	S_R^{\text{bdy}}(A:B)=\frac{c}{3} \left( \cosh ^{-1}\left[\frac{b_2 b_3-b_1^{2}}{b_1 (b_3-b_2)}\right]+ \frac{e^\frac{\sigma_0}{\ell} z_c}{b_1} \sqrt{\frac{(b_2-b_1)(b_3-b_1)}{(b_1+b_2)(b_1+b_3)}}\right)+\mathcal{O}(z_{c}^2).
\end{equation}

\paragraph{The Bulk perspective:}
\begin{figure}[h!]
	\centering
	\includegraphics[scale=0.7]{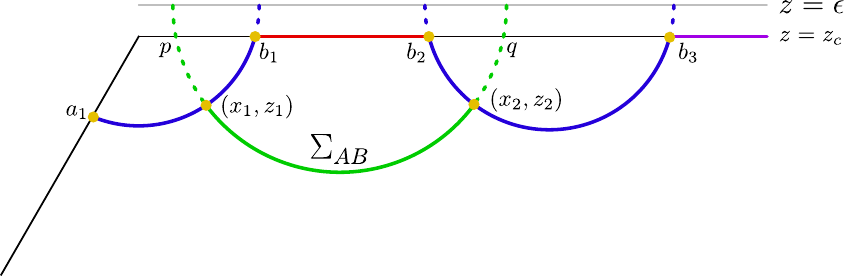}
	\caption{Schematic illustrating the bulk EWCS phase (represented by the solid green curve). } 
	\label{fig:disj2}
\end{figure}
In the bulk description, the curve $\Sigma_{AB}$ lands on the extremal surface joining $b_1$ and $a_1$ and correspondingly the contribution due to the brane matter vanishes identically as the entire island belongs to subsystem $B$: 
\begin{align*}
	S_R^{(\textrm{eff})}(\emptyset:I_B)=0
\end{align*}
Hence, the reflected entropy between the bulk regions $\mathcal{A}$ and $\mathcal{B}$ is determined solely by the (minimal) length of the curve $\Sigma_{AB}$, depicted as solid green curve in \cref{fig:disj2}.

To compute the length of the curve, we consider the setup in \cref{fig:disj2}. 
As discussed in the previous phase, the RT surfaces/geodesics are semi-circles in the bulk geometry centered at $z=\epsilon$. Using this fact the geodesic semi-circle joining the boundary points $(b_2,z_c)$ and $(b_3,z_c)$ is given by
\begin{align}
	\left(x-\frac{b_2+b_3}{2}\right)^2+z^2=\left(\frac{b_3-b_2}{2}\right)^2+z_c^2
\end{align}
In a similar fashion,the radius and centre of the geodesic emanating from $(b_1,z_c)$ and landing on the EOW brane at the location $a_1$ may be obtained by solving the following equations:
\begin{align}
	(b_1-x_0)^2=r^2~~,~~\left(a_1\tanh\left(\frac{\sigma_0}{\ell}\right)+x_0\right)^2+\left(z_c+a_1\textrm{sech}\left(\frac{\sigma_0}{\ell}\right)\right)^2=r^2
\end{align}
resulting in
\begin{align}\label{b1-a1-geodesic}
	&x_0=\frac{b_1^2-a_1^2-2 a_1 z_c\, \text{sech}\left(\frac{\sigma_0}{\ell}\right)}{2 \left(b_1+a_1 \tanh \left(\frac{\sigma_0}{\ell}\right)\right)}\\
	&r=\frac{\sqrt{b_1^2+a_1^2+2 a_1 b_1 \tanh \left(\frac{\sigma_0}{\ell}\right)} \sqrt{b_1^2+a_1^2+2 a_1 b_1 \tanh \left(\frac{\sigma_0}{\ell}\right)+4 a_1 z_c \text{sech}\left(\frac{\sigma_0}{\ell}\right)+4 z_c^2}}{2 \left(b_1+a_1 \tanh \left(\frac{\sigma_0}{\ell}\right)\right)}
\end{align}  
Next, we utilize the fact that the minimal curve $\Sigma_{AB}^\textrm{min}$ is perpendicular to both these circles at their intersection points $(x_1,z_1)$ and $(x_2,z_2)$ as in \cref{fig:disj2}. These geometrical considerations lead to the following constraint equations:
\begin{align}
	&(x_1-x_0)^2+z_1^2=r^2~~,~~\left(x_1-\frac{p+q}{2}\right)^2+z_1^2=\left(\frac{q-p}{2}\right)^2+z_c^2~~,~~z_{1}^2=-(x_1-x_0) \left(x_1-\frac{p+q}{2}\right),\notag\\
	&\left(x_2-\frac{b_2+b_3}{2}\right)^2+z_2^2=\left(\frac{b_3-b_2}{2}\right)^2+z_c^2~~,~~\left(x_2-\frac{p+q}{2}\right)^2+z_2^2=\left(\frac{q-p}{2}\right)^2+z_c^2,\notag\\&z_{2}^2=-\left(x_2-\frac{b_2+b_3}{2}\right) \left(x_2-\frac{p+q}{2}\right),
\end{align}
where, the extension of $\Sigma_{AB}^\textrm{min}$ intersects the boundary at the points $(p,z_c)$ and $(q,z_c)$. The above equations may be solved to fix the endpoints of the curve $\Sigma_{AB}^\textrm{min}$ as follows
\begin{align}
	&x_1=x_0+\frac{r^2 (b_2+b_3-2x_0)}{r^2+(b_2-x_0) (b_3-x_0)-z_c^2}~~,~~x_2=x_0+\frac{(b_2+b_3-2 x_0) \left((b_2-x_0) (b_3-x_0)-r^2-z_c^2\right)}{b_2^2+b_3^2-2 x_0 (b_2+b_3)-2 r^2+2 x_0^2+2 z_c^2}\,,\notag\\
	&z_1=\frac{r \sqrt{\left(r^2-(b_2-x_0)^2\right) \left(r^2-(b_3-x_0)^2\right)-2 z_c^2 \left((x_0-b_2) (x_0-b_3)+r^2\right)+z_c^4}}{(x_0-b_2) (x_0-b_3)+r^2-z_c^2}\,,\notag\\
	&z_2=\frac{\sqrt{(b_2-b_3)^2+4z_c^2}\sqrt{\left(r^2-(b_2-x_0)^2\right) \left(r^2-(b_3-x_0)^2\right)-2 z_c^2 \left((x_0-b_2) (x_0-b_3)+r^2\right)+z_c^4}}{b_2^2+b_3^2-2 x_0 (b_2+b_3)-2 r^2+2 x_0^2+2 z_c^2}\,.
\end{align}
The length of the minimal surface may now be readily computed and the reflected entropy between the disjoint subsystems is given as follows
\begin{align}\label{EW-disj2-NP}
	S_R^{\text{bulk}}(\mathcal{A}:\mathcal{B})&=\frac{1}{2G_N}\cosh^{-1}\left[\frac{(x_2-x_1)^2+z_1^2+z_2^2}{2z_1z_2}\right]\notag\\
	&=\frac{1}{4G_N}\cosh^{-1}\left[\frac{r^2-(b_2-x_0) (b_3-x_0)+z_c^2}{r \sqrt{(b_3-b_2)^2+4 z_c^2}}\right]
\end{align}
Note that with $a_1=a_1(z_c)$ determined through the extremization of the entanglement entropy of $A\cup B$, the above expression is non-perturbative in $z_c$. However, as the extremal solution to $a_1$ is only determined perturbatively as $a_1=b_1-2e^{\frac{\sigma_0}{\ell}}z_c+\mathcal{O}(z_c^2)$, we expand \cref{EW-disj2-NP} in $z_c$ to obtain 
\begin{align}
	S_R^{\text{bulk}}(\mathcal{A}:\mathcal{B})=\frac{1}{2G_N}\cosh^{-1}\left[\frac{b_2b_3-b_1^2}{b_1(b_3-b_2)}\right]+\frac{e^{\frac{\sigma_0}{\ell}}z_c}{2G_N b_1}\sqrt{\frac{(b_2-b_1)(b_3-b_1)}{(b_2+b_1)(b_3+b_1)}}+\mathcal{O}(z_c^2).
\end{align}
Note that when the Brown-Henneaux relation is used, the calculation of the reflected entropy  from both the perspective matches exactly.

\subsubsection*{Phase-II} 
\paragraph{The boundary perspective:}
In this phase we assume that the subsystem $A$ is large enough, so the bulk EWCS lands on the EOW brane and divide the EE island into two parts. 
In the boundary description, the first term of \cref{SR-bdy} may be obtained by computing the three-point twist field correlator $\langle \sigma_{g^{-1}_A}(b_2)\sigma_{g_B}(b_3) \sigma_{g_B g^{-1}_A}(a)\rangle$ as 
\begin{align}
	S^{\text{eff}}_{R}(A\cup I_{S_{R}}(A):B \cup I_{S_{R}}(B))= \frac{c}{3}  \log \left[\frac{4 l (a+b_2)(a+b_3)}{\epsilon_y  (b_3-b_2) (z_c+a \sech \frac{\sigma_0}{\ell} )}\right],
\end{align}
where $\epsilon_y$ is the finite cut-off for the $T\bar{T}$ deformed CFT on the brane and $a$ is the location of the island cross section on the brane. The second term of \cref{SR-bdy} which is the area of the island cross section may be written as follows \cite{Deng:2023pjs} 
\begin{align}\label{area-of-Gamma}
	\text{Area}[\Gamma]=\cosh ^{-1}\left[\frac{\sqrt{a^2+2 a z_c \sech \frac{\sigma_0}{\ell}+z_{c}^2}}{(z_c+a \sech \frac{\sigma_0}{\ell})}\right].
\end{align}
By adding these two equation,  the reflected entropy in the boundary description for this phase is given as
\begin{equation}\label{SR-dis-brane1}
	S_R^{\text{bdy}}(A:B)=	\frac{c}{3}  \log \left[\frac{4 \ell (a+b_2)(a+b_3)}{\epsilon_y  (b_3-b_2) (z_c+a \sech \frac{\sigma_0}{\ell} )}\right]+\frac{c}{3}  \cosh ^{-1}\left[\frac{\sqrt{a^2+2 a z_c \sech \frac{\sigma_0}{\ell}+z_{c}^2}}{z_c+a \sech \frac{\sigma_0}{\ell}}\right].
\end{equation}
The location of the island cross section on the brane may be obtained by extremizing the above expression over $a$ which can only be determined perturbatively as 
\begin{equation}
	a= \sqrt{b_2 b_3}- \frac{(\sqrt{b_2}+\sqrt{b_3})^2}{2 \sqrt{b_2 b_3}} e^\frac{{\sigma_0}}{\ell} z_c+ \mathcal{O}(z_c^2).
\end{equation}
Finally substituting the extremum value of $a$ in \cref{SR-dis-brane1} and then linearly expanding the expression in $z_c$, the reflected entropy for this phase in the boundary description may be obtained as
\begin{equation}\label{SR-dis-brane2}
	S_R^{\text{bdy}}(A:B)=\frac{c}{3}  \left(\frac{\sigma_0}{\ell}+\log \left[\frac{\sqrt{b_3}+\sqrt{b_2}}{\sqrt{b_3}-\sqrt{b_2}}\right]+\log \left[\frac{4 \ell \sech\frac{\sigma_0}{\ell}}{\epsilon_y   }\right]- \frac{e^{\frac{\sigma_0}{\ell}}}{\sqrt{b_2 b_3}}z_c\right)+\mathcal{O}(z_{c}^2).
\end{equation}

\paragraph{The bulk perspective:}
\begin{figure}[h!]
	\centering
	\includegraphics[scale=0.7]{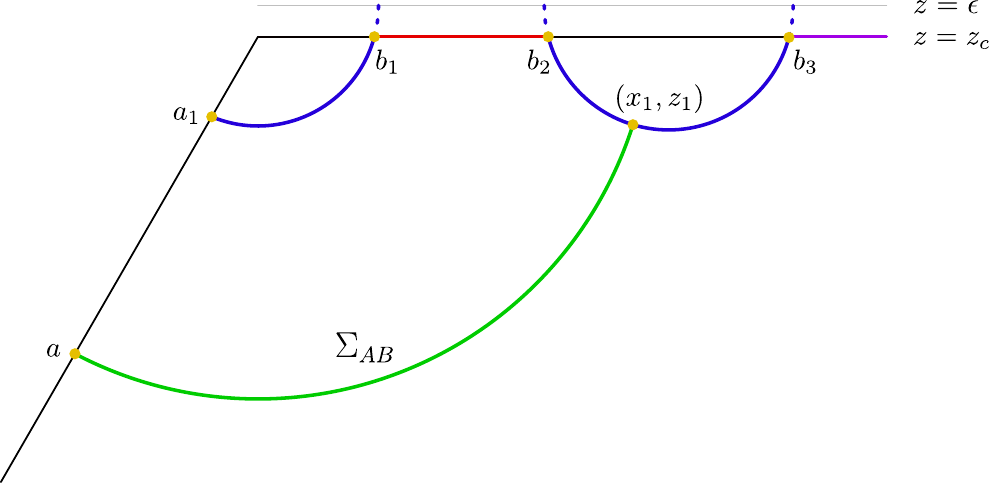}
	\caption{ Schematic illustrating the bulk EWCS phase (represented by the solid green curve) } 
	\label{fig:disj3}
\end{figure}
In the bulk description, the curve $\Sigma_{AB}$ joins dome-type RT surface to the EOW brane at an arbitrary point $a$ and therefore splits the EE island into two parts as shown in \cref{fig:disj3}. Now the first term of \cref{SR-bulk} reduces to $S_{R}^\text{eff}(I_{A}:I_{B})$, which may be computed as
\begin{equation}\label{SR(eff.)_Dis}
	S^{\text{eff}}_{R}(\mathcal{A}:\mathcal{B})=S_{R}^\text{eff}(I_{A}:I_{B})= \lim_{{m,n} \to 1}\frac{1}{1-n}\log \frac{   \Pi_{i} ~ \Omega_{i}^{2 h_{i}}  \langle \sigma_{g^{}_A}(a_1)\sigma_{g^{}_B g_A^{-1}}(a)\sigma_{g^{-1}_B}(a_4)\rangle_{\mathrm{BCFT}^{\bigotimes mn}}}{ \Omega(a_1)^{2 n h_{mn}} \Omega(a_4)^{2 n h_{mn}} \langle\sigma_{g^{}_m}(a_1)\sigma_{g^{-1}_m}(a_4)\rangle^n_{\mathrm{BCFT}^{\bigotimes m}}},
\end{equation}
The BCFT correlator in \cref{SR(eff.)_Dis} may be expended in two possible channels: the boundary operator expansion (BOE) and OPE. 
As explained in \cite{Li:2021dmf}, the OPE channel in this case never dominates in the large central charge limit. In the BOE channel, the three-point twist field correlator of \cref{SR(eff.)_Dis} may be factorized to three one-point twist filed correlators on the BCFT. Now by utilizing the form of one-point twist field correlator in BCFT and appropriate conformal factors, we may obtain the effective reflected entropy as 
\begin{equation}\label{SR(eff.)_dis}
	S_{R}^\text{eff}(I_{A}:I_{B})= \frac{c}{3} \log \frac{2a \ell}{\epsilon_y (z_c + a \sech \frac{\sigma_0}{\ell})}.
\end{equation}
The second term of \cref{SR-bulk} is given by the geodesic length of the curve $\Sigma_{AB}$ which may be computed by considering the setup depicted in \cref{fig:disj3}. As explained in previous phase, the RT surfaces or geodesics are semi-circles in the bulk geometry centred at $z=\epsilon$. Using this fact the radius and center of the geodesic $\Sigma_{AB}$ which connects an arbitrary point $(x_1,z_1)$ to $(-a \tanh\frac{\sigma_0}{\ell}, z_c+ a \sech\frac{\sigma_0}{\ell})$ on the EOW brane may be obtained by solving the following equations
\begin{align}
	(x_1-x_c)^2+z_1^2= R^2,  ~~~~~~~~ (-a \tanh\frac{\sigma_0}{\ell}-x_c)^2+(z_c+ a \sech\frac{\sigma_0}{\ell})=R^2,
\end{align}
which gives 
\begin{align}
	x_c= \frac{-a^2-2 a z_c \sech\frac{\sigma_0}{\ell}+x_1^2+z_1^2-z_c^2}{2 \left(a \tanh \frac{\sigma_0}{\ell}+x_1\right)}.
\end{align}
Utilizing the fact that the curve $\Sigma_{AB}$ is perpendicular to the dome-type RT surface at their intersection point $(x_1,z_1)$, we get the following constraints equations
\begin{align}
	\left(x_1-\frac{b_2+b_3}{2}\right)^2+z_1^2= \left(\frac{b_3-b_2}{2}\right)^2+z_c^2, ~~~~~ z_{1}^2=-\left(x_1-\frac{b_2+b_3}{2}\right)\left(x_1-x_c\right).
\end{align}
Note that the above equation may be solved to determine $(x_1,z_1)$ as follows
\begin{align}
	x_1=\frac{2 b_2 b_3 - (b_2+b_3)x_c- 2 z_c^2}{b_2+b_3- 2 x_c}, ~~~~~~ z_1= \sqrt{\frac{\left((b_2-x_c)(b_3-x_c)-z_c^2\right)((b_3-b_2)^2+4 z_c^2)}{(b_2+b_3-2 x_c^2)}}.
\end{align}
Now the second term of \cref{SR-bulk}  which is the length of the geodesic between points $(x_1,z_1)$ and $(-a \tanh\frac{\sigma_0}{\ell}, z_c+a \sech\frac{\sigma_0}{\ell})$, may be written as 
\begin{align}\label{EW-dis}
	\text{Area}[\Sigma_{AB}]=\cosh ^{-1}\left(\frac{(x_1+a \tanh \frac{\sigma_0}{\ell})^2+(z_c +a \sech\frac{\sigma_0}{\ell})^2+z_1^2}{2 z_1 (z_c+a \sech\frac{\sigma_0}{\ell})}\right).
\end{align}
The reflected entropy may now be written by adding \cref{SR(eff.)_dis} and \cref{EW-dis} as follows
\begin{align}\label{SR-dis-II}
	S_R^{\text{bulk}}(\mathcal{A}:\mathcal{B})= \frac{1}{2 G_N}\left[\cosh ^{-1}\left(\frac{(x_1+a \tanh \frac{\sigma_0}{\ell})^2+(z_c +a \sech\frac{\sigma_0}{\ell})^2+z_1^2}{2 z_1 (z_c+a \sech\frac{\sigma_0}{\ell})}\right)+ \log \frac{2a \ell}{\epsilon_y (z_c + a \sech \frac{\sigma_0}{\ell})}\right].
\end{align}
Now by extremizing the above expression over $a$ perturbatively, we get the extremum value of $a$
as follows
\begin{align}
	a= \sqrt{b_2 b_3}- \frac{(\sqrt{b_2}+\sqrt{b_3})^2}{2 \sqrt{b_2 b_3}} e^\frac{{\sigma_0}}{\ell} z_c+\mathcal{O}(z_c^2).
\end{align}
The reflected entropy for this phase may now be obtained by substituting the extremum value of $a$ in \cref{SR-dis-II} and then expanding the expression perturbatively in $z_c$ as
\begin{align}
	S_R^{\text{bulk}}(\mathcal{A}:\mathcal{B})=\frac{1}{2 G_N}  \left[\frac{\sigma_0}{\ell}+ \log\left[\frac{\sqrt{b_3}+\sqrt{b_2}}{\sqrt{b_3}-\sqrt{b_2}}\right]+\log \left[\frac{4 \ell \sech\frac{\sigma_0}{\ell}}{\epsilon_y   }\right]- \frac{e^{\frac{\sigma_0}{\ell}}}{\sqrt{b_2 b_3}}z_c\right]+\mathcal{O}(z_c^2).
\end{align}
Here also the expression of the reflected entropy form the boundary and the bulk perspective matches exactly upon utilizing the Brown-Henneaux relation.

\subsection*{Phase-III}
\begin{figure}[h!]
	\centering
	\includegraphics[scale=0.7]{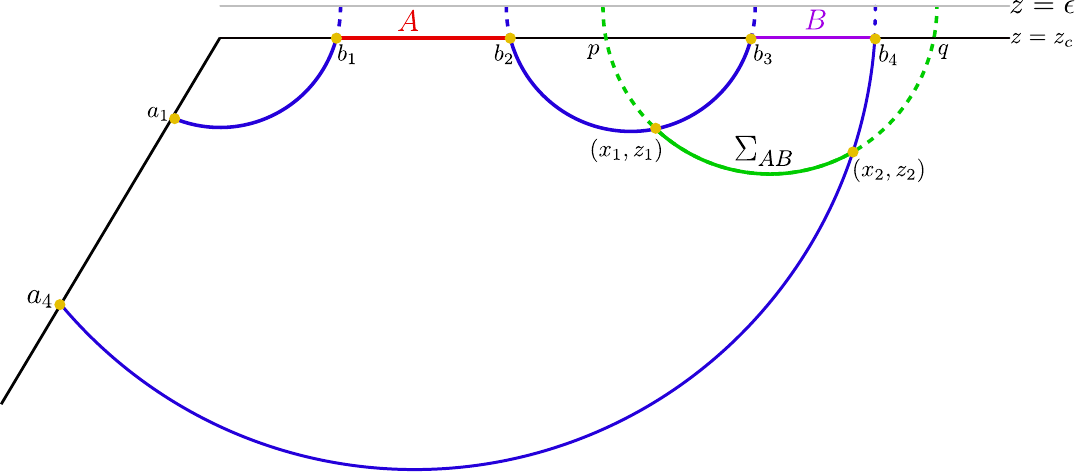}
	\caption{Schematic illustrating the bulk EWCS phase (represented by the solid green curve) } 
	\label{fig:disj4}
\end{figure}
For this phase we assume that the subsystem $B$ is smaller than the subsystem $A$, hence the EWCS lands on the island surface corresponding to subsystem $B$ shown as the solid green curve in \cref{fig:disj4}. In the boundary description, the reflected entropy may be obtained by exchanging $b_1$ and $b_4$ in \cref{SR-dis-2(i)} as follows
\begin{align}\label{SR-dis-2(iii)}
	S_R^{\text{bdy}}(A:B)=\frac{c}{3} \left( \cosh ^{-1}\left[\frac{b_2 b_3-b_4^{2}}{b_4 (b_3-b_2)}\right]+ \frac{e^\frac{\sigma_0}{\ell} z_c}{b_4} \sqrt{\frac{(b_4-b_2)(b_4-b_3)}{(b_4+b_2)(b_4+b_3)}}\right)+\mathcal{O}(z_{c}^2).
\end{align}
The bulk computation in this case may be done similarly to the \hyperref[SR-dis2(i)]{phase-I} and the reflected entropy may be computed by interchanging $b_1$ and $b_4$ which is exactly matches with \cref{SR-dis-2(iii)} upon utilizing the Brown-Henneaux relation.

\subsection{Adjacent Subsystems}
In this subsection we investigate the reflected entropy corresponding to two adjacent subsystems $A\equiv(b_{1},b_{2})$ and $B\equiv (b_{2},b_{3})$ in a fixed time slice 
of the $T\bar{T}$ deformed AdS$_3$/BCFT$_2$ setup described in \cref{sec:review}.
Note that for two adjacent subsystems, there are two possible phases of the EE depending
on the subsystem size and its location. In the following, we explain the computation of the various reflected entropy phases from both the boundary and bulk perspective.

\subsubsection{Entanglement Entropy Phase:1}
\begin{figure}[h!]
	\centering
	\includegraphics[scale=0.7]{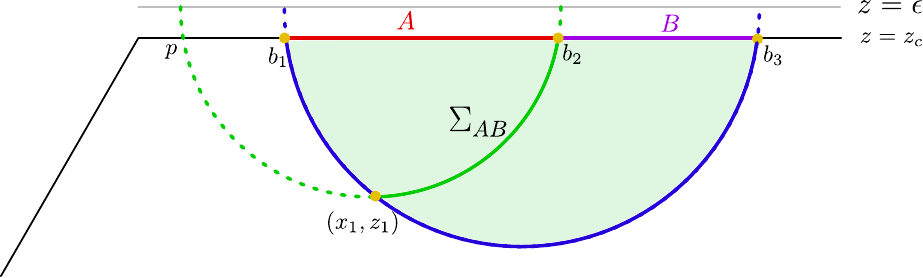}
	\caption{ Schematic diagram of the EE phase $1$ when the RT surface and the EWCS for $A \cup B$ are shown as blue and green curves.  } 
	\label{fig:adj-1}
\end{figure}
In this EE phase, we assume that both the subsystems are very small and close to each other away form the boundary. So the EE is proportional to the dome type RT-surface shown as blue curve in \cref{fig:adj-1}. Now by utilizing the end points of the subsystems in \cref{geodesic-length}, we may obtain the EE for this phase as
\begin{align}
	S_1= \frac{1}{2 G_N} \log\left[\frac{b_3-b_1}{z_c}\right].
\end{align}
In this EE phase, we observe only one phase of the reflected entropy or the bulk EWCS depicted as solid green curve in \cref{fig:adj-1}. Now in the boundary description, the reflected entropy reduces to $S^{\text{(eff)}}_{R}(A:B)$ as there is no island cross section on the brane for this phase. Now by utilizing the three-points twist field correlator the reflected entropy may be obtained as
\begin{equation}\label{Adj-conn-bdy}
	S_R^{\text{bdy}}(A:B)=\frac{c}{3}  \log \left[\frac{2 (b_2-b_1) (b_3-b_2)}{z_c (b_3-b_1)}\right].
\end{equation}
In the bulk perspective, the EWCS for this phase is proportional to the length of the green geodesic shown in \cref{fig:adj-1}. As explained earlier in the bulk geometry the RT surface and geodesic are semi-circles centred at $z=\epsilon$. Utilizing the condition that minimal curve $\Sigma_{AB}^{\text{min}}$ is perpendicular to the RT surface at their intersection point $(x_1,z_1)$ we obtain the following constraint equations 
\begin{align}
	&\left(x_1-\frac{b_1+b_3}{2}\right)^2+z_1^2= \left(\frac{b_3-b_2}{2}\right)^2+z_c^2, ~~~~ \left(x_1-\frac{p+b_2}{2}\right)^2+z_1^2=\left(\frac{b_2-p}{2}\right)^2+z_c^2,\notag\\
	&z_{1}^2=-\left(x_1-\frac{b_1+b_3}{2}\right)\left(x_1-\frac{p+b_2}{2}\right),
\end{align}
where the extension of $\Sigma^{\text{min}}_{AB}$ intersects the boundary at point $(p,z_c)$. Solving the above equations for $(x_1,z_1)$ leads to 
\begin{align}
	&x_1=\frac{b_1^2 b_3+b_1 \left(b_2^2-4 b_2 b_3+b_3^2\right)+b_2 \left(b_2 b_3+4 z_c^2\right)}{b_1^2-2 b_1 b_2+2 b_2^2-2 b_2 b_3+b_3^2+4 z_c^2},\notag\\&
	z_1= \frac{\sqrt{(b_1-b_3)^2+4 z_c^2} \sqrt{(b_1-b_2)^2 (b_2-b_3)^2+z_c^2 (b_1-b_3)^2+4 z_c^4}}{2 b_1 b_2+2 b_2 b_3-2 b_2^2-b_1^2-b_3^2-4 z_c^2}.
\end{align} 
Now substituting the above values in \cref{geodesic-length} and then expanding the expression linearly in $z_c$, we may obtain the bulk EWCS in this phase which is exactly half of the reflected entropy computed in \cref{Adj-conn-bdy} upon utilizing the Brown-Henneaux relation.

\subsubsection{Entanglement Entropy Phase:2}
\begin{figure}[h!]
	\centering
	\includegraphics[scale=0.5]{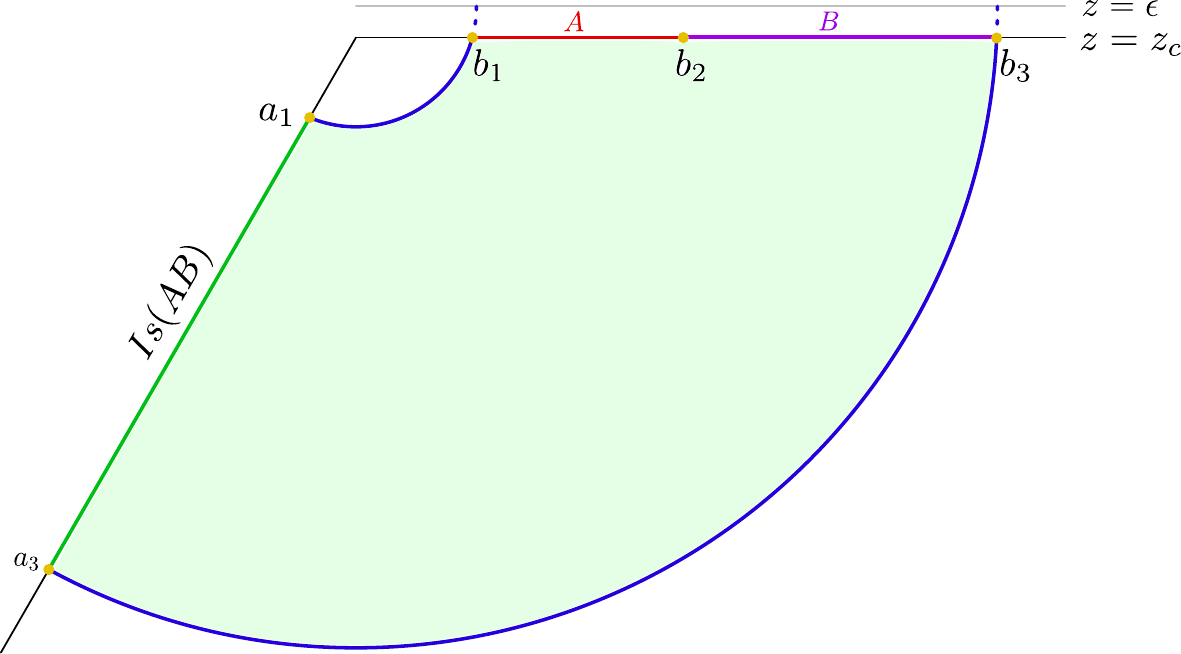}
	\caption{Schematic diagram of the EE phase $2$ when the RT surface for $A \cup B$ are shown as blue curve. } 
	\label{fig:adjacent}
\end{figure}
In this EE phase we consider that the subsystem $A$ and $B$ are close to the boundary, hence the EE is proportional to the sum of length of two island surfaces, depicted as blue curves in \cref{fig:adjacent}. The EE for this configuration is given as 
\begin{equation}
	S_2= \frac{1}{2 G_N} \frac{\sigma_0}{\ell}+\frac{1}{2 G_N}\log\left[\frac{2 \ell \cosh \frac{\sigma_0}{\ell}}{\epsilon_y}\right] +\frac{1}{4 G_N}\log\left[\frac{4 b_1 b_3}{z_c^{2}}\right]- \frac{e^{\frac{\sigma_0}{\ell}}}{4 G_N}\left(\frac{1}{ b_1}+\frac{1}{ b_3}\right)z_c.
\end{equation} 
For this EE phase, there are three possible phases of the reflected entropy or the bulk EWCS, depending on the subsystems size and location. In the following, we explain the computation of the reflected entropy in the boundary and bulk perspective.

\subsubsection*{Phase-I}\label{SR-adj2(i)}
\paragraph{The boundary perspective:}
In this phase we assume that the subsystem $A$ is smaller than the subsystem $B$, so the EWCS lands on the island surface corresponding to the subsystem $A$. 
There is no island cross section for this phase, hence in the boundary description, the reflected entropy is equal to $S^{\text{(eff)}}_{R}(A:B\cup I_{S_{R}}(B))$ which may be computed by utilizing the following expression 
\begin{equation}
	S^{\text{bdy}}_{R}(A:B)= S^{\text{eff}}_{R}(A:B\cup I_{S_{R}}(B))= \lim_{{m,n} \to 1}\frac{1}{1-n}\log \frac{      \langle \sigma_{g^{}_A}(b_1)\sigma_{g^{}_B g_A^{-1}}(b_2)\sigma_{g^{-1}_A}(a_1)\rangle}{ \langle\sigma_{g^{}_m}(b_1)\sigma_{g^{-1}_m}(a_1)\rangle^n}.
\end{equation}  
Here $a_1$ a point on the EOW brane where island surface of $A$ intersects with the EOW brane. As described in \cite{Deng:2023pjs} the intersection point may be obtained by extremizing the EE of subsystems $A\cup B$ as $a_1= b_1- 2 z_c e^{\frac{\sigma_0}{l}}$.  Now by utilizing this value in the form of the three-point twist field correlator and then expending it linearly in $z_c$, we may obtain the reflected entropy for this phase as follows
\begin{equation}\label{SR-adj-2(i)}
	S_R^{\text{bdy}}(A:B)=	\frac{c}{3}  \left(\log \left[\frac{ \left[b_2-b_1\right) \left(b_2+b_1\right)}{b_1 z_c}\right]+\frac{(b_2-b_1)~ e^{\frac{\sigma_0}{\ell}}}{b_1(b_2+b_1)}z_c\right).
\end{equation}

\paragraph{The bulk perspective:}
\begin{figure}[h!]
	\centering
	\includegraphics[scale=0.9]{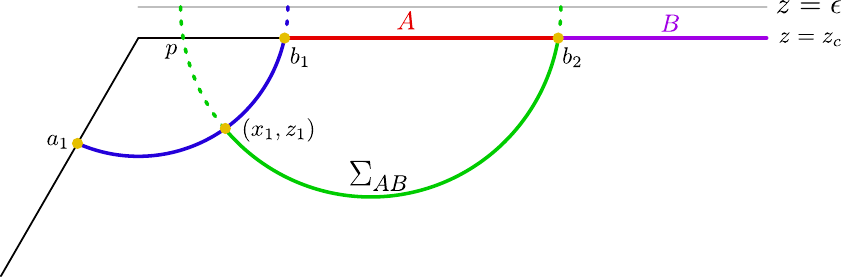}
	\caption{ Diagrammatic illustration of the bulk EWCS between subsystems A and B (depicted as solid green curves.) } 
	\label{fig:adj2}
\end{figure}
In the bulk description, the curve $\Sigma_{AB}$ connects point $b_2$ to a point $(x_1,z_1)$ on the extremal surface joining $b_1$ and $a_1$ and hence the first term of \cref{SR-bulk} vanishes as the entire island belongs to the subsystem $B$. Therefore, the reflected entropy for this case is given by the (minimal) length of the curve $\Sigma_{AB}$ shown as green curve in \cref{fig:adj2}. As explained earlier in the bulk geometry these RT surfaces or geodesics are semi-circles centred at $z=\epsilon$. The radius and centre of the geodesic connecting $b_1$ and $a_1$ is given in \cref{b1-a1-geodesic}. Now by using the fact that minimal curve $\Sigma^{\text{min}}_{AB}$ is perpendicular to this geodesic or circle at their intersection point $(x_1,z_1)$, we get the following set of the constraint equations   
\begin{align}
	&(x_1-x_0)^2+z_1^2= r^2, ~~~ \left(x_1-\frac{p+b_2}{2}\right)^2+z_1^2=\left(\frac{b_2-p}{2}\right)^2+z_c^2,~~z_{1}^2=-(x_1-x_0)\left(x_1-\frac{p+b_2}{2}\right),
\end{align}
where the extension of $\Sigma^{\text{min}}_{AB}$ intersects the boundary at point $(p,z_c)$. Now by solving the above equations we may obtain the point $(x_1, z_1)$ as follows
\begin{align}
	x_1= x_0+ \frac{2 r^2(b_2-x_0)}{r^2+ (b_2-x_0)^2+z_c^2}, ~~~~~ z_1= \frac{r \sqrt{((b_2-x_0)^2-r^2)^2+2((b_2-x_0)^2+r^2)z_c^2+z_c^4}}{(b_2-x_0)^2+r^2+z_c^2},
\end{align}
where $r$ and $x_0$ is given in \cref{b1-a1-geodesic}. Here the length of the minimal surface may be computed directly and therefore the reflected entropy for this phase is given as 
\begin{align}\label{EW-adj2-NP}
	S_R^{\text{bulk}}(\mathcal{A}:\mathcal{B})&=\frac{1}{2G_N}\cosh^{-1}\left[\frac{(x_1-b_2)^2+z_1^2+z_c^2}{2z_1 z_c}\right]\notag\\
	&=\frac{1}{2 G_N}\cosh^{-1}\left[\frac{\sqrt{((b_2-x_0)^2-r^2)^2+2((b_2-x_0)^2+r^2)z_c^2+z_c^4}}{2 r z_c}\right]
\end{align}
Now substituting the value of $x_0$, $r$ and $a_1=b_1-2e^{\frac{\sigma_0}{\ell}}z_c+\mathcal{O}(z_c^2)$ in \cref{EW-adj2-NP} and then expand perturbatively in $z_c$, we obtain
\begin{equation}
	S_R^{\text{bulk}}(\mathcal{A}:\mathcal{B})=	\frac{1}{2 G_N} \left(\log \left[\frac{ \left(b_2-b_1\right) \left(b_2+b_1\right)}{b_1 z_c}\right]+\frac{(b_2-b_1)~ e^{\frac{\sigma_0}{\ell}}}{b_1(b_2+b_1)}z_c\right)+\mathcal{O}(z_c^2).
\end{equation}
It should be noted that the reflected entropy from both the perspective matches exactly when  the Brown-Henneaux relation is used.

\subsubsection*{Phase-II}
In this reflected entropy phase we assume that the subsystem $A$ is large enough, hence the EWCS lands on the EOW brane. For this case there is a non-trivial island cross section on the brane which divides the entanglement island into two parts. 
In the boundary description, \cref{SR-bdy} may be written as
\begin{equation}\label{adj-brane-EW}
	S_R^{\text{bdy}}(A:B)=	\frac{c}{3}  \log \left[\frac{\ell (a+b_2)^2}{\epsilon_y  z_c  (z_c+a \sech \frac{\sigma_0}{\ell} )}\right]+\frac{c}{3}  \cosh ^{-1}\left[\frac{\sqrt{a^2+2 a z_c \sech \frac{\sigma_0}{\ell}+z_{c}^2}}{z_c+a \sech \frac{\sigma_0}{\ell}}\right],
\end{equation}
where $a$ is the location of the island cross section on the brane. The first term of the above expression may be obtained by utilizing two-point twist field correlator $\langle \sigma_{g_A g^{-1}_B}(b_2) \sigma_{g_B g^{-1}_A}(a)\rangle$ and the second term is the area of  island cross section given in \cref{area-of-Gamma}. Note that to obtain the location of the island cross section on the EOW brane, one has to extremize the above expression over $a$. The extremization procedure can only be done perturbatively in the present scenario. By extremizing \cref{adj-brane-EW} over $a$, we get the extremum value of $a$ as 
\begin{equation}\label{adj value of a}
	a= b_2-2 z_c e^{\frac{\sigma_0}{\ell}}+\mathcal{O}(z_{c}^2).
\end{equation}
Substituting \cref{adj value of a} in \cref{adj-brane-EW} and then expanding the expression linearly in $z_c$ the reflected entropy in the boundary description  may be obtained as  
\begin{equation}
	S_R^{\text{bdy}}(A:B)=\frac{c}{3} \left[\frac{\sigma_0}{\ell} + \log \left(\frac{4 b_2 \ell \cosh \frac{\sigma_0}{\ell}}{z_c \epsilon_y }\right)-\frac{ z_c e^{\frac{\sigma_0}{\ell}}}{ b_2}\right]+\mathcal{O}(z_{c}^2).
\end{equation}

\paragraph{The bulk perspective:}
\begin{figure}[h!]
	\centering
	\includegraphics[scale=0.6]{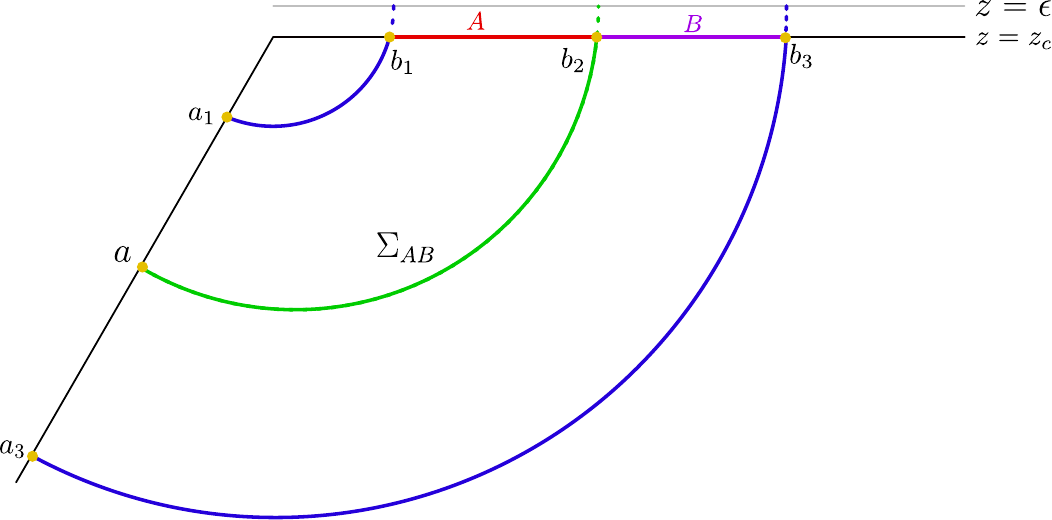}
	\caption{ Diagrammatic illustration of the EWCS between subsystems A and B (depicted as the solid green curves.) } 
	\label{fig:adj3}
\end{figure}
In the bulk description, the curve $\Sigma_{AB}$ connects point $(b_2,z_c)$ to a point $a$ on the EOW brane and thus splits the EE island into two parts, as depicted in \cref{fig:adj3}. The first term of \cref{SR-bulk} i.e. the effective reflected entropy is given in \cref{SR(eff.)_dis}. The second term which is basically the geodesic length between points $(b_2,z_c)$ and $(-a \tanh\frac{\sigma_0}{\ell}, z_c +a \sech\frac{\sigma_0}{\ell})$ is given as follows
\begin{equation}\label{EW_adj}
	\text{Area}[\Sigma_{AB}]=\cosh ^{-1}\left(\frac{(b_{2}+a \tanh \frac{\sigma_0}{\ell})^2+(z_c +a \sech\frac{\sigma_0}{\ell})^2+z_c^2}{2 z_c (z_c+a \sech\frac{\sigma_0}{\ell})}\right).
\end{equation}
By adding \cref{SR(eff.)_dis,EW_adj}, we can write the reflected entropy form the bulk side as
\begin{equation}\label{SR_adj_bulk}
	S^{\text{bulk}}_{R}(\mathcal{A}:\mathcal{B})= \frac{1}{2 G_N}\cosh ^{-1}\left(\frac{(b_{2}+a \tanh \frac{\sigma_0}{\ell})^2+(z_c +a \sech\frac{\sigma_0}{\ell})^2+z_c^2}{2 z_c (z_c+a \sech\frac{\sigma_0}{\ell})}\right)+\frac{c}{3} \log \frac{2a \ell}{\epsilon_y (z_c + a \sech \frac{\sigma_0}{\ell})}.
\end{equation}
Now by extremizing the above expression over $a$ which is only possible perturbatively, we may get the extremum value of $a$ as follows
\begin{equation}
	a= b_2- 2 z_c e^{\frac{\sigma_0}{\ell}}+ \mathcal{O}(z_{c}^2).
\end{equation}
Substituting the value of $a$ in \cref{SR_adj_bulk} and then expanding the expression perturbatively in $z_c$, the reflected entropy for this phase from the bulk side may be obtained as  
\begin{equation}
	S_R^{\text{bulk}}(\mathcal{A}:\mathcal{B})=	\frac{1}{2 G_N}  \left[\frac{\sigma_0}{\ell} + \log \left(\frac{4 b_2 \ell \cosh \frac{\sigma_0}{\ell}}{z_c \epsilon_y }\right)-\frac{ z_c e^{\frac{\sigma_0}{\ell}}}{ b_2}\right]+\mathcal{O}(z_{c}^2).
\end{equation}
Here also the reflected entropy computed from the boundary perspective exactly matches with the bulk perspective upon utilizing the Brown-Henneaux relation.

\subsection*{Phase-III}
\begin{figure}[h!]
	\centering
	\includegraphics[scale=0.6]{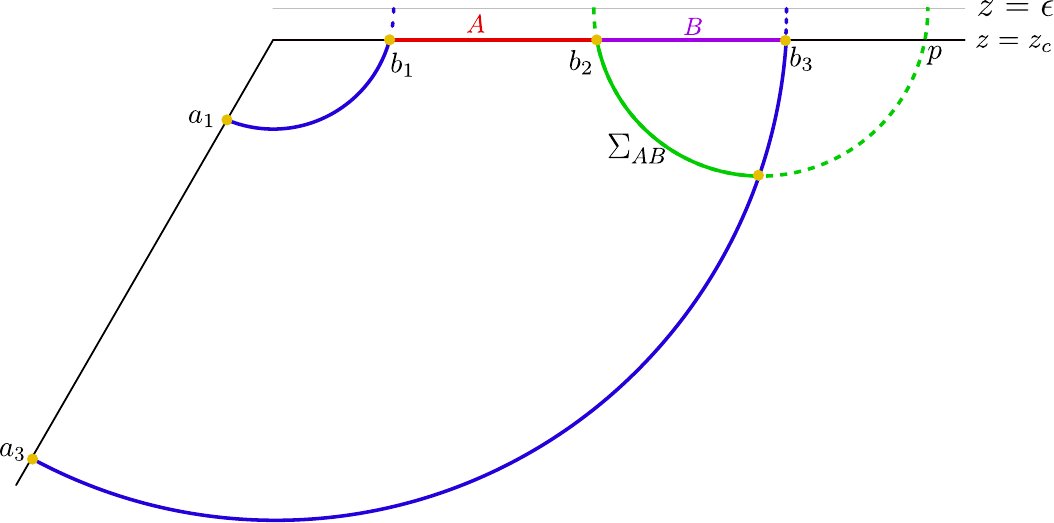}
	\caption{Schematic illustrating the bulk EWCS phase between $A$ and $B$ (represented by the solid green curve.) } 
	\label{fig:adj4}
\end{figure}
In this phase we consider that the subsystem $B$ is smaller than the subsystem $A$, so the bulk EWCS lands on the island surface corresponding to the subsystem $B$, depicted as solid green curve in \cref{fig:adj4}. Now the reflected entropy in the boundary description may be obtained by interchanging $b_1$ and $b_3$ in \cref{SR-adj-2(i)} as 
\begin{align}\label{SR-adj-2(iii)}
	S_R^{\text{bdy}}(A:B)=	\frac{c}{3}  \left(\log \left[\frac{ \left[b_3-b_2\right) \left(b_3+b_2\right)}{b_3 z_c}\right]+\frac{(b_3-b_2)~ e^{\frac{\sigma_0}{\ell}}}{b_3(b_3+b_2)}z_c\right)+\mathcal{O}(z_{c}^2).
\end{align}
In the bulk description, the computation of the reflected entropy may be done similarly to \hyperref[SR-adj2(i)]{phase-I} and the expression for the reflected entropy may be obtained by exchanging the $b_1$ and $b_3$ which exactly matches with \cref{SR-adj-2(iii)} upon utilizing the Brown-Henneaux relation.

\section{Time dependent reflected entropy in black hole evaporation}\label{sec:Finite_Temp}
In this section, we study the reflected entropy between the black hole interiors
and radiation region in a time dependent AdS$_3$/BCFT$_2$ scenario with the $T\bar{T}$ deformation involving an eternal black hole in the effective two-dimensional description \cite{Deng:2023pjs}. We also plot the analogue of the Page
curve for the reflected entropy and observe the correction due to $T\bar{T}$ deformation.

\subsection{The emergence of a 2d eternal black hole}

We begin by reviewing how a 2d eternal black hole emerges from the AdS$_3$/BCFT$_2$ model \cite{Li:2021dmf,Rozali:2019day,Chu:2021gdb}. As discussed in the previous section, the holographic dual to a 2d BCFT is an AdS$_3$ geometry with a codimension-1 EOW brane. The metric of Euclidean AdS$_3$ bulk is given as
\begin{align}
	ds^2 & =\frac{\ell^2}{z^2} (d\tau^2+dz^2+dx^2)   \notag
	\\ & = d\sigma^2+\frac{\ell^2 \cosh^2 \frac{\sigma}{\ell}}{y^2}\left( d\tau^2+dy^2 \right).
\end{align}

We consider the bulk geometry to be bounded by a BCFT defined on a half plane given by $ (x,\tau >0)$\footnote{In Euclidean spacetime, spacelike and timelike coordinates are equivalent, and thus $x$ and $\tau$ can be interchanged as per convenience.}. The EOW brane is located at $\tau=-z\sinh \frac{\sigma_0}{\ell}$, with $\sigma_0$ being a constant.

\begin{figure}[h!]
	\centering
	\includegraphics[scale=0.8]{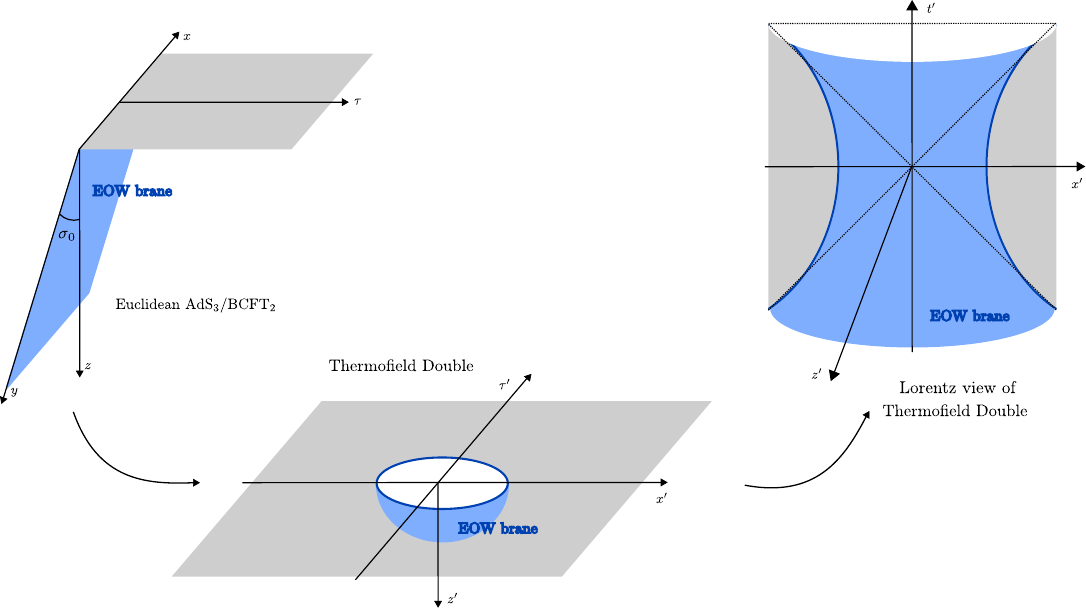}
	\caption{In the first step a global conformal transformation takes us from from Euclidean AdS$_3$/BCFT$_2$ to a thermofield double state. It is followed by analytically continuing $\tau ' \to it'$ which provides us the Lorentz view of the TFD.}
	\label{fig_GCT}
\end{figure}

Using the set of global conformal transformations as follows 
\begin{align}\label{eq_gct}
	& \tau = \frac{2(x'^2+\tau '^2+z'^2-1)}{(\tau '+1)^2+x'^2+z'^2}, \notag
	\\ &  x = \frac{4x'}{(\tau '+1)^2+x'^2+z'^2},
	\\ &  z = \frac{4z'}{(\tau '+1)^2+x'^2+z'^2}, \notag
\end{align}
the boundary of the BCFT and the EOW brane may be mapped to a circle and a part of a sphere respectively as 
\begin{align}\label{eq_2dbh}
	x'^2+\tau '^2 =1, \ \  \ \ \ \  (z'+\sinh \frac{\sigma_0}{\ell})^2 + x'^2 +\tau '^2 = \cosh ^2 \frac{\sigma_0}{\ell}.
\end{align}

Note that the global conformal transformations given by \cref{eq_gct} ensure that the metric of the bulk as well as the metric induced on the EOW brane is preserved. The inverse transformations to \cref{eq_gct} are provided in \cite{Basu:2022reu}.

From the above bulk configuration, a two-sided 2d eternal black hole may be obtained using the partial Randall-Sundrum reduction and AdS$_3$/BCFT$_2$ correspondence \cite{Deng:2020ent}. The black hole appears on the EOW brane, which in the effective 2d description is coupled to the BCFT$_2$ outside the circle in \cref{eq_2dbh}. This hybrid manifold may be described as analytical continuation of $\tau ' \to it'$ followed by introduction of the Rindler coordinates $(X,T)$ as \cite{Almheiri:2019qdq}
\begin{align}\label{eq_rt}
	x' = e^X \cosh T, \ \  \ \ \ \   t'= e^X \sinh T, 
\end{align}
which capture the near-horizon geometry of the black hole.

\subsubsection*{$T\bar{T}$ deformation}

As elaborated in \cref{sec:review}, the boundary of AdS$_3$ is pushed into the bulk to a finite radial cut-off surface when the theory is deformed by the $T\bar{T}$ operator. As a result, the EOW brane is shifted to $\tau=-(z-z_c)\sinh \frac{\sigma_0}{\ell}$. The boundary of the BCFT and the brane profile in \cref{eq_2dbh} is thus modified as
\begin{align}
	x'^2+\tau '^2 =1-z_c'^2, \ \  \ \ \ \  (z'+\sinh \frac{\sigma_0}{\ell})^2 + x'^2 +\tau '^2 = \cosh ^2 \frac{\sigma_0}{\ell}+2z_c'\sinh \frac{\sigma_0}{\ell},
\end{align}
respectively, where the relation between $z_c'$ and $z_c$ is provided by \cref{eq_gct}. Subsequently, we may once again obtain the two-sided 2d eternal black hole via transformations in \cref{eq_rt}, using the partial Randall-Sundrum reduction and AdS$_3$/BCFT$_2$ correspondence. The UV cut-off in the Rindler coordinates $z_R$ can be related to $z_c'$ as
\begin{align}\label{eq_rind_cutoff}
	z_c'=z_Re^X,
\end{align}
where $z_R$ is assumed to be a constant everywhere.

\subsection{Reflected entropy between black hole interiors}

In this section we compute the time dependent reflected entropy between different regions of the black hole interior in presence of $T\bar{T}$ deformation. As illustrated in \cref{fig_BBC,fig_BBDC}, the black hole region is defined by the spacelike interval between points $Q \equiv (\tau_1',-x_1',z_1')$ and $P \equiv (\tau_1',x_1',z_1')$ \cite{Li:2021dmf,Chu:2021gdb}. To simplify the computation we initially work in the unprimed coordinate system. By employing the transformations described in \cref{eq_gct}, the endpoints of the black hole region, given by $(\tau_1',\pm x_1',z_1')$ in the primed coordinates, may be mapped to $(\tau_1,\pm x_1,z_1)$ in the unprimed coordinates.

The configuration of the black hole region—specifically, its size and location determines whether we encounter an island or a no-island phase for the entanglement entropy. Correspondingly, for each EE phase there are various possible reflected entropy phases between the black hole subsystems $B_L = |QO|$ and $B_R = |OP|$ (denoted by the red lines in \cref{fig_BBC} and related illustrations), $O$ being the extremal point on the brane.

\subsubsection{Entanglement Entropy Phase: 1}
\begin{figure}[h!]
	\centering
	\includegraphics[scale=.8]{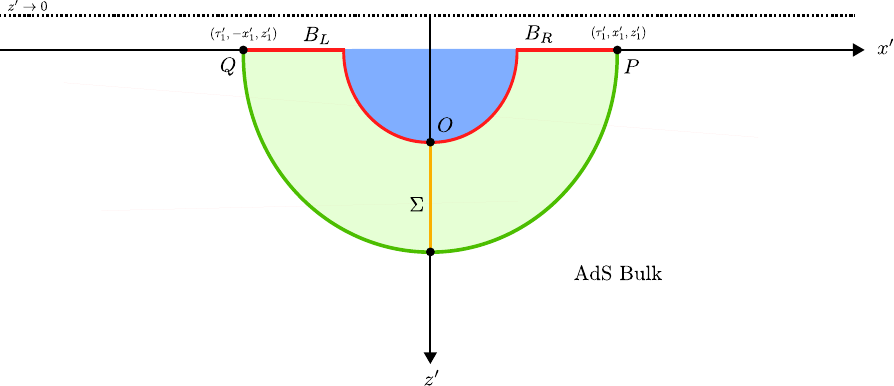}
	\caption{ This diagram illustrates the no-island phase of entanglement entropy of the black hole region. The red lines represent the black hole interior, green line represents the HM surface connecting the endpoints of the black hole region, while the yellow line represents the entanglement wedge cross section.}
	\label{fig_BBC}
\end{figure}

In this phase, EE does not include any contribution from an island, hence the EE is
determined by the length of the Hartman-Maldacena (HM) surface \cite{Hartman:2013qma} $OP$, depicted by the green curve in \Cref{fig_BBC}. As evaluated in \cite{Deng:2023pjs}, in this phase the EE in Rindler coordinates is given by
\begin{align}\label{eq_ee_ni}
	S(B_L \cup B_R)= \frac{c}{3} \log \frac{2 \cosh T}{z_R}.
\end{align}
In this scenario, there exists only a single possible phase of the reflected entropy
or the bulk EWCS between the black hole regions $B_L$ and $B_R$ shown as solid yellow line in \Cref{fig_BBC}. In the following we explain the computation of this both from the boundary and bulk perspectives.

\subsubsection*{Boundary description}

In the boundary description, the reflected entropy between $B_L$ and $B_R$ may be computed using \cref{SR-bdy}, where the first term (the effective term) is given by a three-point twist field correlators as
\begin{align}
	S_R^\text{eff}(B_L:B_R)=\lim_{m, n \to 1} \frac{1}{1-n} \log \frac{\Omega_O^{2 h_{g_A^{-1}g_B}}\langle   \sigma_{g_A}(P) \sigma_{g_A^{-1}g_B}(O) \sigma_{g_B^{-1}}(Q)  \rangle _{CFT^{\otimes mn}}}{\langle   \sigma_{g_m}(P) \sigma_{g_m}(Q)   \rangle^n_{CFT^{\otimes m}}},
\end{align}
where $\Omega_O$ is the conformal factor. By adding the area term to $S_R^{eff}$, which has the same functional form as \cref{area-of-Gamma}, the generalised reflected entropy may be written as
\begin{align}\label{eq_bbc_bdy_gen}
	S_R^\text{bdy}(B_L:B_R)  =&S_R^{eff}(B_L:B_R)+S_{area}(O) \notag
	\\ =&\frac{c}{3} \left( \log \left[ \frac{x_1^2+(\tau_1+y)^2}{ x_1}\right]+\log \frac{ \ell}{\epsilon_y(z_1+y \sech \frac{\sigma_0}{\ell})} \right)\notag\\
	 &+ \frac{c}{3} \cosh ^{-1} \left[ \frac{\sqrt{y^2+z_1^2+2z_1 y \sech \frac{\sigma_0}{\ell}}}{z_1+y \sech \frac{\sigma_0}{\ell}} \right],
\end{align}
with $y$ being the brane coordinate of $O$.
Now by extremizing the above expression over $y$ perturbatively in $z_1$, we get the extremum value of $y$ as
\begin{align}\label{y-ext-bbc}
	y=\sqrt{x_1^2+\tau_1^2}-z_1 e^{\frac{\sigma_0}{\ell}} \left(  \frac{\tau_1+\sqrt{x_1^2+\tau_1^2}}{\sqrt{x_1^2+\tau_1^2}}  \right) +\mathcal{O}(z_{1}^2).
\end{align}
Putting back $y$ into \cref{eq_bbc_bdy_gen} the extremized reflected entropy is obtained upto first order in $z_1$ as
\begin{align}\label{eq_bbc_bdy}
	S_R^\text{bdy}(B_L:B_R)& = \frac{c}{3} \left( \log \left[ \frac{\tau_1+\sqrt{x_1^2+\tau_1^2}}{x_1} \right]+\log \frac{2 \ell}{\epsilon_y \sech \frac{\sigma_0}{\ell}} +\frac{\sigma_0}{\ell} \right)  -\frac{c}{3}\frac{z_1 e^{\frac{\sigma_0}{\ell}}}{\sqrt{x_1^2+\tau_1^2}}+\mathcal{O}(z_{1}^2).
\end{align}

\subsubsection*{Bulk description}

In the bulk description, the reflected entropy between $B_L$ and $B_R$ is computed using \cref{SR-bulk}. The set up for computing the length of the entanglement wedge cross section $\Sigma$ (denoted by the yellow line in \cref{fig_BBC}) is described in \cite{Li:2021dmf}. As discussed previously, the calculation is initially carried out in the unprimed coordinates for simplicity. Following \cite{Li:2021dmf}, the computation is reduced to finding the geodesic length between points $(\tau_1,x,\sqrt{-x^2+x_1^2+z_1^2})$ (see \cref{appFT}) and $O \equiv (-y\tanh \frac{\sigma_0}{\ell},x_y,y\sech \frac{\sigma_0}{\ell}+z_1)$ on the brane, which is given using \cref{geodesic-length} as \cite{Ryu:2006bv,Hubeny:2007xt}
\begin{align}\label{eq_ran1}
	\text{Area}[\Sigma]=\cosh^{-1} \left[ \frac{x_1^2-2xx_y+x_y^2+y^2+2z_1^2+\tau_1^2 +2yz_1\sech \frac{\sigma_0}{\ell} +2y\tau_1\tanh \frac{\sigma_0}{\ell}}{2 \sqrt{-x^2+x_1^2+z_1^2}(z_1+y\sech \frac{\sigma_0}{\ell})} \right].
\end{align}
Adding the defect term \cref{SR(eff.)_dis} to \cref{eq_ran1}, the generalized holographic reflected entropy is given using \cref{SR-bulk} as
\begin{align}\label{eq_bbc_bulk_gen}
	S_R^\text{bulk}(B_L:B_R)  = & \frac{c}{3} \cosh^{-1} \left[ \frac{x_1^2-2xx_y+x_y^2+y^2+2z_1^2+\tau_1^2 +2yz_1\sech \frac{\sigma_0}{\ell} +2y\tau_1\tanh \frac{\sigma_0}{\ell}}{2 \sqrt{-x^2+x_1^2+z_1^2}(z_1+y\sech \frac{\sigma_0}{\ell})} \right] \notag
	\\ & \ \ + \frac{c}{3} \log \frac{2y\ell}{\epsilon_y (z_1+y \sech \frac{\sigma_0}{\ell})}.
\end{align}
Extremizing the above equation over $y$ and $x_y$, we get the extremum values as (up to first order in $z_1$), 
\begin{align}
	y=\sqrt{x_1^2+\tau_1^2}-z_1 \left(  \frac{\tau_1+\sqrt{x_1^2+\tau_1^2}}{\sqrt{x_1^2+\tau_1^2}}  \right) e^{\frac{\sigma_0}{\ell}}+\mathcal{O}(z_{1}^2), \ \ \ \ \ \  x_y=x=0.
\end{align}
Inserting the value of $y$ and $x_y$ in \cref{eq_bbc_bulk_gen}, we obtain the reflected entropy in the bulk description which is identical with the field theory computation outlined in \cref{eq_bbc_bdy}.
Applying the transformations detailed in \cref{eq_gct,eq_rt,eq_rind_cutoff}, the reflected entropy can be expressed in terms of the Rindler coordinates as
\begin{align}\label{eq_ran2}
	S_R^\text{bulk}(B_L:B_R)& = \frac{c}{3} \left( \log \left[ \frac{\sinh X_1+\sqrt{\cosh^2T+\sinh^2X_1}}{\cosh T} \right]  +\log \frac{2 \ell}{\epsilon_y \sech \frac{\sigma_0}{\ell}} +\frac{\sigma_0}{\ell} \right)\notag \\ & -\frac{c}{3} \frac{z_R e^{\frac{\sigma_0}{\ell}}}{\sqrt{\cosh^2T+\sinh^2X_1}}+\mathcal{O}(z_{R}^2),
\end{align}
where $X_1$ is a fixed constant specifying the boundary of the black hole region for a constant $T$ slice in the Rindler coordinates.

\subsubsection{Entanglement Entropy Phase: 2}

\begin{figure}[h!]
	\centering
	\includegraphics[scale=.8]{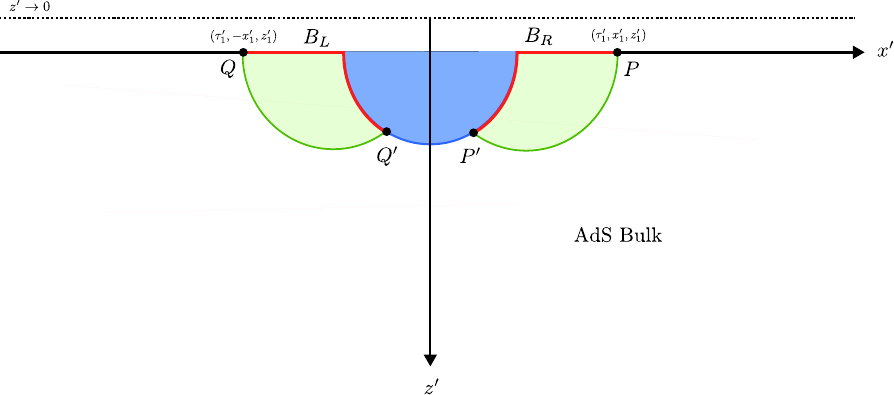}
	\caption{The illustration depicts the island phase of EE of the black hole interior. The green lines represent the disconnected RT surfaces. We have no entanglement wedge cross section in this phase.}
	\label{fig_BBDC}
\end{figure}

As depicted in \cref{fig_BBDC}, this phase of EE includes contributions from the island regions. Consequently, the EE is determined by the combined lengths of the two island surfaces and two RT surfaces denoted by the green curves expressed as \cite{Deng:2023pjs}
\begin{align}\label{eq_ee_is}
	S(B_L \cup B_R)=\frac{c}{3} \left( \frac{\sigma_0}{\ell} +\log \frac{2 \sinh X_1}{z_R} +\log \frac{2 \ell}{\epsilon_y \sech \frac{\sigma_0}{\ell}}   \right)-\frac{c}{3} \frac{z_Re^{\frac{\sigma_0}{\ell}}}{\sinh X_1}.
\end{align}
Similar to the previous EE phase, this configuration leads to a single possible reflected entropy phase between black hole subsystems $B_L=|QQ'|$ and $B_R=|PP'|$, where $Q'$ and $P'$ are the points where the disconnected RT surfaces and the brane intersect. The detailed computation of the reflected entropy in this scenario is as follows.

\subsubsection*{Boundary description}

In this configuration, there is no contribution from the area term since the two regions $B_L$ and $B_R$ do not intersect. Consequently, the second term in \cref{SR-bdy} no longer contributes in this case. Therefore the generalized reflected entropy is simply the effective reflected entropy term between $|QQ'|$ and $|PP'|$, which can be computed  from a four-point correlation function of twist operators at brane points $Q'$ and $P'$, and boundary points $Q$ and $P$ as
\begin{align}\label{eq_bbdc_bdy_gen}
	S_R^\text{bdy}=S_R^\text{eff}=\lim_{m, n \to 1} \frac{1}{1-n} \log \frac{\Omega_{Q'}^{2 h_{g_A}}\Omega_{P'}^{2 h_{g_A}}\langle   \sigma_{g_A}(Q') \sigma_{g_A^{-1}}(Q) \sigma_{g_B}(P') \sigma_{g_B^{-1}}(P)   \rangle _{CFT^{\otimes mn}}}{\Omega_{Q'}^{2nh_m}\Omega_{P'}^{2nh_m}\langle   \sigma_{g_m}(Q') \sigma_{g_m}(Q) \sigma_{g_m}(P') \sigma_{g_m}(P)   \rangle^n_{CFT^{\otimes m}}}.
\end{align}
This four-point correlator can further be factorized into two-point correlators in the large $c$ limit as follows
\begin{align}
	S_R^\text{bdy}(B_L:B_R)=\lim_{m, n \to 1} \frac{1}{1-n} \log \frac{\Omega_{Q'}^{2 h_{g_A}}\Omega_{P'}^{2 h_{g_A}}\langle \sigma_{g_A}(Q') \sigma_{g_A^{-1}}(Q) \rangle _{CFT^{\otimes mn}} \langle \sigma_{g_B}(P') \sigma_{g_B^{-1}}(P') \rangle _{CFT^{\otimes mn}}}{\Omega_{Q'}^{2nh_m}\Omega_{P'}^{2nh_m}\langle \sigma_{g_A}(Q') \sigma_{g_A^{-1}}(Q) \rangle^n_{CFT^{\otimes m}} \langle \sigma_{g_B}(P') \sigma_{g_B^{-1}}(P') \rangle^n_{CFT^{\otimes m}}}.
\end{align}
It is easy to show that the numerator and the denominator in the above expression cancel out in the limit $n \to 1$. Thus the reflected entropy vanishes for this phase.

\subsubsection*{Bulk description}

As shown in \cref{fig_BBDC}, the entanglement wedges of $B_L$ and $B_R$ are naturally disjoint, as a result there is no EWCS in this configuration. Therefore the holographic generalised reflected entropy receives contribution only from the defect term, represented by the first term in \cref{SR-bulk}. This term can be computed using the BCFT two point correlator of twist operators at $Q'$ and $P'$, expressed as
\begin{align}\label{eq_bbdc_bulk_gen}
	S_R^\text{bulk}(B_L:B_R)=S_R^\text{eff}=\lim_{m, n \to 1} \frac{1}{1-n} \log \frac{\Omega_{Q'}^{2 h_{g_A}}\Omega_{P'}^{2 h_{g_A}}\langle   \sigma_{g_A}(Q') \sigma_{g_B}(P')    \rangle _{BCFT^{\otimes mn}}}{\Omega_{Q'}^{2nh_m}\Omega_{P'}^{2nh_m}\langle   \sigma_{g_m}(Q') \sigma_{g_m}(P')    \rangle^n_{BCFT^{\otimes m}}}.
\end{align}
Using the doubling trick \cite{Sully:2020pza} we may write
\begin{align}
	\langle \sigma_{g_A}(Q')\sigma_{g_B}(P') \rangle_{BCFT}= \langle \sigma_{g_A}(Q')\sigma_{g_A}(Q'')\sigma_{g_B}(P')\sigma_{g_B}(P'') \rangle_{CFT},
\end{align}
where $P''$ and $Q''$ are symmetric points to $P'$ and $Q'$ respectively. As a result \cref{eq_bbdc_bulk_gen} is essentially the same as \cref{eq_bbdc_bdy_gen} in the boundary picture, and therefore the reflected entropy vanishes in the limit $n \to 1$.

\subsection{Reflected entropy between black hole interior and radiation}\label{ssec_bhrad}

\begin{figure}[h!]
	\centering
	\includegraphics[scale=.8]{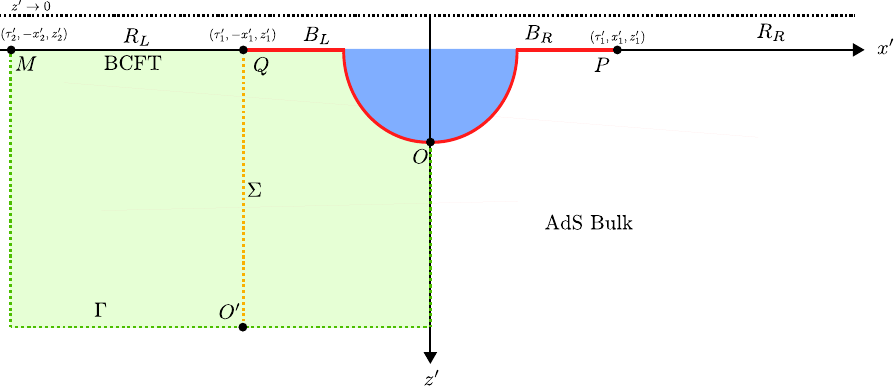}
	\caption{ The diagram represents the no-island phase for reflected entropy between radiation and black hole. The green shaded region represents the entanglement wedge for the left system $R_L \cup B_L$, which is separated from the right system by the $z$ axis. The entanglement wedge cross section is represented by the yellow dashed line $QO'$.}
	\label{fig_BRC}
\end{figure}

In this subsection we compute the reflected entropy between the black hole region and radiation region in presence of $T\bar{T}$ deformation in a time dependent AdS$_3$/BCFT$_2$ setup. As illustrated in \cref{fig_BRC,fig_BRDC}, the black hole region and the radiation region are described as $B_L$ (denoted once again by the red line) and $R_L=|MQ|$ respectively. Unlike the previous case, we only have one entanglement phase denoted by the RT surface $\Gamma$ ending on boundary point $M \equiv (\tau_2',-x_2',z_2')$ and brane point $O$ (the dashed green line in \cref{fig_BRC,fig_BRDC}). At later stages of the computation we use the approximation that $x_2'\to \infty$, assuming that our subsystem extends infinitely to encompass the entirety of the spacetime of the left side. However, within this entanglement entropy phase, there are two possible reflected entropy phases depending on whether reflected entropy islands exist or not. The final reflected entropy between the black hole interior and radiation region is determined by the minimum of these two phases.

\subsubsection{The Entanglement entropy phase}

For simplicity of computation we start again in the unprimed coordinates. Using transformations detailed in \cref{eq_gct} the left subsystem $B_L \cup R_L$ now ends at the boundary point $M \equiv (\tau_2,-x_2,z_2)$. In the limit $x_2'\to \infty$ this point reduces to $(2,0,\epsilon)$, where $\epsilon \to 0$, effectively pushing the endpoint $M$ to spatial infinity. On the other hand, we assume that the Ryu Takayanagi surface $\Gamma$ intersects at the brane at $O\equiv(y,x_y)$. The computations will follow the same line of arguments as given in \cite{Deng:2023pjs}.

\subsubsection*{Boundary description}

The entanglement entropy in the boundary description may be computed by using the island formula \cite{Almheiri:2019hni,Almheiri:2019psf,Almheiri:2019qdq}. The effective EE ($S^{eff}$) in the island formula, which for the left subsystem $B_L \cup R_L$ is given by the two point correlator of twist fields at brane point $O$ and boundary point $M$ as
\begin{align}
	S^\text{eff}(B_L \cup R_L)=\lim_{m \to 1} \frac{1}{1-m} \log \left[ \Omega_O^{2 h_m}\langle   \sigma_{g_m^{-1}}(M) \sigma_{g_m}(O)  \rangle _{CFT^{\otimes m}}\right],
\end{align}
where $h_m=\frac{c}{24} \left( m-\frac{1}{m} \right)$. Including the area term, which in this scenario retains the same functional form as shown in \cref{area-of-Gamma}, the generalized entanglement entropy can be expressed as
\begin{align}\label{eq_ee_bdy_gen}
	S^\text{gen}(B_L \cup R_L)= & \frac{c}{3} \left( \log \left[ \frac{(x_2+x_y)^2+(\tau_2+y)^2}{\epsilon} \right] +\log \left[  \frac{\ell}{\epsilon_y (z_2+y \sech \frac{\sigma_0}{\ell})} \right] \right. \notag \\ &  \left. +\cosh ^{-1} \left[ \frac{\sqrt{y^2+z_2^2+2z_2 y \sech \frac{\sigma_0}{\ell}}}{z_2+y \sech \frac{\sigma_0}{\ell}} \right]   \right).
\end{align}
Extremizing \cref{eq_ee_bdy_gen} for the brane coordinate $y$, we find that $O\equiv(2,0)$ in the limit $x_2' \to \infty$. Thus the entanglement entropy is
\begin{align}\label{eq_ee_bdy}
	S(B_L \cup R_L)= \frac{c}{3} \left( \log \frac{4}{\epsilon}+\log \frac{2\ell}{\epsilon _y \sech \frac{\sigma_0}{\ell}} +\frac{\sigma_0}{\ell} \right).
\end{align}

\subsubsection*{Bulk description}

In the bulk description, the holographic EE may be written by using \cref{DES formula} as 
\begin{align}\label{eq_ee_bulk_gen}
	S(B_L \cup R_L)= & \frac{c}{3} \left( \cosh^{-1}\left[ \frac{(x_2+x_y)^2 +y^2+2z_2^2 +\tau_2^2 +2yz_c\sech \frac{\sigma_0}{\ell} +2y\tau_2\tanh \frac{\sigma_0}{\ell}}{2z_2^2+2yz_2\sech \frac{\sigma_0}{\ell}} \right] \right. \notag \\ & \left. +\log \left[  \frac{2 y \ell}{\epsilon_y (z_2+y \sech \frac{\sigma_0}{\ell})} \right] \right),
\end{align}
where the first term is the geodesic length of the RT surface $\Gamma$ and the second term represents the matter defect contribution on the brane, computed in \cref{SR(eff.)_dis}. Now for the brane coordinate $y$, \cref{eq_ee_bulk_gen} is extremized (in the limit $x_2' \to \infty$) at $y=2$ and $x_y=0$. Putting back these values in \cref{eq_ee_bulk_gen}, we get the holographic entanglement entropy in the limit $x_2' \to \infty$ to be the same as \cref{eq_ee_bdy}.

\subsubsection{Reflected Entropy Phase: 1}

With the endpoints of the left system $B_L \cup R_L$ determined, we can proceed to compute the reflected entropy between subsystems $B_L$ and $R_L$. As depicted in \cref{fig_BRC}, the EWCS intersects with the surface $\Gamma$ for the first reflected entropy phase. Consequently, the radiation region $R_L$ does not have any island on the brane, and the entire black hole region on the brane belongs to $B_L$.

\subsubsection*{Boundary description}

Since the radiation subsystem $R_L$ does not possess any island on the brane, the area term in \cref{SR-bdy} does not contribute to the generalised reflected entropy. As a result, in this phase the reflected entropy between subsystems $B_L$ and $R_L$ is determined by a three point correlation function of brane point $O$, and boundary points $M$ and $Q$ as
\begin{align}
	S_R^\text{bdy}(B_L:R_L) & =\lim_{m, n \to 1} \frac{1}{1-n} \log \frac{ \Omega_{O}^{2 h_{g_B}}  \langle   \sigma_{g_A}(M) \sigma_{g_A^{-1}g_B}(Q) \sigma_{g_B^{-1}}(O)  \rangle _{CFT^{\otimes mn}}}{\langle   \sigma_{g_m}(M) \sigma_{g_m}(O)   \rangle^n_{CFT^{\otimes m}}}.
\end{align} 
In the limit $x'_2 \to \infty$, the reflected entropy has the form
\begin{align}\label{eq_brc_bdy}
	S_R^\text{bdy}(B_L:R_L) = \frac{c}{3} \log \left[ \frac{\sqrt{x_1^2+(\tau_1-2)^2} \sqrt{x_1^2+(\tau_1+2)^2}}{2 z_1}  \right]=\frac{c}{3} \log \left[ \frac{2}{z_R}  \right],
\end{align}
where in the final step we use \cref{eq_gct,eq_rt,eq_rind_cutoff} to express the results in the Rindler coordinates.

\subsubsection*{Bulk description}

In this phase, the radiation region lacks an island on the brane, causing the first term in \cref{SR-bulk} to vanish. Therefore, computing the holographic reflected entropy in this scenario simplifies to evaluating the second term, which is the geodesic length of $\Sigma$. The equation of $\Gamma$ in the limit $x_2' \to \infty$ is given by (check \cref{appRB} for details)
\begin{align}\label{eq_rtgamma}
	z^2+\tau^2=4.
\end{align}
Assuming that $\Sigma$ ends on an arbitrary point on $\Gamma$, its length is obtained using \cref{geodesic-length} as
\begin{align}\label{eq_areasigma}
	\text{Area}[\Sigma]=\cosh^{-1}\left[ \frac{4+x_1^2-2\tau \tau_1+\tau_1^2}{2 z_1 \sqrt{4-\tau^2}} \right].
\end{align}
Extremizing \cref{eq_areasigma} over $\tau$, we get the extremum value of $\tau=\frac{8\tau_1}{4+x_1^1+\tau_1^2}$. Substituting the value of $\tau$ in the above expression the reflected entropy in the bulk description matches exactly with the field theory results in \cref{eq_brc_bdy}.

\subsubsection{Reflected Entropy Phase: 2}

The second reflected entropy phase corresponds to the scenario where $\Sigma$ intersects on the brane, leading to the formation of a radiation island region on the brane. Although the entirety of the island region $Q'P'$ belongs to the radiation subsystem $R_L \cup R_R$, only the portion $Q'O$ belongs to the left radiation subsystem $R_L$.
\begin{figure}[h!]
	\centering
	\includegraphics[scale=.8]{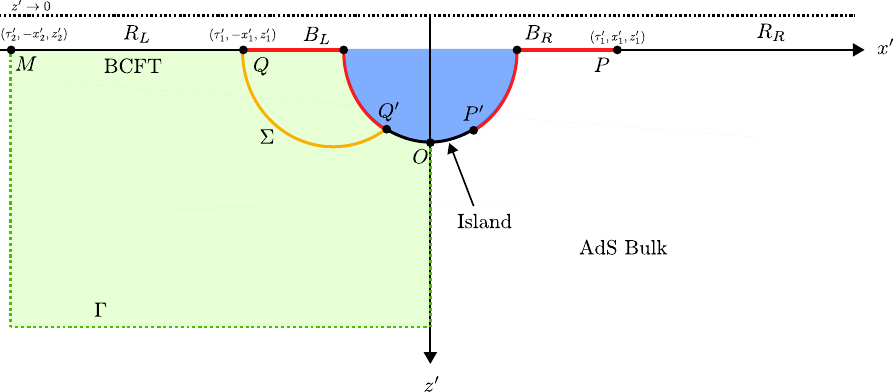}
	\caption{This illustration represents the island phase of reflected entropy between radiation and black hole. The island region (denoted by the black curve $Q'P'$) belongs to the entanglement wedge of radiation, of which the section $Q'O$ belongs to the $R_L$. The EWCS is given by the yellow line $QQ'$.}
	\label{fig_BRDC}
\end{figure}

\subsubsection*{Boundary description}

As evident from \cref{fig_BRDC}, the effective reflected entropy is given by the two point function of twist operators at the boundary point $Q$ and brane point $Q'$ as
\begin{align}
	S_R^\text{eff}(B_L:R_L)=\lim_{m, n \to 1} \frac{1}{1-n} \log \frac{\Omega_{Q'}^{2 h_{g_Ag_B^{-1}}}\langle  \sigma_{g_A^{-1}g_B}(Q) \sigma_{g_Ag_B^{-1}}(Q')  \rangle _{CFT^{\otimes mn}}}{\langle   \sigma_{g_m}(Q) \sigma_{g_m}(Q')   \rangle^n_{CFT^{\otimes m}}}.
\end{align}
Adding to it the area term \cref{area-of-Gamma}, the generalised reflected entropy \cref{SR-bdy} may be obtained as
\begin{align}
	S_R^\text{gen}(B_L:R_L)=\frac{c}{3} \left( \log \frac{(y+\tau_1)^2}{z_1 } +\log \frac{ \ell}{\epsilon_y (z_1+y \sech \frac{\sigma_0}{\ell})} \right)+ \frac{c}{3} \cosh ^{-1} \left[ \frac{\sqrt{y^2+z_1^2+2z_1 y \sech \frac{\sigma_0}{\ell}}}{z_1+y \sech \frac{\sigma_0}{\ell}} \right]
\end{align}
Extremizing $S_R^{gen}$ over the brane coordinate $y$ up to first order in $z_1$, we get $y=\tau_1-2 z_1 e^{\frac{\sigma_0}{\ell}}+\mathcal{O}(z_{1}^2)$. This gives the final reflected entropy for this phase as
\begin{align}\label{eq_brdc_bdy}
	S_R^\text{bdy}(B_L:R_L)&=\frac{c}{3} \left( \log \frac{2 \tau_1}{z_1}+\log \frac{2\ell}{\epsilon _y \sech \frac{\sigma_0}{\ell}} +\frac{\sigma_0}{\ell} \right)-\frac{c}{3} \frac{z_1 e^{\frac{\sigma_0}{\ell}}}{\tau _1}+\mathcal{O}(z_{1}^2).
\end{align}

\subsubsection*{Bulk description}

In this scenario, the area term in the generalised reflected entropy \cref{SR-bulk} is simply the length of the EWCS $\Sigma$, which may be obtained using \cref{area-of-Gamma}. Including the matter defect term, the generalized reflected entropy is obtained as
\begin{align}\label{eq_brdc_bulk_gen}
	S_R^\text{gen}=\frac{c}{3} \cosh^{-1} \left[ \frac{y^2+2z_1^2+\tau_1^2+2yz_1\sech \frac{\sigma_0}{\ell} +2 y \tau_1 \tanh \frac{\sigma_0}{\ell}}{2 z_1^2 +2 y z_1 \sech \frac{\sigma_0}{\ell}} \right]+\frac{c}{3} \log \frac{2y\ell}{\epsilon_y (z_1+y \sech \frac{\sigma_0}{\ell})}.
\end{align}
Extremizing \cref{eq_brdc_bulk_gen} up to first order in $z_1$, we have $y_{min}=\tau_1-2 z_1 e^{\frac{\sigma_0}{\ell}} +\mathcal{O}(z_{1}^2)$, which gives us
\begin{align}\label{eq_brdc_bulk}
	S_R^\text{bulk}(B_L:R_L)&=\frac{c}{3} \left( \log \frac{2 \tau_1}{z_1}+\log \frac{2\ell}{\epsilon _y \sech \frac{\sigma_0}{\ell}} +\frac{\sigma_0}{\ell} \right)-\frac{c}{3} \frac{z_1 e^{\frac{\sigma_0}{\ell}}}{\tau _1}+\mathcal{O}(z_{1}^2) \notag
	\\ & = \frac{c}{3} \left( \log \frac{2\sinh X_1}{z_R} + \log \frac{2 \ell}{\epsilon _y \sech \frac{\sigma_0}{\ell}} + \frac{\sigma_0}{\ell} \right) -\frac{2c}{3} \frac{z_R e^{\frac{\sigma_0}{\ell}}}{\sinh X_1}+\mathcal{O}(z_{R}^2),
\end{align}
where in the last line we use \cref{eq_gct,eq_rt,eq_rind_cutoff} to express the result in Rindler coordinates.

\subsection{Reflected entropy between radiation subsystems}

In this subsection, we analyse the reflected entropy between the left and right radiation region, described as $R_L=|MQ|$ and $R_R=|PN|$ respectively in \cref{fig_RRC}. In this scenario we encounter two possible EE phases, an island and a no-island phase. Similar to previous subsections, we first perform the computation in unprimed coordinates (Euclidean coordinates) and then transform the result to the Rindler coordinates.
The endpoints of $R_L$ are $(\tau_1,-x_1,z_1)$ and $(\tau_2,-x_2,z_2)$, while $R_R$ is defined by endpoints $(\tau_1,x_1,z_1)$ and $(\tau_2,x_2,z_2)$.

\subsubsection{Entanglement Entropy Phase: 1}

\begin{figure}[h!]
	\centering
	\includegraphics[scale=.8]{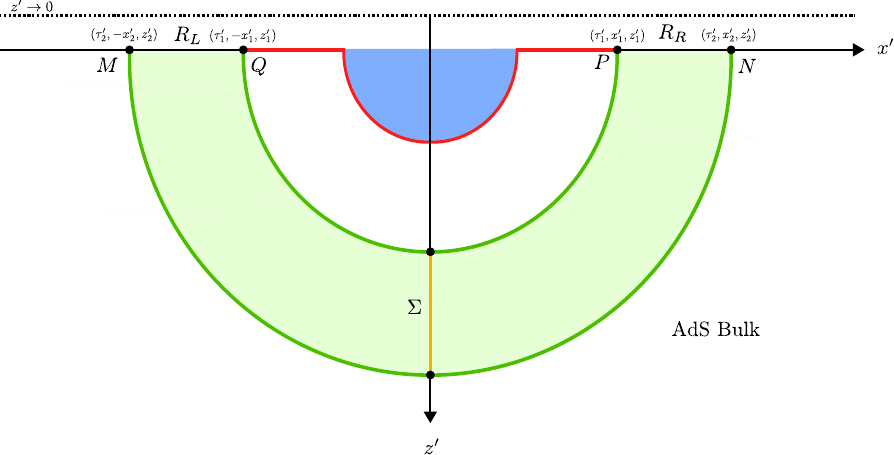}
	\caption{This diagram describes the no-island phase for EE for radiation region $R_L \cup R_R$. The EWCS $\Sigma$, depicted by the yellow line, separates the left and right radiation region in the bulk.}
	\label{fig_RRC}
\end{figure}

In this phase, there are no entanglement entropy islands associated with $R_L \cup R_R$, as illustrated in \cref{fig_RRC}. Consequently the EE may be computed as the sum of the lengths of the two HM surfaces, indicated by the green lines, which may be easily computed using \cref{eq_ee_ni} as
\begin{align}
	S(R_L \cup R_R)= \frac{2c}{3} \log \frac{2 \cosh T}{z_R}.
\end{align}
Corresponding to this EE phase, there is only one possible reflected entropy phase between the disjoint radiation subsystems $R_L$ and $R_R$. The computation of this reflected entropy is detailed as follows.

\subsubsection*{Boundary description}

As seen from the \cref{fig_RRC}, there is no island phase for this EE and reflected entropy for this case, hence the area term in \cref{SR-bdy} does not contribute. Therefore computing the reflected entropy is equivalent to determining the effective reflected entropy between two disjoint intervals $R_L$ and $R_R$, which is given by the four point correlator of twist fields at boundary points $M,Q,P$ and $N$ as 
\begin{align}\label{eq_rrc_bdy}
	S_R^\text{bdy}(R_L:R_R)=\lim_{m, n \to 1} \frac{1}{1-n} \log \frac{\langle   \sigma_{g_A}(Q) \sigma_{g_A^{-1}}(M) \sigma_{g_B}(N) \sigma_{g_B^{-1}}(P)   \rangle _{CFT^{\otimes mn}}}{\langle   \sigma_{g_m}(Q) \sigma_{g_m}(M) \sigma_{g_m}(N) \sigma_{g_m}(P)   \rangle^n_{CFT^{\otimes m}}}=\frac{c}{3} \log \frac{1+\sqrt{\eta}}{1-\sqrt{\eta}},
\end{align}
where in the second step we have used the four point function in the large central charge limit as provided in \cite{Dutta:2019gen,Fitzpatrick:2014vua}. The cross ratio in this case is given as
\begin{align}\label{eq_rrc_cr}
	\eta=\frac{(x_2 - x_1)^2+(\tau_2 - \tau_1)^2}{(x_2 + x_1)^2+(\tau_2 - \tau_1)^2}=\frac{ \sinh^2[\frac{X_2-X_1}{2}]}{ \sinh^2[\frac{X_2-X_1}{2}]+\cosh^2T}.
\end{align}
In this case we observe no linear order correction due to $T\bar{T}$ deformation.\footnote{\Cref{eq_rrc_bdy,eq_rrc_cr} are computed under the approximation that $z_1^2,z_2^2\approx 0$.  Even without making this approximation initially, no linear correction is observed when the final result is expanded in orders of $z_R$. However, we may observe a correction of the order $\mathcal{O}(z_{R}^2)$. The same approximation is considered for computing \cref{eq_rrc_bulk_gen,eq_rrc_bulk}.}

\subsubsection*{Bulk description}

As shown in \cref{fig_RRC}, the absence of any radiation island implies that the matter defect term (the first term in \cref{SR-bulk}) does not contribute. As a result, computing the reflected entropy reduces to finding the length of curve $\Sigma$, represented by the yellow line in \cref{fig_RRC}. By applying a similar approach as \cref{appFT}, we find that computing the the length of $\Sigma$ reduces to finding the geodesic length between points $(\tau_1,p,\sqrt{-p^2+x_1^2+z_1^2})$ and $(\tau_2,q,\sqrt{-q^2+x_2^2+z_2^2})$, which are arbitrary points on the HM surfaces $PQ$ and $MN$ respectively. Thus the reflected entropy may be written by utilizing these end point in \cref{geodesic-length} as  
\begin{align}\label{eq_rrc_bulk_gen}
	S_R^\text{bulk}(R_L:R_R)=\frac{c}{3} \cosh ^{-1} \left[ \frac{-2pq+x_1^2+x_2^2 +(\tau_2-\tau_1)^2}{2 \sqrt{-p^2+x_1^2} \sqrt{-q^2+x_2^2}} \right].
\end{align}
Extremizing \cref{eq_rrc_bulk_gen} over $p$ and $q$ we get $p_{min}=q_{min}=0$. This gives us the reflected entropy as
\begin{align}\label{eq_rrc_bulk}
	S_R^\text{bulk}(R_L:R_R)=\frac{c}{3} \cosh ^{-1} \left[ \frac{x_1^2+x_2^2+(\tau_2-\tau_1)^2}{2x_1x_2} \right]=\frac{c}{3} \log \frac{1+\sqrt{\eta}}{1-\sqrt{\eta}},
\end{align}
which matches exactly with \cref{eq_rrc_bdy}. Once again we observe no linear correction in the holographic reflected entropy due to $T\bar{T}$ deformation.

\subsubsection{Entanglement Entropy Phase: 2}

\begin{figure}[h!]
	\centering
	\includegraphics[scale=.8]{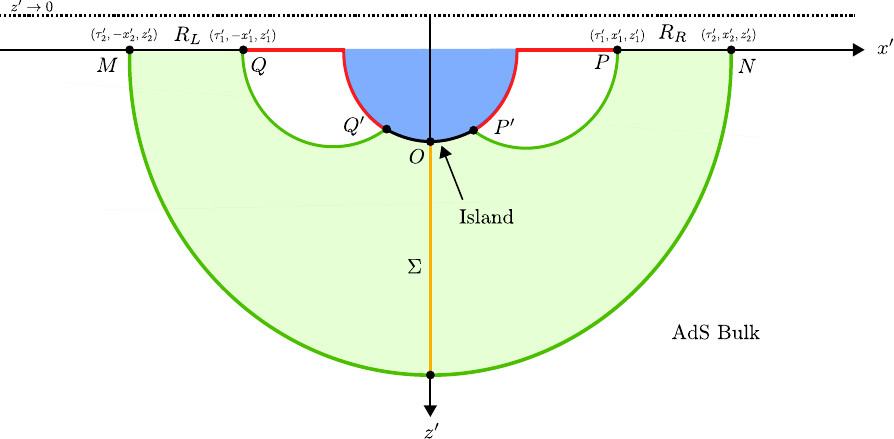}
	\caption{This illustration describes the island phase of EE for the radiation subsystem $R_L \cup R_R$. The entanglement entropy island region is denoted by the black curve $Q'P'$, with point $O$ separating the reflected entropy islands for the radiation subsystems. The yellow line $\Sigma$ represents the EWCS.}
	\label{fig_RRDC}
\end{figure}

In this phase, the presence of entanglement entropy islands corresponding to the radiation regions is observed. In this scenario the EE may be expressed as the sum of the HM surface and two island surfaces (depicted as solid green curves in \cref{fig_RRDC}) as,
\begin{align}
	S(R_L \cup R_R)= \frac{c}{3} \left( \frac{\sigma_0}{\ell}+ \log \frac{2 \cosh T}{z_R} +\log \frac{2 \sinh X_1}{z_R} +\log \frac{2 \ell}{\epsilon_y \sech \frac{\sigma_0}{\ell}}   \right)-\frac{c}{3} \frac{z_Re^{\frac{\sigma_0}{\ell}}}{\sinh X_1},
\end{align}
where we have made use of \cref{eq_ee_ni,eq_ee_is}.
In \cref{fig_RRDC}, the EE island is depicted as the region $Q'P'$. The brane point $O$ segments the EE island into reflected entropy islands $I_L=|OQ'|$ for subsystem $R_L$ and $I_R=|OP'|$ for subsystem $R_R$. The computation of reflected entropy in this phase is as follows.

\subsubsection*{Boundary description}

As illustrated from \cref{fig_RRDC}, the effective reflected entropy is expressed in terms of a seven-point correlation function involving four boundary points $M,Q,P$ and $N$, and three brane points $Q',P,$ and $O$. This correlation function can further be factorized in the large central charge limit into two two-point functions between $Q,Q'$ and $P,P'$, and a three-point function between $M,O$ and $N$. Notable, only the three-point correlator contributes, details of which can be found in \cite{Li:2021dmf} as
\begin{align}
	S_R^\text{eff} & =\lim_{m,n \to 1} \frac{1}{1-n} \log \Biggl[ \frac{ \Omega_{Q'}^{2 h_{g_A}} \Omega_{O}^{2 h_{g_{AB}}} \Omega_{P'}^{2 h_{g_B}}}{\Omega_{Q'}^{2 h_{g_m}} \Omega_{P'}^{2 h_{g_m}}} \notag \\
	& \qquad \qquad \qquad \qquad \qquad \times \frac{  \langle   \sigma_{g_A}(M) \sigma_{g_A^{-1}}(Q) \sigma_{g_A}(Q') \sigma_{g_A^{-1}g_B}(O) \sigma_{g_B^{-1}}(P') \sigma_{g_B}(O) \sigma_{g_B^{-1}}(N)  \rangle _{CFT^{\otimes mn}}}{ \langle \sigma_{g_m}(M) \sigma_{g_m}(Q) \sigma_{g_m}(Q') \sigma_{g_m}(P') \sigma_{g_m}(P) \sigma_{g_m}(N)  \rangle^n_{CFT^{\otimes m}}} \Biggr]\notag \\ 
	& = \lim_{m,n \to 1} \frac{1}{1-n} \log \frac{ \Omega_{O}^{2 h_{g_{AB}}} \langle    \sigma_{g_A^{-1}}(Q) \sigma_{g_A}(Q') \rangle _{mn} \langle \sigma_{g_B^{-1}}(P') \sigma_{g_B}(P) \rangle _{mn} \langle \sigma_{g_A}(M) \sigma_{g_A^{-1}g_B}(O)  \sigma_{g_B^{-1}}(N)  \rangle_{mn}}{\langle \sigma_{g_m}(M) \sigma_{g_m}(N) \rangle ^n _m \langle \sigma_{g_m}(Q) \sigma_{g_m}(Q') \rangle ^n _m \langle \sigma_{g_m}(P') \sigma_{g_m}(P) \rangle ^n _m}\notag \\ 
	& = \lim_{m,n \to 1} \frac{1}{1-n} \log \frac{ \Omega_{O}^{2 h_{g_{AB}}} \langle \sigma_{g_A}(M) \sigma_{g_A^{-1}g_B}(O)  \sigma_{g_B^{-1}}(N)  \rangle_{mn}}{\langle \sigma_{g_m}(M) \sigma_{g_m}(N) \rangle ^n _m }.
\end{align}
Incorporating the area term, the generalized reflected entropy takes the same form as \cref{eq_bbc_bdy_gen} with the substitution $x_1 \to x_2$. Therefore the final reflected entropy for this phase in Rindler coordinates can be directly derived from \cref{eq_ran2} by replacing $X_1 \to X_2$, yielding the expression
\begin{align}\label{eq_rrdc_bdy}
	S_R^\text{bdy}(R_L:R_R) & = \frac{c}{3} \left( \log \left[ \frac{\sinh X_2+\sqrt{\cosh^2T+\sinh^2X_2}}{\cosh T} \right]  +\log \frac{2 \ell}{\epsilon_y \sech \frac{\sigma_0}{\ell}} +\frac{\sigma_0}{\ell} \right)\notag \\ & -\frac{c}{3} \frac{z_R e^{\frac{\sigma_0}{\ell}}}{\sqrt{\cosh^2T+\sinh^2X_2}}+\mathcal{O}(z_{R}^2).
\end{align}

\subsubsection*{Bulk description}

In the bulk description for this phase, the EWCS is represented by the geodesic $\Sigma$ which connects the HM surface $MN$ to the EOW brane. The geodesic length of $\Sigma$ can be computed using \cref{geodesic-length}. Including the matter defect term, the generalized holographic reflected entropy once again has the same functional form of \cref{eq_bbc_bulk_gen} with the substitution $x_1 \to x_2$. As a result the final holographic reflected entropy in this phase is the same as \cref{eq_rrdc_bdy}.

\subsection{Page curve}

In this subsection, we describe the Page curves for the reflected entropy for each of the above discussed cases. To plot the Page curve for the reflected entropy, first we need to determine the phase transitions in the EE phases.
In presence of $T\bar{T}$ deformation the Page time $T_p$ for the EE phases is given as \cite{Deng:2023pjs}
\begin{align}\label{eq_Pagetime}
	T_p=\cosh ^{-1} \left( \sinh X_1 e^{\frac{\sigma_0}{\ell}} e^{-\frac{ z_R e^{\frac{\sigma_0}{\ell}}}{\sinh X_1}} \frac{2 \ell}{\epsilon_y \sech \frac{\sigma_0}{\ell}} \right),
\end{align}
where $X_1$ is the extent of the black hole interior along the spatial direction, $\sigma_0$ is the brane angle, $\ell$ is the AdS radius, $\epsilon_y$ is the UV cutoff on the brane, and $z_R$ denotes the UV cutoff in the Rindler coordinates. Initially no island phase dominates while the island phase, where an EE island appears in the black hole interior, dominates at late times $T>T_p$.

\begin{figure}[h!]
	\centering
	\begin{subfigure}{0.53\linewidth}
		\includegraphics[scale=0.35]{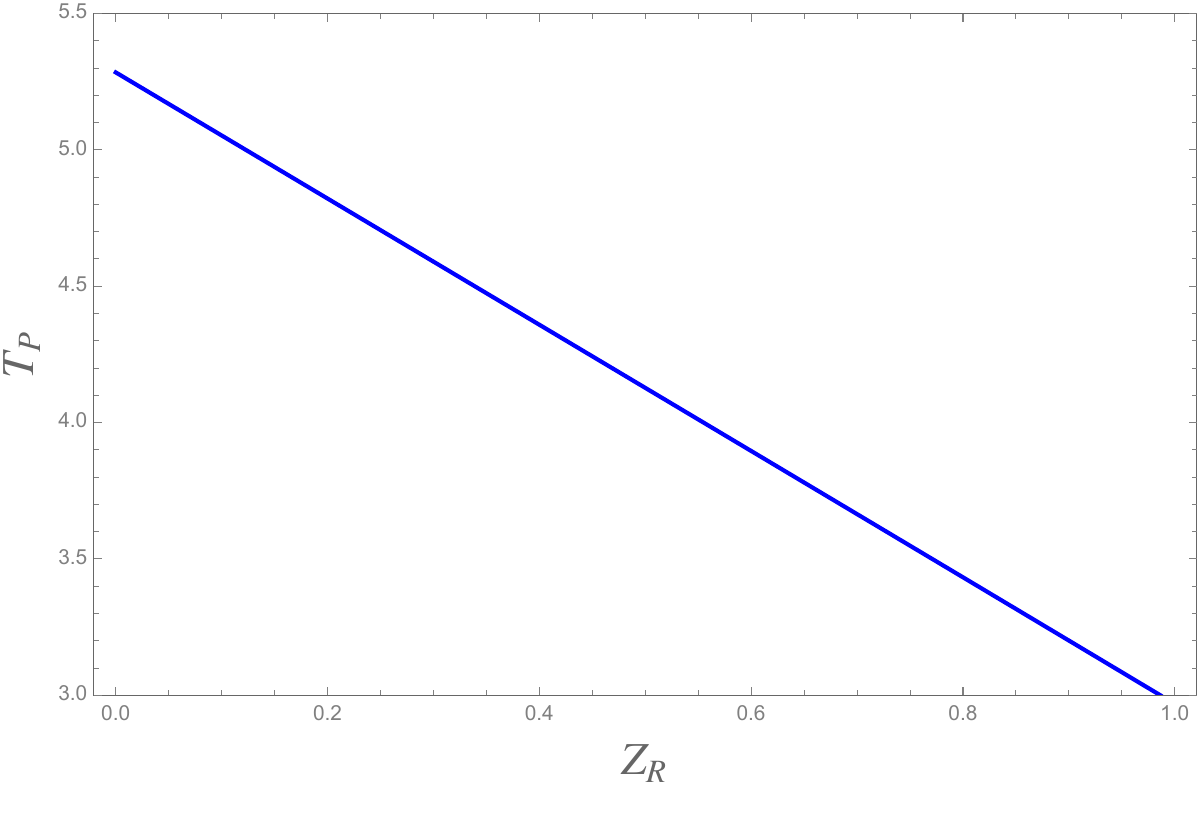}
		\caption{}
	\end{subfigure}
	\begin{subfigure}{0.45\linewidth}
		\includegraphics[scale=0.33]{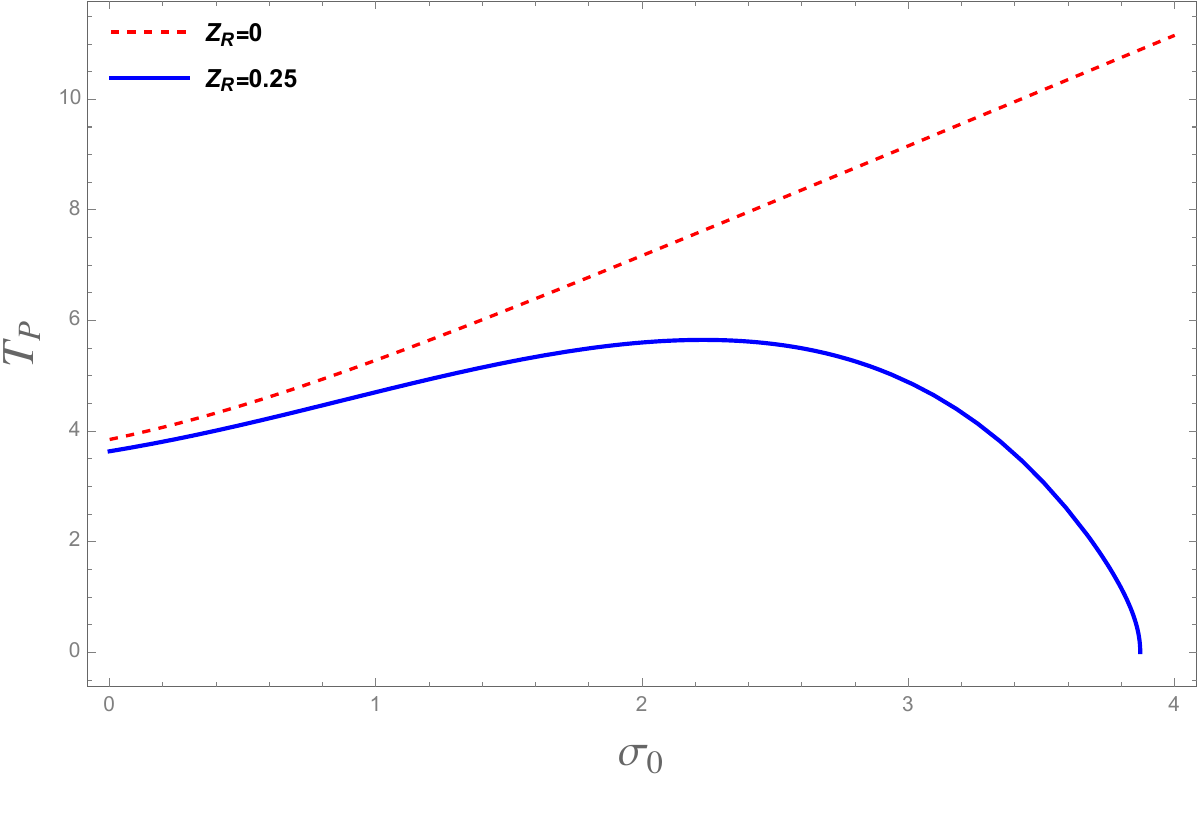}
		\caption{}
	\end{subfigure}
	\caption{(a) Plot of Page time with cut-off $z_R$ with $X_1=1, \epsilon_y=0.1,\ell=1$, and $\sigma_0=1$. (b) Plot of Page time with brane angle $\sigma_0$ with $X_1=1, \epsilon_y=0.1,\ell=1$, for $z_R=(0,0.25)$. Though the red dotted line (the curve representing $z_R=0$) may seem to increase monotonically for all values of $\sigma_0$, it does drop to zero at  $\sigma_0 \to \infty$, where the $T_p$ seems to vanishes.}
	\label{fig_Pagetime}
\end{figure}

As evident from \cref{fig_Pagetime}, the Page time $T_P$ exhibits some interesting behavior when plotted against the cut-off $z_R$ and the brane angle $\sigma_0$. Furthermore, it is stimulating to note that the Page time drops to zero much more rapidly with $\sigma_0$ in presence of $T\bar{T}$ deformation, unlike the scenario with no $T\bar{T}$ deformation where $T_p$ vanishes at $\sigma_0 \to \infty$. This appears to be correlated with the fact that in the presence of $T\bar{T}$ deformation the asymptotic boundary is no longer at $\sigma \to \infty$, but pushed inwards to a finite value.

\begin{figure}[h!]
	\begin{subfigure}{0.54\linewidth}
		\includegraphics[scale=0.33]{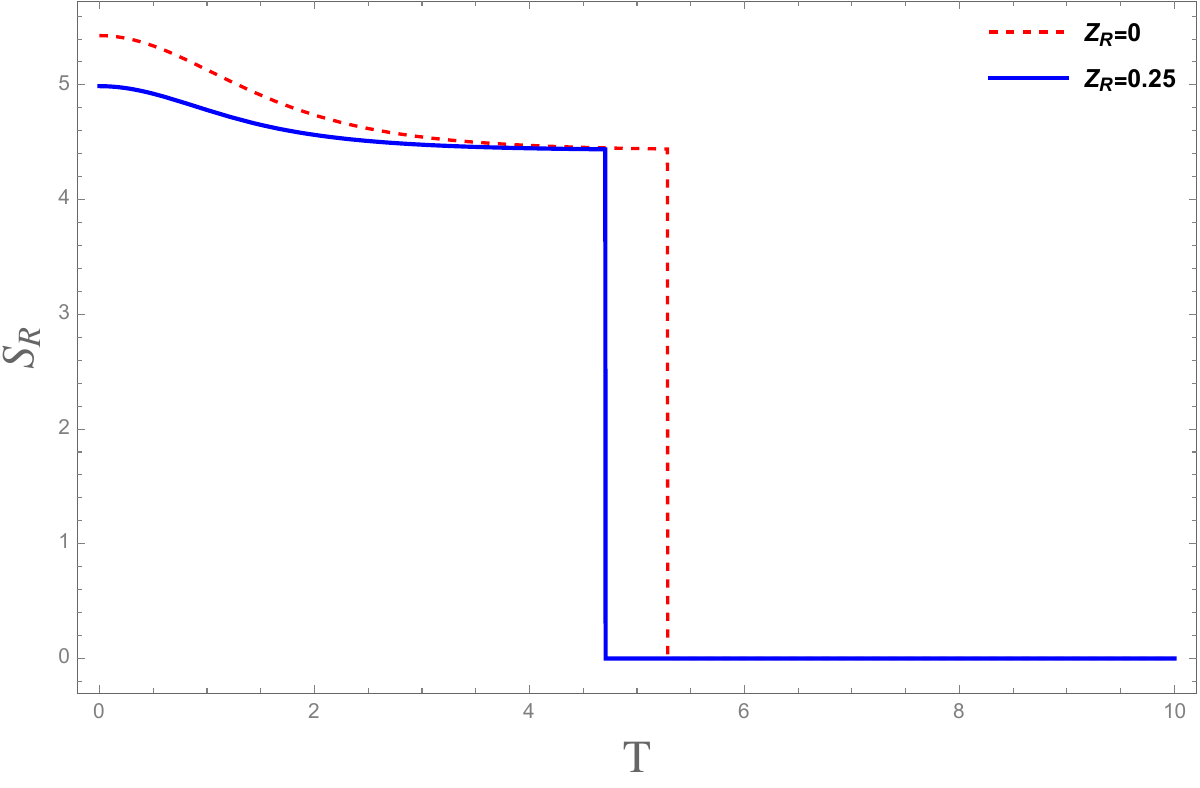}
		\caption{}
	\end{subfigure}
	\begin{subfigure}{0.3\linewidth}
		\includegraphics[scale=0.33]{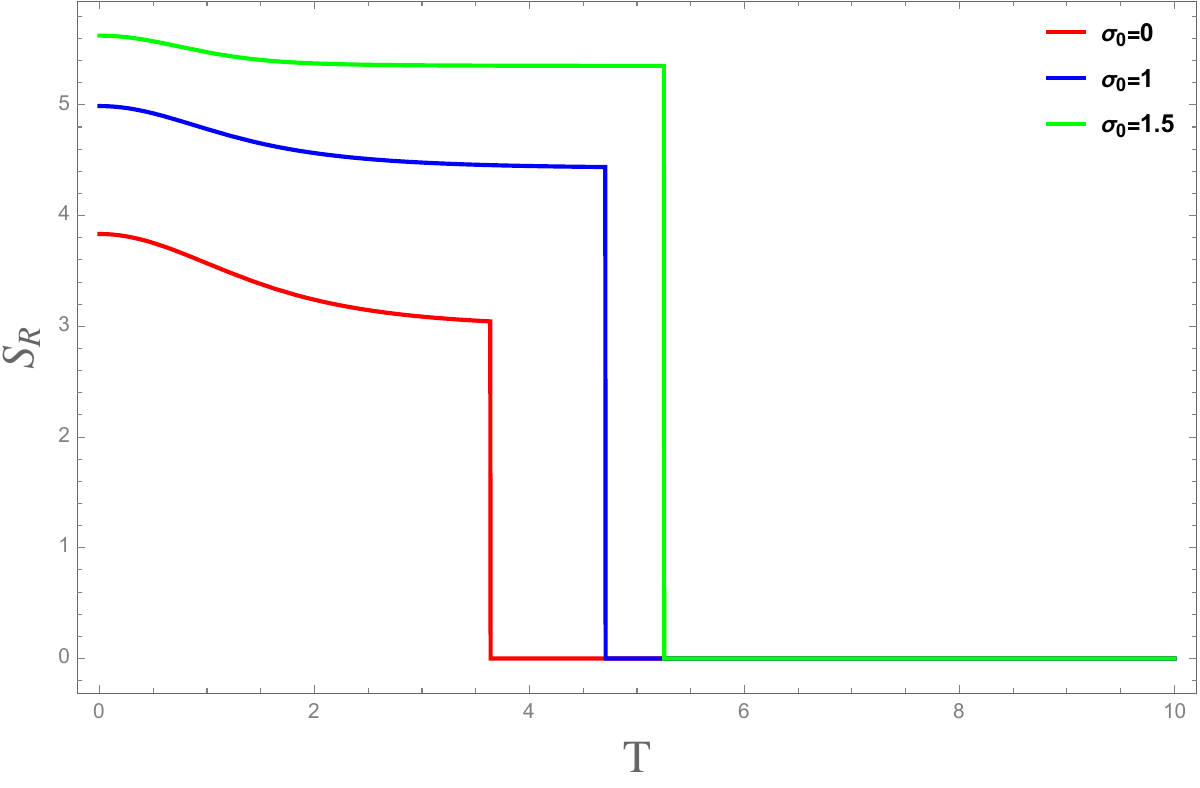}
		\caption{}
	\end{subfigure}
	\caption{ (a) Reflected entropy between black hole interiors with time $T$ in the absence ($z_R=0$) and presence ($z_R=0.25$) of $T\bar{T}$ deformation. Here we set $X_1=1, \epsilon_y=0.1,\ell=1$, and $\sigma_0=1$. (b) Reflected entropy between black hole interiors with time $T$ in the presence of $T\bar{T}$ deformation for different brane angle $\sigma_0= (0, 1, 1.5)$. We have set $X_1=1, \epsilon_y=0.1,\ell=1$, and $z_R=0.25$.}
	\label{fig_bbplot}
\end{figure}

Within the above mentioned EE phases we observe one or more reflected entropy phases depending on the configurations of the subsystems chosen. We now investigate the phase transitions between the various reflected entropy phases, starting with the black hole interior subsystems. We come across two possible EE phases, and within each we observe a single reflected entropy phase. The relevant plots for the reflected entropy between black hole interior subsystems (in units of $c/3$) and time $T$ are given in \cref{fig_bbplot}. The first plot shows how the reflected entropy gets modified in presence of $T\bar{T}$ deformation, while the second plot demonstrates how the reflected entropy depends on the brane angle.

For black hole interior and radiation subsystems we observe only one EE phase (as discussed in \cref{ssec_bhrad}), within which we have two possible reflected entropy phases. As evident from \cref{eq_brc_bdy,,eq_brdc_bulk}, the reflected entropy between black hole interior and radiation for both the phases is independent of time and remains constant for all time. This is expected since no Hartmann-Maldacena surfaces are involved in the computation of the reflected entropy for either phases. As a result, the reflected entropy between black hole interior and radiation subsystems is given by the minimum of the two phases and remains at that constant for all times.

\begin{figure}[h!]
	\begin{subfigure}{0.54\linewidth}
		\includegraphics[scale=0.33]{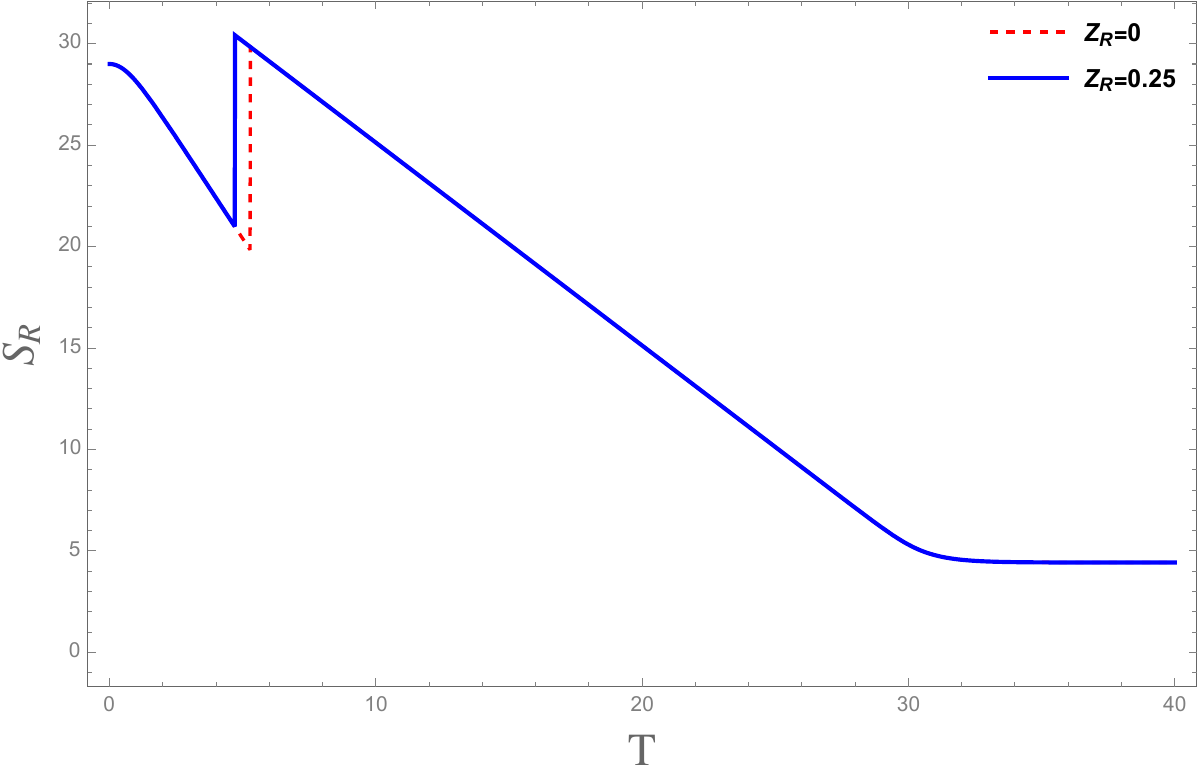}
		\caption{}
	\end{subfigure}
	\begin{subfigure}{0.45\linewidth}
		\includegraphics[scale=0.33]{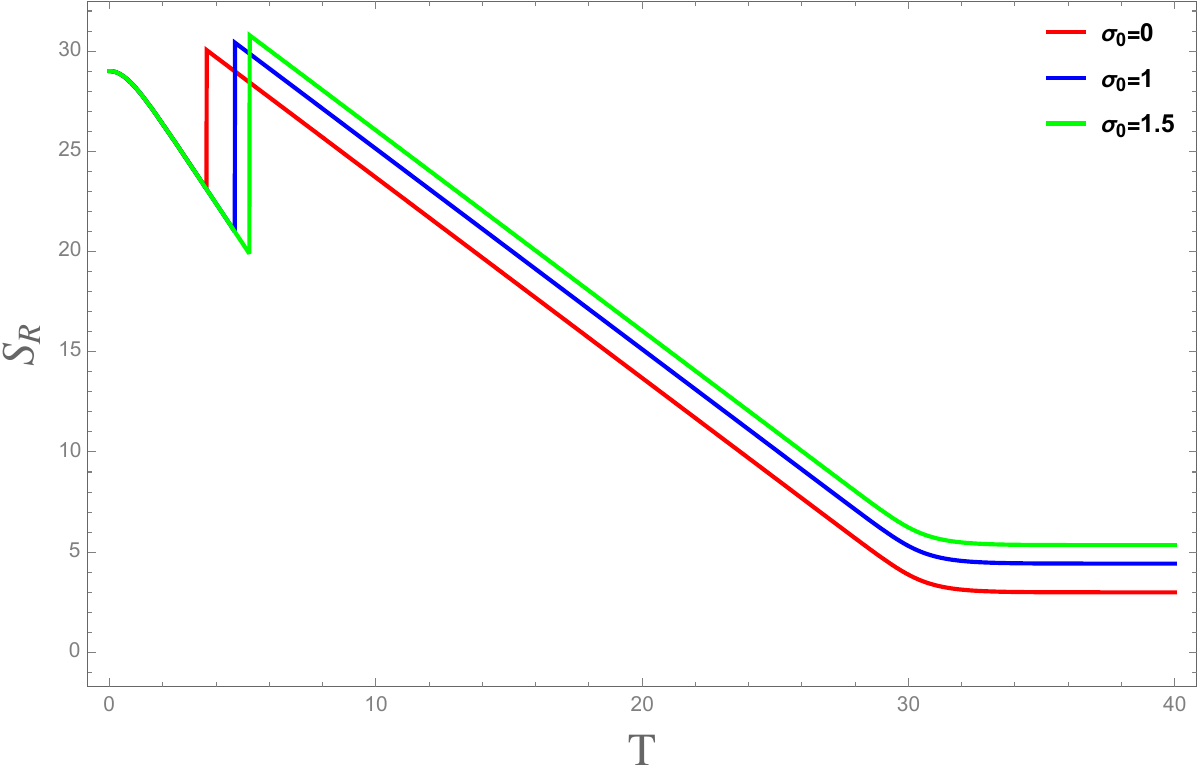}
		\caption{}
	\end{subfigure}
	\caption{ (a) Reflected entropy between left and right intervals of radiation with time $T$ in the absence ($z_R=0$) and presence ($z_R=0.25$) of $T\bar{T}$ deformation. Here we set $X_1=1, X_2=30, \epsilon_y=0.1,\ell=1$, and $\sigma_0=1$. (b) Reflected entropy between left and right intervals of radiation with time $T$ in the presence of $T\bar{T}$ deformation for different brane angle $\sigma_0= (0, 1, 1.5)$. We have set $X_1=1, X_2=30, \epsilon_y=0.1,\ell=1$, and $z_R=0.25$.}
	\label{fig_rrplot}
\end{figure}

For two radiation subsystems, we once again observe two possible EE phases, and within each we have a single reflected entropy phase. The relevant plots are provided in \cref{fig_rrplot}, where we observe the no-island phase dominating at early times and the island phase dominating at late times.

Similar to the reflected entropy between two black hole interior subsystems (\cref{fig_bbplot}), the reflected entropy between two radiation subsystems (\cref{fig_rrplot}) exhibit a sudden discontinuity while transitioning from the no-island to the island phase at the Page time $T_p$, where it gradually decreases with time and eventually plateaus out.
In the limit that the radiation bath extends to spacial infinity $X_2 \to \infty$, the asymptotic behaviour of the reflected entropy is given by 
\begin{align}
	\text{Connected Phase :} S_R ^\text{bdy} = & \frac{c}{3} (X_2-X_1-2\log \cosh T) \notag \\
	\text{Disconnected Phase :} S_R ^\text{bdy} = & \frac{c}{3} \left(X_2 - \log \cosh T +\log \frac{2 \ell}{\epsilon _y \sech \frac{\sigma_0}{\ell}} + \frac{\sigma_0}{\ell} \right).
\end{align}
The gap between the reflected entropy at two phases at time of transition $T_p$ in this limit is
\begin{align}
	S_\text{gap}=\frac{c}{3} \left( 2\log \frac{2\ell}{\epsilon_y \sech \frac{\sigma_0}{\ell}}+2 \frac{\sigma_0}{\ell} +\log (e^{X_1}\sinh X_1) \right) -\frac{c}{3}\frac{ z_R e^{\frac{\sigma_0}{\ell}}}{\sinh X_1} .
\end{align}

\section{Conclusions}\label{sec:conc}
In this article, we investigate the structure of mixed state entanglement and correlation through the reflected entropy in an AdS$_3$ braneworld model equipped with a finite distance Dirichlet wall, putatively dual to some boundary conformal field theory with $T\bar{T}$ deformation in the spirit of \cite{McGough:2016lol}. In particular, we focus on an AdS$_3$ geometry characterized by a finite radial cut-off, induced by the $T\bar{T}$ deformation, which is truncated by an end-of-the-world (EOW) brane containing defect matter. As discussed in \cite{Deng:2023pjs}, such a model can naturally arise from a partial reduction, and the resulting braneworld gravity is governed by Neumann boundary conditions on the EOW brane, while we apply AdS/CFT duality for Dirichlet boundary conditions on the finite cut-off surface. In this context, we extend the defect extremal surface (DES) prescription, essentially a double-holographic counterpart to the island formula, for computing the reflected entropy of bipartite mixed states within the defect AdS/BCFT correspondence. The DES formula offers a novel approach to understanding entanglement in such deformed setups, bridging insights from both holographic duality and boundary theories.

We begin by computing the reflected entropy for both adjacent and disjoint intervals in static scenarios, considering various possible entanglement entropy phases. Our findings demonstrate that the results from the DES formula are equivalent to those obtained via the island formula, up to linear orders in the radial cut-off $z_c$, thereby reinforcing the validity of our approach. We then extend our analysis to time dependent configurations involving an eternal black hole coupled to a radiation bath in the effective lower-dimensional setup. In this context, we compute the reflected entropy between subsystems in the black hole interior, between black hole and radiation subsystems, and between subsystems in the radiation bath utilizing both the island and DES prescriptions. We also examine the time evolution of reflected entropy for these configurations and plot the analogous of the Page curves for the reflected entropy, considering scenarios with and without the $T\bar{T}$ deformation. Once again, we find that the results from boundary and bulk calculations are consistent up to the leading order in the radial cut-off for time-dependent cases, providing a strong consistency check for the proposal discussed earlier.

It is important to note that, as shown in   \cite{He:2019vzf,Chen:2018eqk,Jeong:2019ylz} the effects of the $T\bar{T}$ deformation in the boundary theory is captured by modifying the correlation functions by appropriate insertions of stress-tensor components within the framework of conformal perturbation theory. However, in this article, we take a different route and impose that the effects of deformation are encapsulated in the induced metric on the radial cut-off surface, similar to \cite{Deng:2023pjs}. It will be interesting to explicitly verify the credibility of this approach by developing a perturbation theory for the $T\bar{T}$-deformed BCFT from scratch.

Although our doubly holographic computations conform to the results from the island formula in the leading order, a critical issue remains to be addressed. Namely, the well-known breakdown of the holographic proposal in \cite{McGough:2016lol} upon the inclusions of bulk conformal matter. A possible way out is to consider the alternative proposal of mixed boundary conditions \cite{Guica:2019nzm} which can address the presence of matter in the bulk. 
We leave a critical re-examination of the present work in the context of modified asymptotic boundary conditions for the future. 

A potential higher-dimensional extension of the present work, as well as that of \cite{Deng:2023pjs}, possibly in the spirit of \cite{Hartman:2018tkw}, warrants further investigation due to the subtleties associated with incorporating conformal matter on the defect brane. Moreover, even in the undeformed case, the DES and island prescriptions do not align consistently in higher dimensions, and the process of partial dimensional reduction necessitates the inclusion of all Kaluza–Klein modes \cite{Chu:2021gdb}. Resolving these challenges is crucial for developing a robust higher-dimensional framework that accurately captures the effects of $T\bar{T}$ deformations and the dynamics of conformal matter on defect branes.

There are several other future directions to explore. 
One potential avenue is to investigate the correction to reflected entropy at all orders in the radial cutoff for $T \bar{T}$ deformed BCFT and to verify the equivalence between the island  and the DES formula in this scenarios.
Furthermore it would be interesting to investigate other mixed state entanglement and correlation measures such as entanglement negativity, odd entanglement entropy, balance partial entanglement and multipartite entanglement in the context of such $T\bar{T}$ deformed BCFT.
We leave these interesting open issue for future consideration.

\appendix
\section{On geodesics in cut-off AdS$_3$}
\subsection{Geodesic on a constant $\tau$-slice}\label{appFT}

\begin{figure}[h]
	\centering
	\begin{subfigure}{0.54\linewidth}
		\centering
		\includegraphics[scale=0.6]{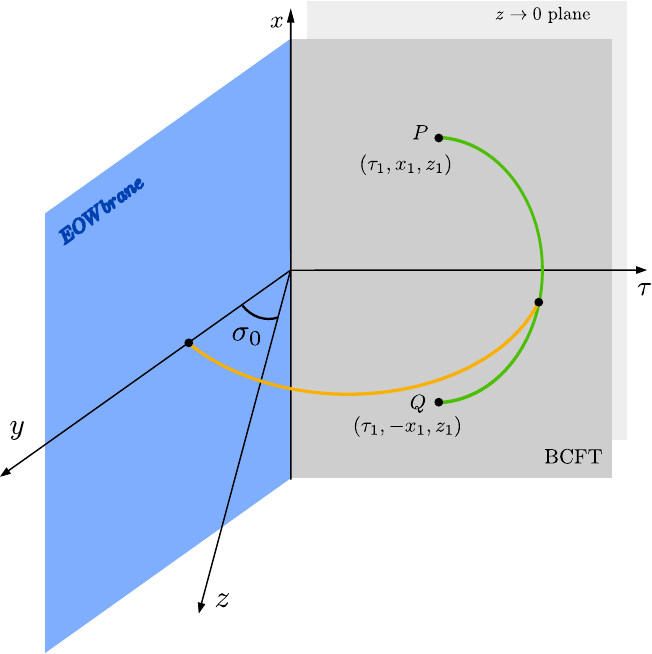}
		\caption{}
	\end{subfigure}
	\begin{subfigure}{0.45\linewidth}
		\centering
		\includegraphics[scale=.85]{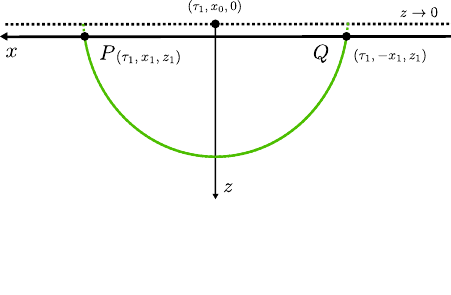}
		\caption{}
	\end{subfigure}
	\caption{\it (a) Representation of no-island phase of the entanglement entropy of black hole region in the unprimed coordinates. (b) Schematic for determining the equation of the Hartmann Maldacena surface $PQ$ on a constant $\tau$-slice. }
	\label{fig_constt}
\end{figure}

The equation of geodesic in AdS$_3$ on a constant time slice (as described in \cref{fig_constt}) is that of a circle given by 
\begin{align}
	(x+x_0)^2+(z-z_0)^2=R^2,
\end{align}
where $(\tau_1,x_0,z_0)$ is the coordinates of the center  and $R$ is the radius of the circle. Given that the circle must be centred at $z_0=0$, solving for the points $(\pm x_1, z_1)$, i.e., solving
\begin{align}
	(x_1+x_0)^2+z_1^2=R^2, \ \  \ \ (x_1-x_0)^2+z_1^2=R^2
\end{align}
gives us the $x_0=0$ and $R=\sqrt{x_1^2+z_1^2}$. Considering that the EWCS lands on this Ryu Takayanagi surface at arbitrary value $x$, this gives us the value of $z$ as
\begin{align}
	z=\sqrt{-x^2+x_1^2+z_1^2}
\end{align}

\subsection{Geodesic on a constant $x$-slice}\label{appRB}

\begin{figure}[h]
	\centering
	\includegraphics[scale=1.5]{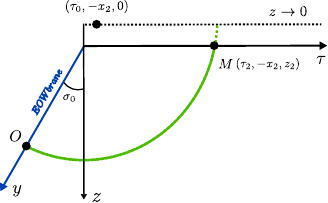}
	\caption{\it Schematic for determining the equation of the Ryu Takayanagi surface $MO$ on a constant $x$-slice. }
	\label{fig_constx}
\end{figure}

In this case the geodesic in AdS$_3$ is on a constant $x$-slice (refer to \cref{fig_constx}), which is described by the circle
\begin{align}
	(\tau-\tau_0)^2+(z-z_0)^2=R^2.
\end{align}
Here $(\tau_0,-x_2,z_0)$ is the coordinates of the center and $R$ is the radius of the circle. Given that the circle must be centred at $z_0=0$, then solving for boundary point $M \equiv (\tau_2, -x_2, z_2)$ and brane point $O \equiv (-\tau_2 \tanh \sigma_0, -x_2, z_2+\tau_2 \sech \sigma_0)$, i.e., solving
\begin{align}
	z_2^2+(\tau_2-\tau_0)^2=R^2, \ \  \ \ (z_2+\tau_2 \sech \frac{\sigma_0}{\ell})^2+(\tau_0+\tau_2 \tanh \frac{\sigma_0}{\ell})^2=R^2
\end{align}
gives us
\begin{align}
	R=\sqrt{\frac{2z_2(z_2+\tau_2 \sech \frac{\sigma_0}{\ell})+\tau_2^2 (1+\tanh \frac{\sigma_0}{\ell})}{1+\tanh \frac{\sigma_0}{\ell}}}, \ \  \ \ \tau_0=-\frac{z_2 \sech \frac{\sigma_0}{\ell}}{1+\tanh \frac{\sigma_0}{\ell}}.
\end{align}
In the limit $x'_2 \to \infty$ (and by employing \cref{eq_gct}) it reduces to $R \to 2$ and $\tau_0 \to 0$. Thus the equation of the circle in the above mentioned limit is given as
\begin{align}
	z^2+\tau^2=4.
\end{align}

\bibliographystyle{utphys}
	
\bibliography{ref}

\providecommand{\href}[2]{#2}\begingroup\raggedright\begin{thebibliography}{10}

\bibitem{Almheiri:2019hni}
A.~Almheiri, R.~Mahajan, J.~Maldacena, and Y.~Zhao, ``{The Page curve of
  Hawking radiation from semiclassical geometry},''
  \href{http://dx.doi.org/10.1007/JHEP03(2020)149}{{\em JHEP} {\bfseries 03}
  (2020) 149}, \href{http://arxiv.org/abs/1908.10996}{{\ttfamily
  arXiv:1908.10996 [hep-th]}}.

\bibitem{Almheiri:2019psf}
A.~Almheiri, N.~Engelhardt, D.~Marolf, and H.~Maxfield, ``{The entropy of bulk
  quantum fields and the entanglement wedge of an evaporating black hole},''
  \href{http://dx.doi.org/10.1007/JHEP12(2019)063}{{\em JHEP} {\bfseries 12}
  (2019) 063}, \href{http://arxiv.org/abs/1905.08762}{{\ttfamily
  arXiv:1905.08762 [hep-th]}}.

\bibitem{Almheiri:2019qdq}
A.~Almheiri, T.~Hartman, J.~Maldacena, E.~Shaghoulian, and A.~Tajdini,
  ``{Replica Wormholes and the Entropy of Hawking Radiation},''
  \href{http://dx.doi.org/10.1007/JHEP05(2020)013}{{\em JHEP} {\bfseries 05}
  (2020) 013}, \href{http://arxiv.org/abs/1911.12333}{{\ttfamily
  arXiv:1911.12333 [hep-th]}}.

\bibitem{Almheiri:2019psy}
A.~Almheiri, R.~Mahajan, and J.~E. Santos, ``{Entanglement islands in higher
  dimensions},'' \href{http://dx.doi.org/10.21468/SciPostPhys.9.1.001}{{\em
  SciPost Phys.} {\bfseries 9} no.~1, (2020) 001},
  \href{http://arxiv.org/abs/1911.09666}{{\ttfamily arXiv:1911.09666
  [hep-th]}}.

\bibitem{Almheiri:2019yqk}
A.~Almheiri, R.~Mahajan, and J.~Maldacena, ``{Islands outside the horizon},''
  \href{http://arxiv.org/abs/1910.11077}{{\ttfamily arXiv:1910.11077
  [hep-th]}}.

\bibitem{Almheiri:2020cfm}
A.~Almheiri, T.~Hartman, J.~Maldacena, E.~Shaghoulian, and A.~Tajdini, ``{The
  entropy of Hawking radiation},''
  \href{http://arxiv.org/abs/2006.06872}{{\ttfamily arXiv:2006.06872
  [hep-th]}}.

\bibitem{Page:1993wv}
D.~N. Page, ``{Information in black hole radiation},''
  \href{http://dx.doi.org/10.1103/PhysRevLett.71.3743}{{\em Phys. Rev. Lett.}
  {\bfseries 71} (1993) 3743--3746},
  \href{http://arxiv.org/abs/hep-th/9306083}{{\ttfamily arXiv:hep-th/9306083}}.

\bibitem{Page:1993df}
D.~N. Page, ``{Average entropy of a subsystem},''
  \href{http://dx.doi.org/10.1103/PhysRevLett.71.1291}{{\em Phys. Rev. Lett.}
  {\bfseries 71} (1993) 1291--1294},
  \href{http://arxiv.org/abs/gr-qc/9305007}{{\ttfamily arXiv:gr-qc/9305007}}.

\bibitem{Page:2013dx}
D.~N. Page, ``{Time Dependence of Hawking Radiation Entropy},''
  \href{http://dx.doi.org/10.1088/1475-7516/2013/09/028}{{\em JCAP} {\bfseries
  09} (2013) 028}, \href{http://arxiv.org/abs/1301.4995}{{\ttfamily
  arXiv:1301.4995 [hep-th]}}.

\bibitem{Engelhardt:2014gca}
N.~Engelhardt and A.~C. Wall, ``{Quantum Extremal Surfaces: Holographic
  Entanglement Entropy beyond the Classical Regime},''
  \href{http://dx.doi.org/10.1007/JHEP01(2015)073}{{\em JHEP} {\bfseries 01}
  (2015) 073}, \href{http://arxiv.org/abs/1408.3203}{{\ttfamily arXiv:1408.3203
  [hep-th]}}.

\bibitem{Ryu:2006bv}
S.~Ryu and T.~Takayanagi, ``{Holographic derivation of entanglement entropy
  from AdS/CFT},'' \href{http://dx.doi.org/10.1103/PhysRevLett.96.181602}{{\em
  Phys. Rev. Lett.} {\bfseries 96} (2006) 181602},
  \href{http://arxiv.org/abs/hep-th/0603001}{{\ttfamily arXiv:hep-th/0603001}}.

\bibitem{Hubeny:2007xt}
V.~E. Hubeny, M.~Rangamani, and T.~Takayanagi, ``{A Covariant holographic
  entanglement entropy proposal},''
  \href{http://dx.doi.org/10.1088/1126-6708/2007/07/062}{{\em JHEP} {\bfseries
  07} (2007) 062}, \href{http://arxiv.org/abs/0705.0016}{{\ttfamily
  arXiv:0705.0016 [hep-th]}}.

\bibitem{Faulkner:2013ana}
T.~Faulkner, A.~Lewkowycz, and J.~Maldacena, ``{Quantum corrections to
  holographic entanglement entropy},''
  \href{http://dx.doi.org/10.1007/JHEP11(2013)074}{{\em JHEP} {\bfseries 11}
  (2013) 074}, \href{http://arxiv.org/abs/1307.2892}{{\ttfamily arXiv:1307.2892
  [hep-th]}}.

\bibitem{Geng:2024xpj}
H.~Geng, ``{Replica Wormholes and Entanglement Islands in the Karch-Randall
  Braneworld},'' \href{http://arxiv.org/abs/2405.14872}{{\ttfamily
  arXiv:2405.14872 [hep-th]}}.

\bibitem{Sully:2020pza}
J.~Sully, M.~V. Raamsdonk, and D.~Wakeham, ``{BCFT entanglement entropy at
  large central charge and the black hole interior},''
  \href{http://dx.doi.org/10.1007/JHEP03(2021)167}{{\em JHEP} {\bfseries 03}
  (2021) 167}, \href{http://arxiv.org/abs/2004.13088}{{\ttfamily
  arXiv:2004.13088 [hep-th]}}.

\bibitem{Rozali:2019day}
M.~Rozali, J.~Sully, M.~Van~Raamsdonk, C.~Waddell, and D.~Wakeham,
  ``{Information radiation in BCFT models of black holes},''
  \href{http://dx.doi.org/10.1007/JHEP05(2020)004}{{\em JHEP} {\bfseries 05}
  (2020) 004}, \href{http://arxiv.org/abs/1910.12836}{{\ttfamily
  arXiv:1910.12836 [hep-th]}}.

\bibitem{Chen:2020uac}
H.~Z. Chen, R.~C. Myers, D.~Neuenfeld, I.~A. Reyes, and J.~Sandor, ``{Quantum
  Extremal Islands Made Easy, Part I: Entanglement on the Brane},''
  \href{http://dx.doi.org/10.1007/JHEP10(2020)166}{{\em JHEP} {\bfseries 10}
  (2020) 166}, \href{http://arxiv.org/abs/2006.04851}{{\ttfamily
  arXiv:2006.04851 [hep-th]}}.

\bibitem{Chen:2020hmv}
H.~Z. Chen, R.~C. Myers, D.~Neuenfeld, I.~A. Reyes, and J.~Sandor, ``{Quantum
  Extremal Islands Made Easy, Part II: Black Holes on the Brane},''
  \href{http://dx.doi.org/10.1007/JHEP12(2020)025}{{\em JHEP} {\bfseries 12}
  (2020) 025}, \href{http://arxiv.org/abs/2010.00018}{{\ttfamily
  arXiv:2010.00018 [hep-th]}}.

\bibitem{Grimaldi:2022suv}
G.~Grimaldi, J.~Hernandez, and R.~C. Myers, ``{Quantum Extremal Islands Made
  Easy, Part IV: Massive Black Holes on the Brane},''
  \href{http://arxiv.org/abs/2202.00679}{{\ttfamily arXiv:2202.00679
  [hep-th]}}.

\bibitem{Suzuki:2022xwv}
K.~Suzuki and T.~Takayanagi, ``{BCFT and Islands in Two Dimensions},''
  \href{http://arxiv.org/abs/2202.08462}{{\ttfamily arXiv:2202.08462
  [hep-th]}}.

\bibitem{Geng:2020qvw}
H.~Geng and A.~Karch, ``{Massive islands},''
  \href{http://dx.doi.org/10.1007/JHEP09(2020)121}{{\em JHEP} {\bfseries 09}
  (2020) 121}, \href{http://arxiv.org/abs/2006.02438}{{\ttfamily
  arXiv:2006.02438 [hep-th]}}.

\bibitem{Geng:2020fxl}
H.~Geng, A.~Karch, C.~Perez-Pardavila, S.~Raju, L.~Randall, M.~Riojas, and
  S.~Shashi, ``{Information Transfer with a Gravitating Bath},''
  \href{http://dx.doi.org/10.21468/SciPostPhys.10.5.103}{{\em SciPost Phys.}
  {\bfseries 10} no.~5, (2021) 103},
  \href{http://arxiv.org/abs/2012.04671}{{\ttfamily arXiv:2012.04671
  [hep-th]}}.

\bibitem{Geng:2021iyq}
H.~Geng, S.~L\"ust, R.~K. Mishra, and D.~Wakeham, ``{Holographic BCFTs and
  Communicating Black Holes},''
  \href{http://dx.doi.org/10.1007/JHEP08(2021)003}{{\em jhep} {\bfseries 08}
  (2021) 003}, \href{http://arxiv.org/abs/2104.07039}{{\ttfamily
  arXiv:2104.07039 [hep-th]}}.

\bibitem{Geng:2021mic}
H.~Geng, A.~Karch, C.~Perez-Pardavila, S.~Raju, L.~Randall, M.~Riojas, and
  S.~Shashi, ``{Entanglement Phase Structure of a Holographic BCFT in a Black
  Hole Background},'' \href{http://arxiv.org/abs/2112.09132}{{\ttfamily
  arXiv:2112.09132 [hep-th]}}.

\bibitem{Geng:2021hlu}
H.~Geng, A.~Karch, C.~Perez-Pardavila, S.~Raju, L.~Randall, M.~Riojas, and
  S.~Shashi, ``{Inconsistency of Islands in Theories with Long-Range
  Gravity},'' \href{http://arxiv.org/abs/2107.03390}{{\ttfamily
  arXiv:2107.03390 [hep-th]}}.

\bibitem{Karch:2022rvr}
A.~Karch, H.~Sun, and C.~F. Uhlemann, ``{Double holography in string theory},''
  \href{http://dx.doi.org/10.1007/JHEP10(2022)012}{{\em JHEP} {\bfseries 10}
  (2022) 012}, \href{http://arxiv.org/abs/2206.11292}{{\ttfamily
  arXiv:2206.11292 [hep-th]}}.

\bibitem{Cardy:2004hm}
J.~L. Cardy, ``{Boundary conformal field theory},''
  \href{http://arxiv.org/abs/hep-th/0411189}{{\ttfamily arXiv:hep-th/0411189}}.

\bibitem{Takayanagi:2011zk}
T.~Takayanagi, ``{Holographic Dual of BCFT},''
  \href{http://dx.doi.org/10.1103/PhysRevLett.107.101602}{{\em Phys. Rev.
  Lett.} {\bfseries 107} (2011) 101602},
  \href{http://arxiv.org/abs/1105.5165}{{\ttfamily arXiv:1105.5165 [hep-th]}}.

\bibitem{Fujita:2011fp}
M.~Fujita, T.~Takayanagi, and E.~Tonni, ``{Aspects of AdS/BCFT},''
  \href{http://dx.doi.org/10.1007/JHEP11(2011)043}{{\em JHEP} {\bfseries 11}
  (2011) 043}, \href{http://arxiv.org/abs/1108.5152}{{\ttfamily arXiv:1108.5152
  [hep-th]}}.

\bibitem{Kastikainen:2021ybu}
J.~Kastikainen and S.~Shashi, ``{Structure of holographic BCFT correlators from
  geodesics},'' \href{http://dx.doi.org/10.1103/PhysRevD.105.046007}{{\em Phys.
  Rev. D} {\bfseries 105} no.~4, (2022) 046007},
  \href{http://arxiv.org/abs/2109.00079}{{\ttfamily arXiv:2109.00079
  [hep-th]}}.

\bibitem{Deng:2020ent}
F.~Deng, J.~Chu, and Y.~Zhou, ``{Defect extremal surface as the holographic
  counterpart of Island formula},''
  \href{http://dx.doi.org/10.1007/JHEP03(2021)008}{{\em JHEP} {\bfseries 03}
  (2021) 008}, \href{http://arxiv.org/abs/2012.07612}{{\ttfamily
  arXiv:2012.07612 [hep-th]}}.

\bibitem{Chu:2021gdb}
J.~Chu, F.~Deng, and Y.~Zhou, ``{Page curve from defect extremal surface and
  island in higher dimensions},''
  \href{http://dx.doi.org/10.1007/JHEP10(2021)149}{{\em JHEP} {\bfseries 10}
  (2021) 149}, \href{http://arxiv.org/abs/2105.09106}{{\ttfamily
  arXiv:2105.09106 [hep-th]}}.

\bibitem{Li:2021dmf}
T.~Li, M.-K. Yuan, and Y.~Zhou, ``{Defect extremal surface for reflected
  entropy},'' \href{http://dx.doi.org/10.1007/JHEP01(2022)018}{{\em JHEP}
  {\bfseries 01} (2022) 018}, \href{http://arxiv.org/abs/2108.08544}{{\ttfamily
  arXiv:2108.08544 [hep-th]}}.

\bibitem{Basu:2022reu}
D.~Basu, H.~Parihar, V.~Raj, and G.~Sengupta, ``{Defect extremal surfaces for
  entanglement negativity},'' \href{http://arxiv.org/abs/2205.07905}{{\ttfamily
  arXiv:2205.07905 [hep-th]}}.

\bibitem{Shao:2022wrm}
Y.~Shao, M.-K. Yuan, and Y.~Zhou, ``{Entanglement negativity and defect
  extremal surface},''
  \href{http://dx.doi.org/10.21468/SciPostPhysCore.7.2.027}{{\em SciPost Phys.
  Core} {\bfseries 7} no.~2, (2024) 027},
  \href{http://arxiv.org/abs/2206.05951}{{\ttfamily arXiv:2206.05951
  [hep-th]}}.

\bibitem{Dutta:2019gen}
S.~Dutta and T.~Faulkner, ``{A canonical purification for the entanglement
  wedge cross-section},'' \href{http://dx.doi.org/10.1007/JHEP03(2021)178}{{\em
  JHEP} {\bfseries 03} (2021) 178},
  \href{http://arxiv.org/abs/1905.00577}{{\ttfamily arXiv:1905.00577
  [hep-th]}}.

\bibitem{Akers:2021pvd}
C.~Akers, T.~Faulkner, S.~Lin, and P.~Rath, ``{Reflected entropy in random
  tensor networks},'' \href{http://arxiv.org/abs/2112.09122}{{\ttfamily
  arXiv:2112.09122 [hep-th]}}.

\bibitem{Vidal:2002zz}
G.~Vidal and R.~F. Werner, ``{Computable measure of entanglement},''
  \href{http://dx.doi.org/10.1103/PhysRevA.65.032314}{{\em Phys. Rev. A}
  {\bfseries 65} (2002) 032314},
  \href{http://arxiv.org/abs/quant-ph/0102117}{{\ttfamily
  arXiv:quant-ph/0102117}}.

\bibitem{Plenio:2005cwa}
M.~B. Plenio, ``{Logarithmic Negativity: A Full Entanglement Monotone That is
  not Convex},'' \href{http://dx.doi.org/10.1103/PhysRevLett.95.090503}{{\em
  Phys. Rev. Lett.} {\bfseries 95} no.~9, (2005) 090503},
  \href{http://arxiv.org/abs/quant-ph/0505071}{{\ttfamily
  arXiv:quant-ph/0505071}}.

\bibitem{Takayanagi:2017knl}
T.~Takayanagi and K.~Umemoto, ``{Entanglement of purification through
  holographic duality},''
  \href{http://dx.doi.org/10.1038/s41567-018-0075-2}{{\em Nature Phys.}
  {\bfseries 14} no.~6, (2018) 573--577},
  \href{http://arxiv.org/abs/1708.09393}{{\ttfamily arXiv:1708.09393
  [hep-th]}}.

\bibitem{Nguyen:2017yqw}
P.~Nguyen, T.~Devakul, M.~G. Halbasch, M.~P. Zaletel, and B.~Swingle,
  ``{Entanglement of purification: from spin chains to holography},''
  \href{http://dx.doi.org/10.1007/JHEP01(2018)098}{{\em JHEP} {\bfseries 01}
  (2018) 098}, \href{http://arxiv.org/abs/1709.07424}{{\ttfamily
  arXiv:1709.07424 [hep-th]}}.

\bibitem{Wen:2021qgx}
Q.~Wen, ``{Balanced Partial Entanglement and the Entanglement Wedge Cross
  Section},'' \href{http://dx.doi.org/10.1007/JHEP04(2021)301}{{\em JHEP}
  {\bfseries 04} (2021) 301}, \href{http://arxiv.org/abs/2103.00415}{{\ttfamily
  arXiv:2103.00415 [hep-th]}}.

\bibitem{Li:2020ceg}
T.~Li, J.~Chu, and Y.~Zhou, ``{Reflected Entropy for an Evaporating Black
  Hole},'' \href{http://dx.doi.org/10.1007/JHEP11(2020)155}{{\em JHEP}
  {\bfseries 11} (2020) 155}, \href{http://arxiv.org/abs/2006.10846}{{\ttfamily
  arXiv:2006.10846 [hep-th]}}.

\bibitem{Chandrasekaran:2020qtn}
V.~Chandrasekaran, M.~Miyaji, and P.~Rath, ``{Including contributions from
  entanglement islands to the reflected entropy},''
  \href{http://dx.doi.org/10.1103/PhysRevD.102.086009}{{\em Phys. Rev. D}
  {\bfseries 102} no.~8, (2020) 086009},
  \href{http://arxiv.org/abs/2006.10754}{{\ttfamily arXiv:2006.10754
  [hep-th]}}.

\bibitem{Jeong:2019xdr}
H.-S. Jeong, K.-Y. Kim, and M.~Nishida, ``{Reflected Entropy and Entanglement
  Wedge Cross Section with the First Order Correction},''
  \href{http://dx.doi.org/10.1007/JHEP12(2019)170}{{\em JHEP} {\bfseries 12}
  (2019) 170}, \href{http://arxiv.org/abs/1909.02806}{{\ttfamily
  arXiv:1909.02806 [hep-th]}}.

\bibitem{KumarBasak:2020ams}
J.~Kumar~Basak, D.~Basu, V.~Malvimat, H.~Parihar, and G.~Sengupta, ``{Islands
  for entanglement negativity},''
  \href{http://dx.doi.org/10.21468/SciPostPhys.12.1.003}{{\em SciPost Phys.}
  {\bfseries 12} no.~1, (2022) 003},
  \href{http://arxiv.org/abs/2012.03983}{{\ttfamily arXiv:2012.03983
  [hep-th]}}.

\bibitem{Zamolodchikov:2004ce}
A.~B. Zamolodchikov, ``{Expectation value of composite field T anti-T in
  two-dimensional quantum field theory},''
  \href{http://arxiv.org/abs/hep-th/0401146}{{\ttfamily arXiv:hep-th/0401146}}.

\bibitem{Cavaglia:2016oda}
A.~Cavagli\`a, S.~Negro, I.~M. Sz\'ecs\'enyi, and R.~Tateo, ``{$T
  \bar{T}$-deformed 2D Quantum Field Theories},''
  \href{http://dx.doi.org/10.1007/JHEP10(2016)112}{{\em JHEP} {\bfseries 10}
  (2016) 112}, \href{http://arxiv.org/abs/1608.05534}{{\ttfamily
  arXiv:1608.05534 [hep-th]}}.

\bibitem{Smirnov:2016lqw}
F.~A. Smirnov and A.~B. Zamolodchikov, ``{On space of integrable quantum field
  theories},'' \href{http://dx.doi.org/10.1016/j.nuclphysb.2016.12.014}{{\em
  Nucl. Phys. B} {\bfseries 915} (2017) 363--383},
  \href{http://arxiv.org/abs/1608.05499}{{\ttfamily arXiv:1608.05499
  [hep-th]}}.

\bibitem{McGough:2016lol}
L.~McGough, M.~Mezei, and H.~Verlinde, ``{Moving the CFT into the bulk with $
  T\overline{T} $},'' \href{http://dx.doi.org/10.1007/JHEP04(2018)010}{{\em
  JHEP} {\bfseries 04} (2018) 010},
  \href{http://arxiv.org/abs/1611.03470}{{\ttfamily arXiv:1611.03470
  [hep-th]}}.

\bibitem{Hirano:2020nwq}
S.~Hirano and M.~Shigemori, ``{Random boundary geometry and gravity dual of $
  T\overline{T} $ deformation},''
  \href{http://dx.doi.org/10.1007/JHEP11(2020)108}{{\em JHEP} {\bfseries 11}
  (2020) 108}, \href{http://arxiv.org/abs/2003.06300}{{\ttfamily
  arXiv:2003.06300 [hep-th]}}.

\bibitem{Kraus:2018xrn}
P.~Kraus, J.~Liu, and D.~Marolf, ``{Cutoff AdS$_{3}$ versus the $ T\overline{T}
  $ deformation},'' \href{http://dx.doi.org/10.1007/JHEP07(2018)027}{{\em JHEP}
  {\bfseries 07} (2018) 027}, \href{http://arxiv.org/abs/1801.02714}{{\ttfamily
  arXiv:1801.02714 [hep-th]}}.

\bibitem{Guica:2019nzm}
M.~Guica and R.~Monten, ``{$T\bar T$ and the mirage of a bulk cutoff},''
  \href{http://dx.doi.org/10.21468/SciPostPhys.10.2.024}{{\em SciPost Phys.}
  {\bfseries 10} no.~2, (2021) 024},
  \href{http://arxiv.org/abs/1906.11251}{{\ttfamily arXiv:1906.11251
  [hep-th]}}.

\bibitem{Dubovsky:2018bmo}
S.~Dubovsky, V.~Gorbenko, and G.~Hern\'andez-Chifflet, ``{$ T\overline{T} $
  partition function from topological gravity},''
  \href{http://dx.doi.org/10.1007/JHEP09(2018)158}{{\em JHEP} {\bfseries 09}
  (2018) 158}, \href{http://arxiv.org/abs/1805.07386}{{\ttfamily
  arXiv:1805.07386 [hep-th]}}.

\bibitem{Tolley:2019nmm}
A.~J. Tolley, ``{$ T\overline{T} $ deformations, massive gravity and
  non-critical strings},''
  \href{http://dx.doi.org/10.1007/JHEP06(2020)050}{{\em JHEP} {\bfseries 06}
  (2020) 050}, \href{http://arxiv.org/abs/1911.06142}{{\ttfamily
  arXiv:1911.06142 [hep-th]}}.

\bibitem{Shyam:2017znq}
V.~Shyam, ``{Background independent holographic dual to $T\bar{T}$ deformed CFT
  with large central charge in 2 dimensions},''
  \href{http://dx.doi.org/10.1007/JHEP10(2017)108}{{\em JHEP} {\bfseries 10}
  (2017) 108}, \href{http://arxiv.org/abs/1707.08118}{{\ttfamily
  arXiv:1707.08118 [hep-th]}}.

\bibitem{Cottrell:2018skz}
W.~Cottrell and A.~Hashimoto, ``{Comments on $T \bar T$ double trace
  deformations and boundary conditions},''
  \href{http://dx.doi.org/10.1016/j.physletb.2018.09.068}{{\em Phys. Lett. B}
  {\bfseries 789} (2019) 251--255},
  \href{http://arxiv.org/abs/1801.09708}{{\ttfamily arXiv:1801.09708
  [hep-th]}}.

\bibitem{Taylor:2018xcy}
M.~Taylor, ``{TT deformations in general dimensions},''
  \href{http://arxiv.org/abs/1805.10287}{{\ttfamily arXiv:1805.10287
  [hep-th]}}.

\bibitem{Hartman:2018tkw}
T.~Hartman, J.~Kruthoff, E.~Shaghoulian, and A.~Tajdini, ``{Holography at
  finite cutoff with a $T^2$ deformation},''
  \href{http://dx.doi.org/10.1007/JHEP03(2019)004}{{\em JHEP} {\bfseries 03}
  (2019) 004}, \href{http://arxiv.org/abs/1807.11401}{{\ttfamily
  arXiv:1807.11401 [hep-th]}}.

\bibitem{Shyam:2018sro}
V.~Shyam, ``{Finite Cutoff AdS$_{5}$ Holography and the Generalized Gradient
  Flow},'' \href{http://dx.doi.org/10.1007/JHEP12(2018)086}{{\em JHEP}
  {\bfseries 12} (2018) 086}, \href{http://arxiv.org/abs/1808.07760}{{\ttfamily
  arXiv:1808.07760 [hep-th]}}.

\bibitem{Caputa:2019pam}
P.~Caputa, S.~Datta, and V.~Shyam, ``{Sphere partition functions
  \textbackslash{}\& cut-off AdS},''
  \href{http://dx.doi.org/10.1007/JHEP05(2019)112}{{\em JHEP} {\bfseries 05}
  (2019) 112}, \href{http://arxiv.org/abs/1902.10893}{{\ttfamily
  arXiv:1902.10893 [hep-th]}}.

\bibitem{Giveon:2017myj}
A.~Giveon, N.~Itzhaki, and D.~Kutasov, ``{A solvable irrelevant deformation of
  AdS$_{3}$/CFT$_{2}$},'' \href{http://dx.doi.org/10.1007/JHEP12(2017)155}{{\em
  JHEP} {\bfseries 12} (2017) 155},
  \href{http://arxiv.org/abs/1707.05800}{{\ttfamily arXiv:1707.05800
  [hep-th]}}.

\bibitem{Asrat:2017tzd}
M.~Asrat, A.~Giveon, N.~Itzhaki, and D.~Kutasov, ``{Holography Beyond AdS},''
  \href{http://dx.doi.org/10.1016/j.nuclphysb.2018.05.005}{{\em Nucl. Phys. B}
  {\bfseries 932} (2018) 241--253},
  \href{http://arxiv.org/abs/1711.02690}{{\ttfamily arXiv:1711.02690
  [hep-th]}}.

\bibitem{Donnelly:2018bef}
W.~Donnelly and V.~Shyam, ``{Entanglement entropy and $T \overline{T}$
  deformation},'' \href{http://dx.doi.org/10.1103/PhysRevLett.121.131602}{{\em
  Phys. Rev. Lett.} {\bfseries 121} no.~13, (2018) 131602},
  \href{http://arxiv.org/abs/1806.07444}{{\ttfamily arXiv:1806.07444
  [hep-th]}}.

\bibitem{Lewkowycz:2019xse}
A.~Lewkowycz, J.~Liu, E.~Silverstein, and G.~Torroba, ``{$ T\overline{T} $ and
  EE, with implications for (A)dS subregion encodings},''
  \href{http://dx.doi.org/10.1007/JHEP04(2020)152}{{\em JHEP} {\bfseries 04}
  (2020) 152}, \href{http://arxiv.org/abs/1909.13808}{{\ttfamily
  arXiv:1909.13808 [hep-th]}}.

\bibitem{Chen:2018eqk}
B.~Chen, L.~Chen, and P.-X. Hao, ``{Entanglement entropy in
  $T\overline{T}$-deformed CFT},''
  \href{http://dx.doi.org/10.1103/PhysRevD.98.086025}{{\em Phys. Rev. D}
  {\bfseries 98} no.~8, (2018) 086025},
  \href{http://arxiv.org/abs/1807.08293}{{\ttfamily arXiv:1807.08293
  [hep-th]}}.

\bibitem{Banerjee:2019ewu}
A.~Banerjee, A.~Bhattacharyya, and S.~Chakraborty, ``{Entanglement Entropy for
  $TT$ deformed CFT in general dimensions},''
  \href{http://dx.doi.org/10.1016/j.nuclphysb.2019.114775}{{\em Nucl. Phys. B}
  {\bfseries 948} (2019) 114775},
  \href{http://arxiv.org/abs/1904.00716}{{\ttfamily arXiv:1904.00716
  [hep-th]}}.

\bibitem{Jeong:2019ylz}
H.-S. Jeong, K.-Y. Kim, and M.~Nishida, ``{Entanglement and R\'enyi entropy of
  multiple intervals in $T\overline{T}$-deformed CFT and holography},''
  \href{http://dx.doi.org/10.1103/PhysRevD.100.106015}{{\em Phys. Rev. D}
  {\bfseries 100} no.~10, (2019) 106015},
  \href{http://arxiv.org/abs/1906.03894}{{\ttfamily arXiv:1906.03894
  [hep-th]}}.

\bibitem{Murdia:2019fax}
C.~Murdia, Y.~Nomura, P.~Rath, and N.~Salzetta, ``{Comments on holographic
  entanglement entropy in $TT$ deformed conformal field theories},''
  \href{http://dx.doi.org/10.1103/PhysRevD.100.026011}{{\em Phys. Rev. D}
  {\bfseries 100} no.~2, (2019) 026011},
  \href{http://arxiv.org/abs/1904.04408}{{\ttfamily arXiv:1904.04408
  [hep-th]}}.

\bibitem{Park:2018snf}
C.~Park, ``{Holographic Entanglement Entropy in Cutoff AdS},''
  \href{http://dx.doi.org/10.1142/S0217751X18502263}{{\em Int. J. Mod. Phys. A}
  {\bfseries 33} no.~36, (2019) 1850226},
  \href{http://arxiv.org/abs/1812.00545}{{\ttfamily arXiv:1812.00545
  [hep-th]}}.

\bibitem{Asrat:2019end}
M.~Asrat, ``{Entropic $c$\textendash{}functions in $T{\bar T}, J{\bar T},
  T{\bar J}$ deformations},''
  \href{http://dx.doi.org/10.1016/j.nuclphysb.2020.115186}{{\em Nucl. Phys. B}
  {\bfseries 960} (2020) 115186},
  \href{http://arxiv.org/abs/1911.04618}{{\ttfamily arXiv:1911.04618
  [hep-th]}}.

\bibitem{He:2019vzf}
S.~He and H.~Shu, ``{Correlation functions, entanglement and chaos in the $
  T\overline{T}/J\overline{T} $-deformed CFTs},''
  \href{http://dx.doi.org/10.1007/JHEP02(2020)088}{{\em JHEP} {\bfseries 02}
  (2020) 088}, \href{http://arxiv.org/abs/1907.12603}{{\ttfamily
  arXiv:1907.12603 [hep-th]}}.

\bibitem{Grieninger:2019zts}
S.~Grieninger, ``{Entanglement entropy and $ T\overline{T} $ deformations
  beyond antipodal points from holography},''
  \href{http://dx.doi.org/10.1007/JHEP11(2019)171}{{\em JHEP} {\bfseries 11}
  (2019) 171}, \href{http://arxiv.org/abs/1908.10372}{{\ttfamily
  arXiv:1908.10372 [hep-th]}}.

\bibitem{Khoeini-Moghaddam:2020ymm}
S.~Khoeini-Moghaddam, F.~Omidi, and C.~Paul, ``{Aspects of Hyperscaling
  Violating Geometries at Finite Cutoff},''
  \href{http://dx.doi.org/10.1007/JHEP02(2021)121}{{\em JHEP} {\bfseries 02}
  (2021) 121}, \href{http://arxiv.org/abs/2011.00305}{{\ttfamily
  arXiv:2011.00305 [hep-th]}}.

\bibitem{Basu:2024bal}
D.~Basu and V.~Raj, ``{Reflected entropy and timelike entanglement in
  $\textrm{T}\bar{\textrm{T}}$ deformed CFT$_2$s},''
  \href{http://arxiv.org/abs/2402.07253}{{\ttfamily arXiv:2402.07253
  [hep-th]}}.

\bibitem{Basu:2023aqz}
D.~Basu, S.~Biswas, A.~Dey, B.~Paul, and G.~Sengupta, ``{Odd entanglement
  entropy in TT\textasciimacron{} deformed CFT2s and holography},''
  \href{http://dx.doi.org/10.1103/PhysRevD.108.126013}{{\em Phys. Rev. D}
  {\bfseries 108} no.~12, (2023) 126013},
  \href{http://arxiv.org/abs/2307.04832}{{\ttfamily arXiv:2307.04832
  [hep-th]}}.

\bibitem{Basu:2023bov}
D.~Basu, Lavish, and B.~Paul, ``{Entanglement negativity in
  $\text{T}\bar{\text{T}}$-deformed CFT$_2$s},''
  \href{http://dx.doi.org/10.1103/PhysRevD.107.126026}{{\em Phys. Rev. D}
  {\bfseries 107} no.~12, (2023) 126026},
  \href{http://arxiv.org/abs/2302.11435}{{\ttfamily arXiv:2302.11435
  [hep-th]}}.

\bibitem{Basu:2024enr}
D.~Basu and S.~Biswas, ``{Entanglement, $\textrm{T}\bar{\textrm{T}}$ and
  rotating black holes},'' \href{http://arxiv.org/abs/2410.06363}{{\ttfamily
  arXiv:2410.06363 [hep-th]}}.

\bibitem{Grieninger:2023knz}
S.~Grieninger, K.~Ikeda, and D.~E. Kharzeev, ``{Temporal entanglement entropy
  as a probe of renormalization group flow},''
  \href{http://dx.doi.org/10.1007/JHEP05(2024)030}{{\em JHEP} {\bfseries 05}
  (2024) 030}, \href{http://arxiv.org/abs/2312.08534}{{\ttfamily
  arXiv:2312.08534 [hep-th]}}.

\bibitem{Chang:2024voo}
J.-C. Chang, S.~He, Y.-X. Liu, and L.~Zhao, ``{The holographic $T\bar{T}$
  deformation of the entanglement entropy in (A)dS$_3$/CFT$_2$},''
  \href{http://arxiv.org/abs/2409.08198}{{\ttfamily arXiv:2409.08198
  [hep-th]}}.

\bibitem{Deng:2023pjs}
F.~Deng, Z.~Wang, and Y.~Zhou, ``{End of the world brane meets $ T\overline{T}
  $},'' \href{http://dx.doi.org/10.1007/JHEP07(2024)036}{{\em JHEP} {\bfseries
  07} (2024) 036}, \href{http://arxiv.org/abs/2310.15031}{{\ttfamily
  arXiv:2310.15031 [hep-th]}}.

\bibitem{Deng:2024dct}
F.~Deng and Z.~Wang, ``{Holographic boundary conformal field theory with $T\bar
  T$ deformation},'' \href{http://arxiv.org/abs/2411.06345}{{\ttfamily
  arXiv:2411.06345 [hep-th]}}.

\bibitem{Aharony:2010ay}
O.~Aharony, D.~Marolf, and M.~Rangamani, ``{Conformal field theories in anti-de
  Sitter space},'' \href{http://dx.doi.org/10.1007/JHEP02(2011)041}{{\em JHEP}
  {\bfseries 02} (2011) 041}, \href{http://arxiv.org/abs/1011.6144}{{\ttfamily
  arXiv:1011.6144 [hep-th]}}.

\bibitem{Karch:2000ct}
A.~Karch and L.~Randall, ``{Locally localized gravity},''
  \href{http://dx.doi.org/10.1088/1126-6708/2001/05/008}{{\em JHEP} {\bfseries
  05} (2001) 008}, \href{http://arxiv.org/abs/hep-th/0011156}{{\ttfamily
  arXiv:hep-th/0011156}}.

\bibitem{Randall:1999vf}
L.~Randall and R.~Sundrum, ``{An Alternative to compactification},''
  \href{http://dx.doi.org/10.1103/PhysRevLett.83.4690}{{\em Phys. Rev. Lett.}
  {\bfseries 83} (1999) 4690--4693},
  \href{http://arxiv.org/abs/hep-th/9906064}{{\ttfamily arXiv:hep-th/9906064}}.

\bibitem{Maldacena:1997re}
J.~M. Maldacena, ``{The Large N limit of superconformal field theories and
  supergravity},'' \href{http://dx.doi.org/10.1023/A:1026654312961}{{\em Adv.
  Theor. Math. Phys.} {\bfseries 2} (1998) 231--252},
  \href{http://arxiv.org/abs/hep-th/9711200}{{\ttfamily arXiv:hep-th/9711200}}.

\bibitem{Brown:1986nw}
J.~D. Brown and M.~Henneaux, ``{Central Charges in the Canonical Realization of
  Asymptotic Symmetries: An Example from Three-Dimensional Gravity},''
  \href{http://dx.doi.org/10.1007/BF01211590}{{\em Commun. Math. Phys.}
  {\bfseries 104} (1986) 207--226}.

\bibitem{Kusuki:2019evw}
Y.~Kusuki and K.~Tamaoka, ``{Entanglement Wedge Cross Section from CFT:
  Dynamics of Local Operator Quench},''
  \href{http://dx.doi.org/10.1007/JHEP02(2020)017}{{\em JHEP} {\bfseries 02}
  (2020) 017}, \href{http://arxiv.org/abs/1909.06790}{{\ttfamily
  arXiv:1909.06790 [hep-th]}}.

\bibitem{Akers:2022max}
C.~Akers, T.~Faulkner, S.~Lin, and P.~Rath, ``{The Page Curve for Reflected
  Entropy},'' \href{http://arxiv.org/abs/2201.11730}{{\ttfamily
  arXiv:2201.11730 [hep-th]}}.

\bibitem{Fitzpatrick:2014vua}
A.~L. Fitzpatrick, J.~Kaplan, and M.~T. Walters, ``{Universality of
  Long-Distance AdS Physics from the CFT Bootstrap},''
  \href{http://dx.doi.org/10.1007/JHEP08(2014)145}{{\em JHEP} {\bfseries 08}
  (2014) 145}, \href{http://arxiv.org/abs/1403.6829}{{\ttfamily arXiv:1403.6829
  [hep-th]}}.

\bibitem{Hartman:2013qma}
T.~Hartman and J.~Maldacena, ``{Time Evolution of Entanglement Entropy from
  Black Hole Interiors},''
  \href{http://dx.doi.org/10.1007/JHEP05(2013)014}{{\em JHEP} {\bfseries 05}
  (2013) 014}, \href{http://arxiv.org/abs/1303.1080}{{\ttfamily arXiv:1303.1080
  [hep-th]}}.

\end{thebibliography}\endgroup
	
\end{document}